# *Colloquium:* Quantum information processing with ensembles of identical particles

Matteo Piccolini,[1,2] Marcin Karczewski,[3] Jin-Shi Xu,[4,5,6] Chuan-Feng Li,[4,5,6] Guang-Can Guo,[4,5,6] Giuseppe Compagno,[7] Farzam Nosrati,[1,8,9] and Rosario Lo Franco[1]

[1] *Dipartimento di Ingegneria, Università degli Studi di Palermo, Viale delle Scienze, 90128 Palermo, Italy*
[2] *Dipartimento di Fisica, Università di Pavia, Via Bassi 6, 1-27100, Pavia, Italy*
[3] *Institute of Spintronics and Quantum Information, Faculty of Physics, Adam Mickiewicz University, 61-614 Poznań, Poland*
[4] *Laboratory of Quantum Information, University of Science and Technology of China, Hefei, Anhui 230026, China*
[5] *Anhui Province Key Laboratory of Quantum Network, University of Science and Technology of China, Hefei, Anhui 230026, China*
[6] *Hefei National Laboratory, University of Science and Technology of China, Hefei, Anhui 230088, China*
[7] *Dipartimento di Fisica e Chimica - Emilio Segrè, Università di Palermo, via Archirafi 36, 90123 Palermo, Italy*
[8] *IMDEA Networks Institute, Madrid, Spain*
[9] *Nexus for Quantum Technologies, University of Ottawa, K1N 5N6, Ottawa, ON, Canada*

Identical particles underpin most quantum technologies, but the role of indistinguishability in these processes extends far beyond the symmetrization of many-particle states. This Colloquium presents a unified operational perspective on how particle identity, indistinguishability, exchange statistics, and many-particle interference can instead be harnessed as resources for quantum information processing. We first clarify the notions of entanglement and indistinguishability for identical particles, emphasizing their dependence on the accessible measurements and information. We then review how spatial overlap, multiparticle interference, and path identity enable the generation, extraction, protection, and distillation of coherence and entanglement. Their applications to quantum metrology and imaging, state discrimination, teleportation, entanglement swapping, and quantum cloning are examined. Finally, we discuss the thermodynamic role of indistinguishability, from the Gibbs paradox to quantum thermal machines and statistics-driven engines. Across these settings, indistinguishability emerges as a controllable, task-dependent resource whose benefits are determined by state preparation, exchange statistics, interference, and measurement constraints.

## Contents

## I. INTRODUCTORY ASPECTS ON PARTICLE IDENTITY

Quantum information processing (QIP) is one of the forefronts of modern physics (Spiller *et al.*, 2005). Its core idea, consisting in exploiting inherent quantum effects in information processing tasks, offers the premise of technological breakthroughs and major advancements in fields such as computation (Cerezo *et al.*, 2021; Ladd *et al.*, 2010; Memon *et al.*, 2024; Preskill, 2018; Steane, 1998), simulation (Altman *et al.*, 2021; Cornish *et al.*, 2024; Daley *et al.*, 2022; Fauseweh, 2024; Georgescu *et al.*, 2014), communication (Gisin and Thew, 2007; Sangouard *et al.*, 2011; Sidhu *et al.*, 2021; Wehner *et al.*, 2018), cryptography (Gisin *et al.*, 2002; Pirandola *et al.*, 2020), metrology (Braun *et al.*, 2018; Giovannetti *et al.*, 2004, 2006, 2011; Paris, 2009; Pezze *et al.*, 2018; Polino *et al.*, 2020), sensing (Degen *et al.*, 2017; DeMille *et al.*, 2024; Pirandola *et al.*, 2018), and thermodynamics (Campbell *et al.*, 2026; Goold *et al.*, 2016b; Kosloff, 2013; Vinjanampathy and Anders, 2016). The interest towards the potential practical advantages of quantum technologies has led to the fast development of different physical platforms to host QIP protocols, each characterized by distinctive properties. These include, but are not limited to, superconducting circuits (Krantz *et al.*, 2019; Wendin, 2017), photonic setups (Flamini *et al.*, 2018; Giordani *et al.*, 2023), neutral atoms (Henriet *et al.*, 2020), trapped ions (Bruzewicz *et al.*, 2019; Strohm *et al.*, 2024; Wineland *et al.*, 2003), solid state devices (Burkard *et al.*, 2023; Cory *et al.*, 2000; Sato *et al.*, 2009; Yao *et al.*, 2012), and hybrid configurations (Monroe, 2002; Xu *et al.*, 2022).

Various quantum phenomena have been identified as resources in this effort to develop more efficient QIP schemes, most notably entanglement (Horodecki *et al.*, 2009; Laflorencie, 2016; Wootters, 1998b) and coherence (Streltsov *et al.*, 2017, 2015). Crucially, the majority of such protocols involves the manipulation of ensembles of *identical* particles, either bosons or fermions. In contrast to nonidentical ones, identical constituents are allowed by the laws of quantum mechanics to be *indistinguishable*, a property which plays a central role in many QIP protocols.

From a historical point of view, the terms *identical particles* and *indistinguishable particles* have been used with a variety of different meanings, leading to confusion between the two concepts and blurring the relation between them (De Muynck, 1975). Currently, a general consensus seems to have been established about *particle identity*, with identical constituents being defined as *particles characterized by the same intrinsic properties* such as mass, spin, and charge (Cohen-Tannoudji *et al.*, 2006; Huggett and Imbo, 2009; Paunkovic, 2004). Nonetheless, different definitions of *particle indistinguishability* are still in use, sometimes addressing only the single-particle state of the constituents involved, and sometimes exploring the relation between the system and the employed measurement setup. In this *Colloquium*, we discuss different meanings of indistinguishability, though paying particular attention to its measurement-based treatment.

In measurement-based connotation, particle indistinguishability can be achieved by manipulating identical constituents in proper measurement settings and used to access quantum resources such as coherence and entanglement (Castellini *et al.*, 2019; Killoran *et al.*, 2014; Lo Franco and Compagno, 2016, 2018; Piccolini *et al.*, 2023b; Tichy *et al.*, 2013a). Furthermore, the symmetry properties of identical particle systems can cause the related many-body probability amplitudes to interfere during the system's dynamics, linking indistinguishability and interference (Barzel and Lämmerzahl, 2023; Piccolini, 2024; Tichy, 2014).

Interferometric and indistinguishability effects arising from properly tailoring the dynamics of identical entities are the basis of many measurement-based entanglement generation protocols exploiting either postselection or heralding (Barros *et al.*, 2020; Chin *et al.*, 2024a,b; Dudarev *et al.*, 2003; Fiurášek, 2002; Hong *et al.*, 1987; Karczewski *et al.*, 2019; Killoran *et al.*, 2014; Lee *et al.*, 2002a; Lo Franco and Compagno, 2018; Mahdavipour *et al.*, 2024), sometimes additionally providing correlation protection from noisy environments (Ferrara *et al.*, 2026; Nosrati *et al.*, 2024, 2020; Perez-Leija *et al.*, 2018; Piccolini, 2024; Piccolini *et al.*, 2023a, 2024a,b, 2023b, 2021; Siltanen *et al.*, 2021). Moreover, such peculiar phenomena are frequently employed in protocols to achieve practical tasks in quantum metrology (Chen *et al.*, 2022; Su *et al.*, 2017; Sun *et al.*, 2022; Triggiani *et al.*, 2023), communication (Mattle *et al.*, 1996; Stahlke, 2014; Sun *et al.*, 2020), computing (Brod *et al.*, 2019a; Hamilton *et al.*, 2017; Knill *et al.*, 2001; Kok *et al.*, 2007; Laibacher and Tamma, 2015; Tillmann *et al.*, 2013), in schemes to investigate the foundations of quantum mechanics (Dittel *et al.*, 2018a,b; Müntinga *et al.*, 2013; Pittman *et al.*, 1996; Tschernig *et al.*, 2021; Wang *et al.*, 2022a), and find further potential applications in quantum thermodynamics (Ferreri *et al.*, 2025; Gherardini and De Chiara, 2024; Hernández-Gómez *et al.*, 2024; Holmes *et al.*, 2020a,b; Jarzyna and Zwierz, 2015; Koch *et al.*, 2023; Latune *et al.*, 2019b; Marzolino, 2024; Myers and Deffner, 2020; Myers *et al.*, 2022; Plesch *et al.*, 2014; Yadin *et al.*, 2021).

Although the foundational aspects characterizing identical particle systems have been widely discussed in the literature (Benatti *et al.*, 2020; Caulton, 2013; Compagno *et al.*, 2018; Ghirardi and Marinatto, 2004; Leinaas and Myrheim, 1977; Lo Franco and Compagno, 2016; Tichy, 2014), the exploitation of the related indistinguishability and interferometric effects in practical applications is often overlooked. In this *Colloquium*, we aim at filling this gap by discussing how such characteristics can be exploited for QIP, fueled by the emergence of experimental platforms offering unprecedented control over multi-particle states such as quantum processors and simulators. In view of these developments, we shall present the main experimental applications, either consolidated or potential, resulting from these intrinsic quantum traits.

The entanglement of identical subsystems has been widely debated and studied in recent years, leading to the introduction of different approaches to deal with it. In Section II, we review the essential aspects behind this concept, highlighting the crucial dependence of entanglement on the operational framework considered. In Section III, we discuss the main schemes for the generation, extraction, distillation, or recovery of useful multipartite quantum states for QIP via the manipulation of identical particles. In Section IV, we elaborate on the role of identical particle interferometry and indistinguishability in quantum metrology, where identical constituents are typically used as probe states in quantum-enhanced detection and sensing protocols. In Section V, we review how identical elements are utilized in relevant quantum communication protocols, including quantum teleportation, entanglement swapping, and optimal quantum cloning. In Section VI, we present the crucial role that particle identity and indistinguishability play in the most relevant quantum thermodynamic processes and machines. In Section VII, we conclude by discussing how the theoretical study of the above mentioned topics is strategic for the scientific progress, as they impact both the advancement of our basic understanding of the quantum world and the development of quantum-enhanced technologies.

## II. IDENTICAL PARTICLES: ENTANGLEMENT, INDISTINGUISHABILITY, INTERFEROMETRY

This *Colloquium* is dedicated to the discussion of practical QIP schemes with ensembles of identical particles. However, we find it useful to begin with a discursive theoretical introduction to the central concepts underlying the manipulation of identical particles, thereby laying the foundation for the subsequent sections.

### A. Entanglement in identical particle systems

We first briefly discuss the differences in treating entanglement between identical constituents with respect to nonidentical ones.

#### 1. The symmetrization postulate: is everything entangled?

The standard first quantization approach to quantum mechanics deals with identical particles by assigning to each constituent composing the system an individual label by hand. However, since all the involved entities are characterized by the same intrinsic properties, there is no preferred way to attribute such labels. As a consequence, the same physical multiparticle state is identified by different equivalent kets (state vectors), which differ from each other only in the labeling choice. However, such kets may be orthogonal with respect to each other, leading to different physical predictions. To eliminate such a redundancy, known as *exchange degeneracy*, the standard approach introduces the *symmetrization postulate*, stating that: "Physical kets are, depending on the nature of the identical particles, either completely symmetric or completely antisymmetric with respect to permutation of these particles. Those particles for which the physical kets are symmetric are called bosons, and those for which they are antisymmetric, fermions" (Cohen-Tannoudji *et al.*, 2006). Nonetheless, a straightforward application of the symmetrization postulate leads to problematic conclusions when trying to evaluate the entanglement of identical particle systems using the same tools employed for nonidentical ones. Consider, for example, a bipartite system of two identical spinless particles, one in the state $|\varphi\rangle$ and the other in the state $|\chi\rangle$. According to the symmetrization postulate, the global state is given by

$$|\Psi^{(2)}\rangle = \frac{1}{\sqrt{2}}\Big( |\varphi\rangle_A \otimes |\chi\rangle_B + \eta\ |\chi\rangle_A \otimes |\varphi\rangle_B \Big), \qquad (1)$$

where $A$ and $B$ are the labels mentioned above, while $\eta = 1$ for bosons and $\eta = -1$ for fermions. We immediately notice that, unless we are dealing with identical bosons in the same single-particle state ($|\varphi\rangle = |\chi\rangle$), this approach makes a many-body state never separable. So, the overall state looks always formally entangled according to the typical entanglement evaluation criteria employed for nonidentical constituents, such as those relying on the Schmidt number or the von Neumann entropy (Horodecki *et al.*, 2009). Luckily, some considerations prevent us from accepting this result blindly and to conclude that "all identical particles are inherently correlated from the outset, regardless of how far apart their creation took place" (Blasiak and Markiewicz, 2019).

#### 2. Avoiding the symmetrization postulate: ambiguous and unambiguous measurement settings

Consider two spin-1/2 particles, one with spatial wave function $\psi_1$ and spin $\sigma_1$ and the other with spatial wave function $\psi_2$ and spin $\sigma_2$. Their properly symmetrized bipartite state is

$$|\Psi^{(2)}\rangle = \frac{|\psi_1\,\sigma_1\rangle_A \otimes |\psi_2\,\sigma_2\rangle_B + \eta\ |\psi_2\,\sigma_2\rangle_A \otimes |\psi_1\,\sigma_1\rangle_B}{\sqrt{2}}. \qquad (2)$$

We now take a simultaneous measurement of both particles performed via two distinct local measurement devices, probing non-overlapping regions of space. It can be shown that the application of the symmetrization postulate does not introduce any additional contribution to the two-particle probability amplitudes if the two-particle measurement falls within the following scenarios (Cohen-Tannoudji *et al.*, 2006; Piccolini, 2024):

1. there is at least one detection region where it is impossible to find one of the two particles;
2. the spin of at least one of the two particles is orthogonal to the one probed by one of the detectors.

In scenario 1, the impossibility to measure one of the two constituents in a specific region allows to ascribe the detection in such a region to the other constituent. Scenario 2 behaves analogously, ruling out one constituent due to its spin state being orthogonal to the probed one. The common denominator of the two cases is the possibility to associate each detector with exactly one of the particles *without any ambiguity*: for this reason, these scenarios are deemed *unambiguous settings* (Tichy *et al.*, 2013a). In contrast, situations where this is not possible are said *ambiguous* (see Fig. 1): the related impossibility of identifying which constituent has triggered which detector constitutes a lack of *which-particle information* (Piccolini, 2024), and provides an initial, later refined definition of *spatial indistinguishability* for the particles involved (here, the indistinguishability is meant *to the eyes of the detectors*).

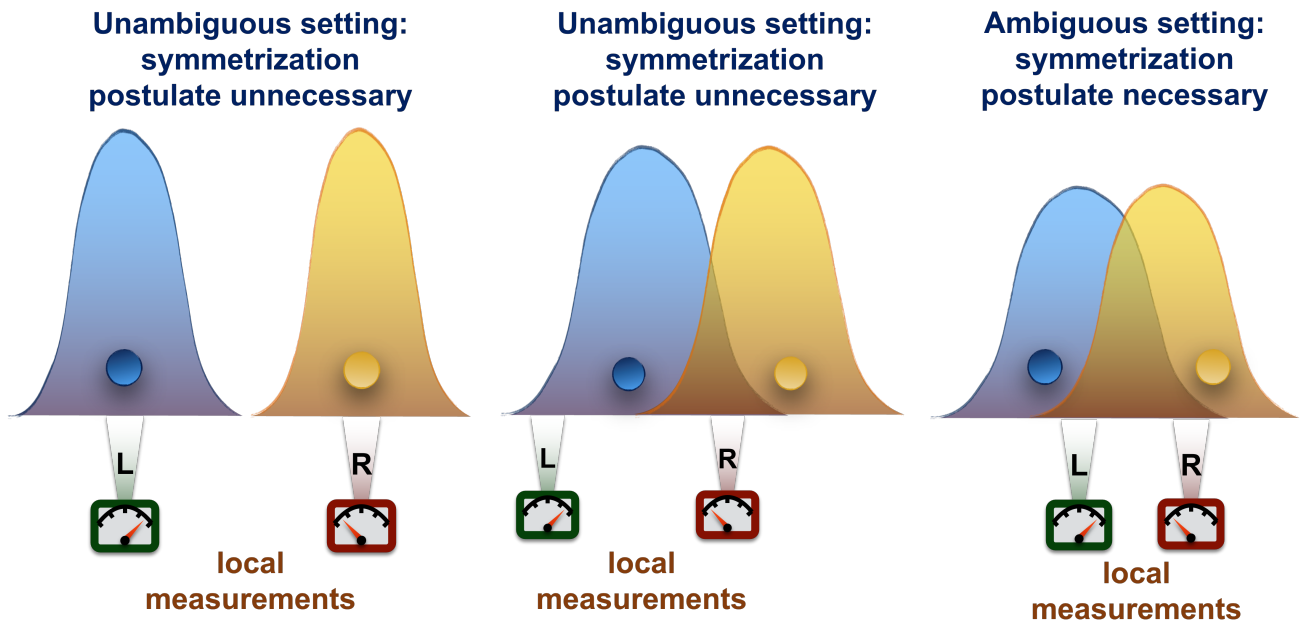

FIG. 1: Unambiguous and ambiguous settings according to scenario 1, discussed in the main text, and their relation with the symmetrization postulate.

In unambiguous settings it is therefore possible to assign a physical meaning to the first-quantization labels: we could, for example, assign label $A$ to "the only particle detectable in that specific region" or to "the only particle detectable by that specific detector due to its internal state". This operation assigns to each constituent an individual identity, allowing to effectively treat them as nonidentical particles within the standard *local operations and classical communication* (LOCC) paradigm (Chitambar *et al.*, 2014; Horodecki *et al.*, 2009),where the term *local* denotes *particle locality*. In such unambiguous situations, the symmetrization postulate is therefore unnecessary, in the sense that we can obtain the same results for probability amplitudes by directly reasoning as if the labels denote constituents of different natures. The apparent entanglement binding all identical particles in the universe is therefore deemed unphysical, as only due to the unnecessary redundancy of the label formalism (Tichy *et al.*, 2013a, 2011), allowing to conclude that "the existence of identical particles does not prevent the separate study of restricted systems, composed of a small number of particles" (Cohen-Tannoudji *et al.*, 2006), since "no quantum prediction, referring to an atom located in our laboratory, is affected by the mere presence of similar atoms in remote parts of the universe" (Peres, 2002).

We shall see in Section II.B.3 how interference, one of the most peculiar effects of identical particles, hinders the given description of ambiguous settings in terms of no which-particle information by making it ambiguous itself in certain scenarios, consequently affecting the definition of spatial indistinguishability with the same flaw. We shall then discuss how a more accurate analysis of ambiguous settings leads to a shift in the point of view and to a refinement of the definition of ambiguous measurements and of spatial indistinguishability, fixing this problem and providing useful prospects for QIP schemes.

Nevertheless, the discussion given thus far already highlights a first crucial difference with respect to nonidentical particles systems, where the algebraic structure of a system's state does not depend on its relation with some measurement setting. Standard algebraic tools for evaluating entanglement in nonidentical particle systems, such as the Von Neumann entropy or the Schmidt number, are consequently affected by the same dependence if applied straightforwardly to identical particles, requiring further investigation of their implications.

### 3. Approaching entanglement between identical particles

*a. Changing formalism: the no-label approach.* We have seen that the method to decide on the necessity of state symmetrization does not depend on the system's state alone, but rather on its relationship with the employed measurement setting. However, sometimes the measurement setup is not specified, or it is simply unknown. One could therefore prefer to employ a formalism where the measurement context does not directly affect the way a many-body state is written. An approach pursuing this goal is the no-label formalism (Compagno *et al.*, 2018; Lo Franco and Compagno, 2016). Within this framework, the state of Eq. (2) is simply written as the single ket $|\Psi^{(2)}\rangle = |\phi_1, \phi_2\rangle = |\psi_1\,\sigma_1, \psi_2\,\sigma_2\rangle$, where $|\phi_j\rangle = |\psi_j\,\sigma_j\rangle$ ($j = 1, 2$) is a single-particle state containing the relevant degrees of freedom. The symmetrization postulate is here encoded in the rule for computing many-body probability amplitudes, which for two bipartite states $|\phi_1, \phi_2\rangle$, $|\phi_1', \phi_2'\rangle$ reads

$$\langle\phi_1', \phi_2'|\phi_1, \phi_2\rangle := \langle\phi_1'|\phi_1\rangle\,\langle\phi_2'|\phi_2\rangle + \eta\,\langle\phi_1'|\phi_2\rangle\,\langle\phi_2'|\phi_1\rangle\,, \quad (3)$$

implying that $|\phi_1, \phi_2\rangle = \eta\,|\phi_2, \phi_1\rangle$. Although the discussion of its mathematical features are beyond the scope of this *Colloquium*, hereafter we employ the no-label formalism where not specified otherwise.

The adoption of the no-label formalism conveniently circumvents the necessity of prior information on the measurement setting to write down a many-body state. However, we stress that it shifts it to the calculation of probability amplitudes in Eq. (3), as the two terms in the right-hand side are both present in ambiguous measurement settings, while one of the two does not contribute in unambiguous ones: this is the same behavior of Eq. (2).

*b. Addressing ambiguous measurement settings.* While the entanglement appearing in unambiguous settings in the standard formalism can be deemed as unphysical, establishing the nature of quantum correlations in ambiguous settings is a matter which cannot be solved by simply changing the employed formalism: it requires a different physical reasoning on the entanglement of identical particle systems. Roughly speaking, three general approaches can be adopted to provide a wider and more comprehensive framework of quantum correlations between identical constituents:

1. preserving the standard approach in the first quantization formalism, imposing the symmetrization postulate regardless of the detection setting, and providing a *measurement-independent* definition of entanglement for identical particles capable to filter out unphysical correlations emerging from the postulate application;

2. renouncing to first quantization altogether, defining entanglement within a different framework where the symmetrization postulate and ambiguous settings do not give rise to hindrances, as in second quantization;

3. preserving the standard, first-quantization approach, but defining quantum correlations as a *measurement-*

*dependent*, *operational* physical resource exploitable by distinct parties in quantum information protocols.

We now provide a brief overview of the three approaches. Detailed discussions of the mathematical structure, advantages, and weaknesses of each method can be found, e.g., in Refs. (Benatti *et al.*, 2020; Tichy *et al.*, 2011).

#### 4. Redefining entanglement in the first quantization formalism: a measurement-independent approach

A framework of the first kind has been developed, e.g, in Ref. (Ghirardi and Marinatto, 2004). Here, the authors initially define a bipartite system of two identical particles as *non-entangled* when "both constituents possess a complete set of properties". This heuristic definition catches the physical meaning typically attributed to non-entangled states of non-identical particles, linking it to the possibility of computing any physical property of one constituent without having to account for the other. The authors later show that this criterion amounts to deem a bipartite fermionic state as non-entangled if and only if it is obtained by antisymmetrizing a factorized state. Similarly, a bosonic bipartite state is deemed non-entangled either if it is obtained by symmetrizing a factorized product of two orthogonal states or if it is the product of the same state for the two constituents. A quantitative method to evaluate such conditions on bipartite states is then provided in terms of a combination of the von Neumann entropy and the Slater rank (Eckert *et al.*, 2002; Schliemann *et al.*, 2001), playing the role of the Schmidt rank in nonidentical scenarios.

While this method allows for an entanglement evaluation which is measurement-independent and only algebraic, it treats bosons and fermions on a different footage and requires many-body states to be non-separable (except for identical bosons in the same single-particle state) even when entanglement is absent, as the symmetrization postulate is here always applied.

#### 5. Mode-entanglement in the second quantization formalism

An approach of the second kind is given via second quantization (Barnum *et al.*, 2004; Benatti *et al.*, 2017, 2020; Tichy *et al.*, 2011; Zanardi, 2001). Here, each state vector spanning the single-particle Hilbert space defines a *mode*. Building on this, the concept of (mode) *locality* is introduced with respect to a chosen bipartition of the algebra of operators acting on the related many-body Fock space. Locality is the fundamental building block to discuss nonlocal correlations, therefore allowing for a rigorous definition of *mode entanglement*, which is out of the scope of this *Colloquium*.

Although this approach may seem to avoid relying on a measurement choice, mode entanglement strictly depends on the choice of a bipartition: states which could be entangled with respect to a bipartition may be non-entangled with respect to another. However, the specific choice of a bipartition is typically induced by the experimental setting (Tichy *et al.*, 2011), so that the mode entanglement approach ends up being measurement-dependent in practical contexts.

#### 6. Measurement-based approach: detector-level entanglement and the sLOCC framework

A different solution to the problem of evaluating the entanglement of identical particles in ambiguous settings is approaching entanglement as a property which can be exploited as a resource to realize quantum information protocols unachievable with only classical correlations. Crucially, any protocol of this kind involves a set of operations and measurements performed by distinct parties in local regions of space, thus deeming this method as measurement-dependent.

A key concept to face quantum correlations from this viewpoint is *entanglement extraction* (Killoran *et al.*, 2014; Tichy *et al.*, 2011). This process consists in transforming the ignorance about internal states in ambiguous measurement settings into quantum correlations between identical constituents in many-body states emerging from proper detections. Such detections are performed on spatially distinct regions of space, so that they provide an unambiguous setting for the projected state, allowing to disregard the symmetrization postulate and to evaluate the resulting quantum correlations using the standard tools for nonidentical elements. For this reason, such correlations are also called *detector-level entanglement* (Tichy *et al.*, 2013a). Detector-level entanglement can therefore be readily exploited by distinct parties to perform QIP protocols within the standard LOCC framework.

Overall, this paradigm treats quantum correlations between identical constituents in relation to the ones transferrable to operationally accessible spatial modes (Killoran *et al.*, 2014). It answers questions such as *what happens within this region of space* rather than *what happens to this particle*, and is therefore named *spatially localized operations and classical communication* (sLOCC) framework (Lo Franco and Compagno, 2018). The sLOCC framework expands beyond the analysis of entanglement extraction processes by introducing mathematical tools, such as a spatially localized partial trace and von Neumann entropy, to address identical particles from a perspective which focuses on spatial regions rather than individual constituents. In this *Colloquium* we focus on the operationally useful features of this approach, redirecting the interested readers to Refs. (Compagno *et al.*, 2018; Lo Franco and Compagno, 2016, 2018; Piccolini *et al.*, 2023b) for a more complete analysis.

Examples of how the sLOCC paradigm can be exploited to access useful quantum correlations in operational protocols are the subject of the following sections.

### B. Identical particle interferometry and indistinguishability

We now examine how particle statistics govern interference and how, in a multiphoton regime, indistinguishability becomes a genuinely many-body property that can be probed through interference and certified by suitable witnesses. We then turn to its information-theoretic interpretation via no-which-way information. These complementary notions of indistinguishability will serve as operational resources for the quantum information protocols discussed in this *Colloquium*.

### 1. Interference and particle statistics

Quantum many-body interference is probably the most peculiar phenomenon characterizing identical particles. In one of the most famous experiments building on this effect, namely the Hong-Ou-Mandel (HOM) experiment (Hong *et al.*, 1987), two photons are let to impinge on the two input ports of a beamsplitter, subsequently undergoing a coincidence measurement at the output. For completeness, we briefly review two-photon, or HOM, interference (for a detailed treatment, see Ref. (Bouchard *et al.*, 2021)).

We assume the two particles are in the initial state $|\Psi^{(2)}\rangle_0 = |L\,\chi_1, R\,\chi_2\rangle$. Although the original experiment focused on the arrival time of the two photons at the output detectors, here we allow $\chi_j$ $(j = 1, 2)$ to encode any additional degree of freedom, such as polarization, frequency, or momentum. Applying a balanced beamsplitter transformation $|L\rangle \to |D\rangle := (|L\rangle + |R\rangle)/\sqrt{2}$ and $|R\rangle \to |A\rangle := (|L\rangle - |R\rangle)/\sqrt{2}$, the output state is

$$|\Psi^{(2)}\rangle_{\rm BS} = \Big( |L\chi_1, L\chi_2\rangle - |R\chi_1, R\chi_2\rangle - |L\chi_1, R\chi_2\rangle + |L\chi_2, R\chi_1\rangle \Big)/2. \tag{4}$$

We notice that if the two photons are in the same state, that is, if $|\chi_1\rangle = |\chi_2\rangle$, the last two terms interfere destructively and cancel out. The two particles thus both exit from the same output port, either L or R, thus showing a *bunching* behavior which manifests itself in a dip of the coincidence count against the variation of the degrees of freedom encoded in $\chi_j$. The two photons in this scenario are typically called *indistinguishable* and the HOM experiment is nowadays considered as the standard experiment to probe the indistinguishability of identical particles. This concept of indistinguishability, however, is not to be confused with the one of *spatial indistinguishability* mentioned above. While the latter probes the relation between a many-body state and a measurement setup, the former gives information on the possibility of discriminating between the two single-particle states, that is, the inner product $\langle\chi_1|\chi_2\rangle$. Spatial indistinguishability is usually involved in addressing quantum correlations between identical particles in ambiguous measurement settings, whereas indistinguishability in the traditional interferometric sense is a property analyzed within the HOM experiment to characterize photon sources where the generation of particles in the same state is desired.

Bunching and antibunching at a beamsplitter are input-state-dependent processes, rather than signatures fixed solely by the particle species. For example, two identical two-component bosons (bosonic qubits) prepared in the Bell singlet state $|\Psi^{(2)}\rangle_- = \big(|L\uparrow, R\downarrow\rangle - |L\downarrow, R\uparrow\rangle\big)/\sqrt{2}$ antibunch at a balanced beamsplitter, with the state remaining invariant $|\Psi^{(2)}\rangle_{\rm BS} = \big(|D\uparrow, A\downarrow\rangle - |D\downarrow, A\uparrow\rangle\big)/\sqrt{2} = |\Psi^{(2)}\rangle_-$. Conversely, for two identical fermions, the Bell singlet has the appropriate exchange symmetry to bunch, whereas the three fermionic triplet states antibunch and remain invariant under the same transformation. Finally, one can consider the state $|\Psi^{(2)}\rangle_0 = |L, R\rangle$ of two fermions with the same spin (here omitted for simplicity). It can be easily shown that, due to destructive interference, this state antibunches under the action of the beamsplitter and is left invariant, that is $|\Psi^{(2)}\rangle_{\rm BS} = |D, A\rangle = |L, R\rangle$. These effects are, for instance, exploited in entanglement protection protocols to filter out noise-induced terms on target states within the sLOCC framework (Nosrati *et al.*, 2020; Piccolini *et al.*, 2023a, 2024a,b, 2021). Interference effects play a fundamental role in many QIP schemes based on the manipulation of identical particles, from the generation of measurement-induced entanglement (either heralded or postselected) to its protection from detrimental noise. These topics shall be treated in the next sections.

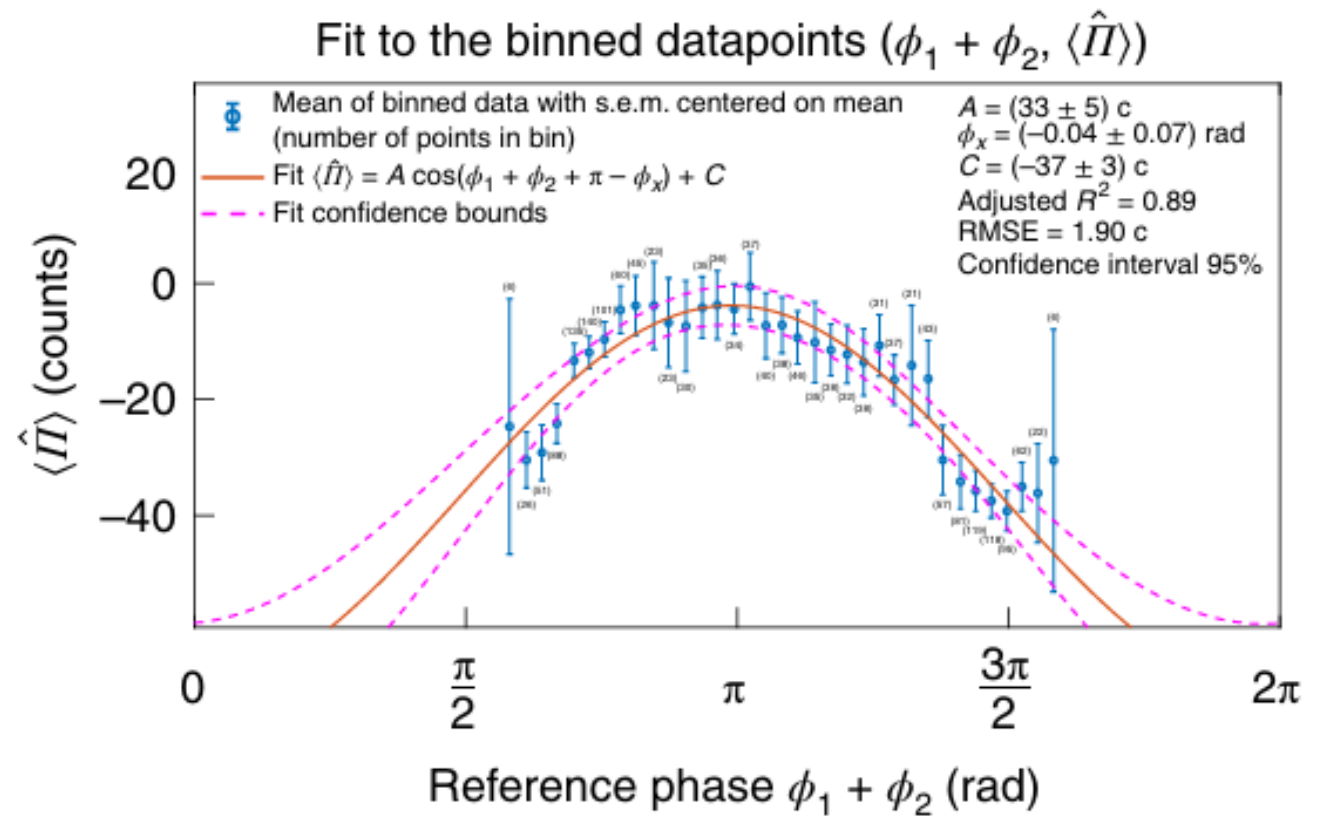

FIG. 2: Experimental results for the confirmation of the symmetry of the two-photon wave function. Figure adapted from Ref. (Tschernig *et al.*, 2021).

The HOM experiment demonstrates that quantum interference provides an operational probe of particle statistics. In this picture, interference arises from the exchange phase $\phi_x$, defined as the phase acquired by a many-particle wavefunction under particle exchange. For bosons $\phi_x = 0$, while for fermions $\phi_x = \pi$. Traditionally, this phase had only been inferred indirectly by verifying the absence of particular states which are forbidden by the symmetrization postulate (de Angelis *et al.*, 1996; Hilborn and Yuca, 1996; Kekez *et al.*, 1990; Modugno *et al.*, 1998). Tschernig *et al.* (Tschernig *et al.*, 2021) reported the first direct measurement of this exchange phase by realizing an interferometer in which the two indistinguishable histories corresponding to particle exchange interfere coherently. The resulting coincidence signal is

$$\langle\hat{\Pi}\rangle = A\cos(\phi_1 + \phi_2 + \pi - \phi_x) + c, \tag{5}$$

so that the exchange phase is extracted directly from the shift of the interference fringes (other parameters of Eq. (5) are not relevant here). The measured value, $\phi_x = -0.04 \pm 0.07$ rad, is fully consistent with the bosonic prediction $\phi_x = 0$ (Fig. 2), providing a direct verification of the symmetry of the two-photon wavefunction. As emphasized by Lo Franco (Lo Franco, 2021), this experiment transforms exchange symmetry from a fundamental postulate into a directly measurable physical quantity.

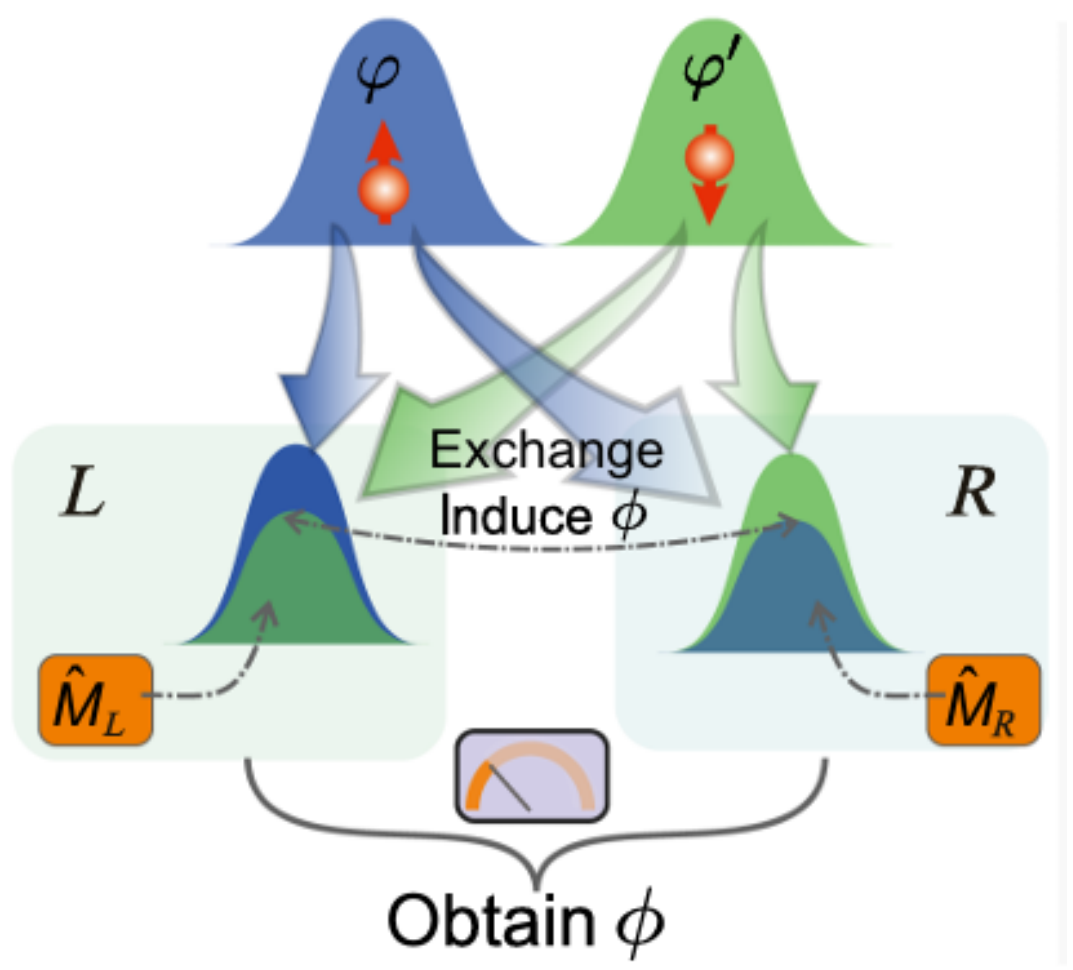

FIG. 3: Conceptual scheme for measuring exchange statistics using sLOCC. Two photons are distributed over regions $L$ and $R$ with controlled overlap, a tunable relative phase $\phi$ is introduced, and local rotations followed by coincidence detection allow the exchange phase to be extracted. Adapted from Ref. (Wang *et al.*, 2022a)

A complementary approach was proposed by Wang *et al.* (Wang *et al.*, 2022a), who inferred the exchange phase within the sLOCC framework from correlations between spatially separated detection regions. The protocol employs two photons distributed over regions $L$ and $R$, followed by local rotations and sLOCC measurements that map the global exchange phase onto measurable coincidence probabilities. As illustrated in Fig. 3, the method is universal, applying to bosons ($\phi = 0$), fermions ($\phi = \pi$), simulated anyons ($0 < \phi < \pi$), and even mixtures of particles with different exchange statistics. Experimentally, high-visibility ($\sim 97.7\%$) HOM interference first verifies photon indistinguishability, while a quartz-plate phase shifter simulates arbitrary exchange phases. The measured values, $\phi_b = 0.04 \pm 0.06$ rad for bosons and $\phi_f = 3.12 \pm 0.05$ rad for fermions, agree with the expected phases. By continuously varying the inserted phase, the protocol further reproduces arbitrary anyonic exchange phases and mixtures of different statistics.

### 2. Multiphoton interference as an indistinguishability metric

As mentioned above, the HOM experiment enables to discriminate between two single-particle states through the inner product $\langle \chi_1 | \chi_2 \rangle$, which originates from the spectrotemporal overlap of the two photons. Indeed, this inner product describes distinguishability stemming from mismatches in any photonic degree of freedom (Tillmann *et al.*, 2015). When the overlap is equal to one, the photons are regarded as indistinguishable. The degree of indistinguishability is usually quantified by the HOM interference visibility

$$\mathcal{V} = \frac{C_{\max} - C_{\min}}{C_{\max}}, \tag{6}$$

where $C_{\max}$ and $C_{\min}$ are, respectively, the maximum and minimum coincidence count rates measured at the output of the beamsplitter as the relative arrival time between the two input photons is varied. In the perfect limit, bosons exhibit complete bunching, giving rise to the characteristic HOM dip in the coincidence rate. Consequently, HOM visibility has become the standard figure of merit for characterizing indistinguishability across a wide range of single-photon sources (Flagg *et al.*, 2010; Kaltenbaek *et al.*, 2006; Kiraz *et al.*, 2005; Maunz *et al.*, 2007a; Mosley *et al.*, 2008). The same interferometric configuration can be realized with fermions, where antisymmetric exchange statistics cause perfect antibunching (Pauli blocking), producing a coincidence peak instead of the characteristic bosonic dip (Chakraborti *et al.*, 2025; Freulon *et al.*, 2015). So, the interference visibility is a direct measure of particle indistinguishability for both bosons and fermions.

The concept extends naturally to the multiphoton regime, where the degree of indistinguishability among many photons determines the structure and visibility of multiparticle interference (Tillmann *et al.*, 2015). A common approach to characterizing multiphoton indistinguishability is to express it in terms of pairwise indistinguishability between all photon pairs (Jones *et al.*, 2020). Building on this idea, Brod *et al.* proposed linear-optical witnesses for genuine multiphoton indistinguishability (Brod *et al.*, 2019b). In their framework, the measured witness provides lower and upper bounds on the degree $c_1$ of genuine $N$-boson indistinguishability

$$\frac{p_b - p_b^{\perp}}{1 - p_b^{\perp}} \leq c_1 \leq 2p_b - 1, \tag{7}$$

where $p_b = (N-1)^{-1} \sum_{j=1}^{N-1} p_b^{\mathrm{HOM}}(r, j)$ is the average HOM bunching probability between a reference photon $r$ and each of the remaining $n - 1$ photons, and $p_b^{\perp} = 1 - [2(N-1)]^{-1}$ is the threshold associated with states that are not genuinely $n$-particle indistinguishable.

The concept of the indistinguishability witness was then generalized to quantify the overlap between an arbitrary multiphoton state and the ideal pure state of $N$ perfectly indistinguishable photons (Pont *et al.*, 2022). By employing low-depth cyclic $2N$-mode interferometers, Pont *et al.* devised a protocol that produces a quantum interference fringe whose visibility directly quantifies genuine $N$-photon indistinguishability. These advances motivated a rigorous mathematical formulation of multiphoton indistinguishability based on the bosonic mode formalism. Within this framework, the degree of indistinguishability is quantified by the expectation value of the projector onto the symmetric subspace, $p_N = \mathrm{Tr}[P^{(N)}\rho]$, which measures the overlap of a multiphoton state with the fully symmetric sector (Englbrecht *et al.*, 2024). Perfectly indistinguishable states give $p_N = 1$, whereas perfectly distinguishable states give $p_N = 1/N!$. Also, the two-particle quantity $p_2$, which measures the average symmetry of all two-particle reduced states, supplies a powerful witness because sufficiently large values of $p_2$ impose tight lower bounds on the genuine indistinguishability $p_N$ (Englbrecht *et al.*, 2024).

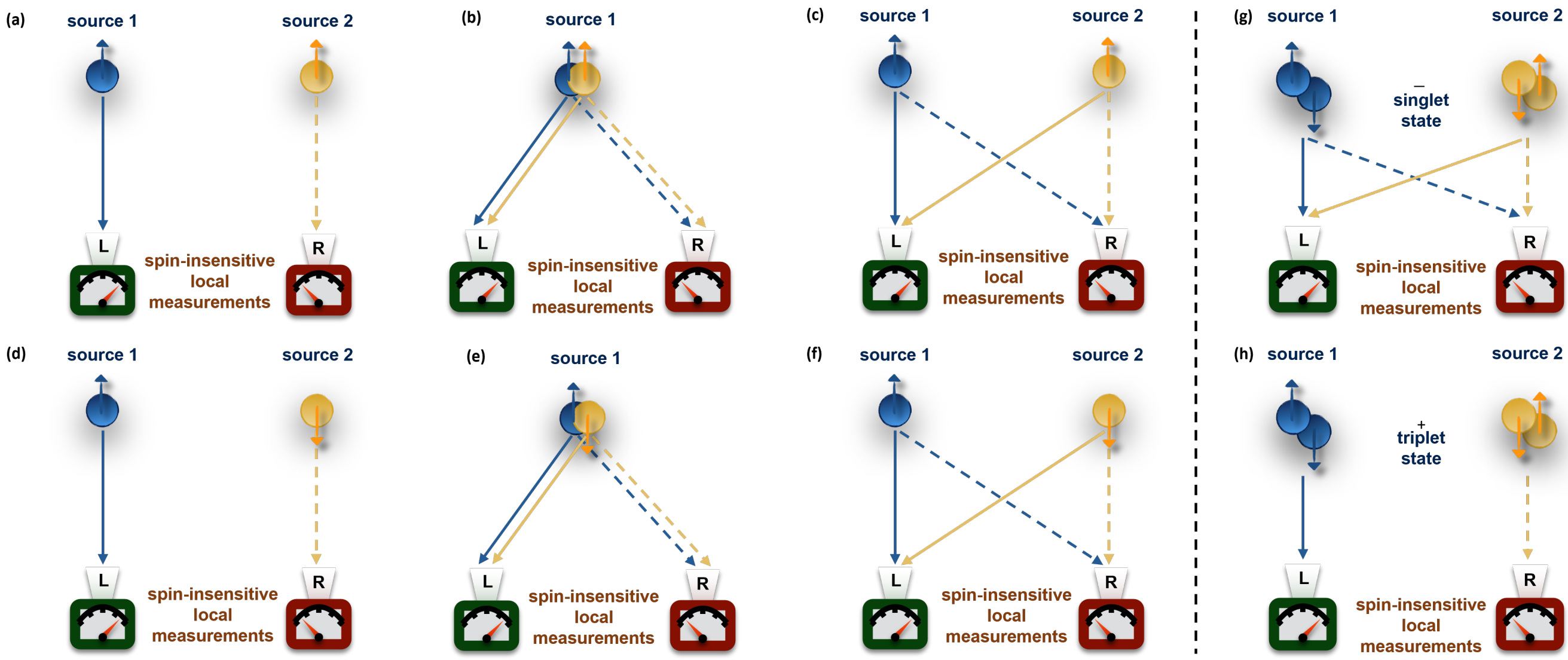

FIG. 4: Different relations between identical bosons path, pseudospin, and their overlap over detection regions, leading to different types of missing information. **(a)**,**(d)** Unambiguous setting: all the information are known. **(b)** Ambiguous setting: no which-particle information, source and spin known. **(c)** Ambiguous setting: no which-particle and no which-way information, spin known. **(e)** Ambiguous setting: no which-particle and no which-spin information, path known. **(f)** Ambiguous setting: no information is known. **(g)** Ambiguity undecidable (interference): which-particle and which-path information undecidable, no which-spin information. **(h)** Unambiguous setting: no which-spin information, particle and path known.

### 3. Indistinguishability and missing information

Beyond interference visibility, indistinguishability can also be characterized from an information-theoretic perspective. Within the sLOCC framework, spatial overlap of identical particles, which were initially separated, removes which-particle information at the detection stage. This operational ambiguity, corresponding to the ambiguous setting introduced in Section II.A.2, is then converted into quantum correlations (entanglement) through localized measurements.

Motivated by this operational picture, Nosrati *et al.* (Nosrati *et al.*, 2020) introduced an entropic measure of spatial indistinguishability that quantifies the uncertainty in assigning the detected particles to the available spatial regions

$$\mathcal{I} := -\frac{P_{\rm L}P'_{\rm R}}{P_{\{L,R\}}}\log_2\frac{P_{\rm L}P'_{\rm R}}{P_{\{L,R\}}} - \frac{P'_{\rm L}P_{\rm R}}{P_{\{L,R\}}}\log_2\frac{P'_{\rm L}P_{\rm R}}{P_{\{L,R\}}}, \qquad (8)$$

where $P_{\{L,R\}} := P_{\rm L}P'_{\rm R} + P'_{\rm L}P_{\rm R}$, $P_{\rm X} = |\langle X|\psi\rangle|^2$, $P'_{\rm X} = |\langle X|\psi'\rangle|^2$ for $X = L, R$, and $P_{\rm L} + P_{\rm R} = P'_{\rm L} + P'_{\rm R} = 1$. Notice that this quantity ranges from $\mathcal{I} = 0$ for spatially distinguishable particles, as in the state $|\Psi^{(2)}\rangle_0 = |L\uparrow, R\downarrow\rangle$, to $\mathcal{I} = 1$ when maximum spatial indistinguishability is achieved, as in the state of Eq. (4) with $\chi_1 =\uparrow$, $\chi_2 =\downarrow$.

The concepts of ambiguous measurements and spatial indistinguishability introduced in Section II.A.2, together with the entropic measure $\mathcal{I}$ of Eq. (8), were originally motivated by the absence of *which-particle information*: spatial overlap prevents the measurement from identifying which particle is detected. When the particles originate from different sources, the same overlap also removes *which-way information*, i.e., the ability to determine the source-to-detector path. Building on Hardy's original proposal (Hardy, 1992), Krenn *et al.* (Krenn *et al.*, 2017b) showed that such no which-way information, also known as path identity (Hochrainer *et al.*, 2022), can be used to create multipartite and high-dimensional entangled states.

This particle-centric interpretation becomes problematic in the presence of many-body interference. Different single-particle descriptions can correspond to the same two-particle state, so that attributing labels or trajectories to individual particles loses operational meaning. For example, two identical fermions with the same spin localized in different sites are indistinguishable, under two-particle measurements, from two fermions both delocalized over the two sites, even if they were independently generated and never interacted. More strikingly, the state $|\rm L, R\rangle = |\rm D, A\rangle$ remains invariant under a beamsplitter, as seen above, and is therefore fully separable despite the absence of an unambiguous particle picture.

We stress that what fails here is the *particle-label* interpretation of which-particle and which-way information, together with their quantifiers built from single-particle probabilities. Mode-based descriptions are not affected. For instance, path identity (Gu *et al.*, 2019b; Krenn *et al.*, 2017a), formulated in terms of modes and coherent sums over perfect matchings, natively incorporates multiparticle interference, since all source alternatives leading to the same output configuration contribute coherently to a single amplitude.

To account for this, Piccolini *et al.* (Piccolini, 2024) proposed replacing particle-based no which-particle and no which-way information with *no which-spin information*. The relevant ambiguity is no longer the identity or trajectory of the detected particle, but the assignment of its unprobed in-ternal degree of freedom (e.g., pseudospin). As illustrated in Fig. 4, only detection scenarios lacking which-spin information generate entanglement after coincidence measurements and postselection. Accordingly, indistinguishability should be quantified through two-particle probability amplitudes rather than single-particle probabilities, since the former remain well defined in interferometric settings. This leads to the entropic measure of pseudospin indistinguishability as (Piccolini, 2024)

$$\tilde{\mathcal{I}}(|\Psi^{(2)}\rangle) := -\frac{P_{\mathrm{L}\uparrow\mathrm{R}\downarrow}}{P_{\{\uparrow,\downarrow\}}}\log_2\frac{P_{\mathrm{L}\uparrow\mathrm{R}\downarrow}}{P_{\{\uparrow,\downarrow\}}} - \frac{P_{\mathrm{L}\downarrow\mathrm{R}\uparrow}}{P_{\{\uparrow,\downarrow\}}}\log_2\frac{P_{\mathrm{L}\downarrow\mathrm{R}\uparrow}}{P_{\{\uparrow,\downarrow\}}}, \quad (9)$$

where $P_{\{\uparrow,\downarrow\}} := P_{\mathrm{L}\uparrow\mathrm{R}\downarrow} + P_{\mathrm{L}\downarrow\mathrm{R}\uparrow}$ and $P_{\mathrm{L}\sigma\mathrm{R}\tau} = |\langle L\sigma, R\tau|\Psi^{(2)}\rangle|^2$ for $\sigma,\tau = \uparrow,\downarrow$. We notice that the above measure coincides with $\mathcal{I}$ of Eq. (8) whenever each particle carries a definite and orthogonal spin state, as for the state $|L\uparrow, R\downarrow\rangle$. Here, indeed, interference effects do not occur, and the spatial deformation generates ambiguity about all the types of information considered (see Fig. 4(f)). There are, however, cases where the two measures diverge, even when the original formulation is physically grounded. This is the case, e.g., of the triplet state $|\Psi^{(2)}\rangle_+ = \big(|L\uparrow, R\downarrow\rangle + |L\downarrow, R\uparrow\rangle\big)/\sqrt{2}$, where $\tilde{\mathcal{I}}(|\Psi^{(2)}\rangle_+) = 1$ due to its no which-spin information, although there is no spatial overlap as the two particles are well localized one in $L$ and one in $R$ without interferometric ambiguity.

This refined viewpoint also sharpens the understanding of the entanglement protection mechanisms mentioned before. Rather than attributing entanglement protection directly to spatial indistinguishability, we recognize that the underlying trait is the *symmetry-based filtering* enabled by many-body interference. Formation of spatial overlap activates bunching/antibunching effects that redistribute different terms of a noisy state according to their exchange symmetry, and the final coincidence measurement postselects the desired entangled components. We will go back to this topic in Sec. III.A.3.

## III. COHERENCE AND ENTANGLEMENT GENERATION

After discussing the theoretical foundations of entanglement in ensembles of identical particles—including the sLOCC framework, entropic measure of spatial indistinguishability, and interference mechanisms underlying bunching and antibunching—we now turn to their first major application: the generation of entanglement and coherence. Rather than providing a comprehensive survey of existing protocols, for which we refer the reader to the recent reviews on path identity (Hochrainer *et al.*, 2022) and heralded entanglement generation (Forbes *et al.*, 2025), we focus on the distinct physical roles played by indistinguishability.

### A. Postselected schemes

Entanglement generation schemes based on single particles can be categorized into two main types: postselected and heralded ones. In the postselected case, it is not known whether the scheme succeeded in producing the target entangled state until all particles are detected. For example, a scheme might require that exactly one particle is detected in each pair of modes forming a dual-rail qubit. In such cases, every generation attempt is initially treated as successful and the produced state can be used as input for some experiments. Only later, upon detecting all particles, it can be determined whether the attempt was indeed successful and the results should be kept.

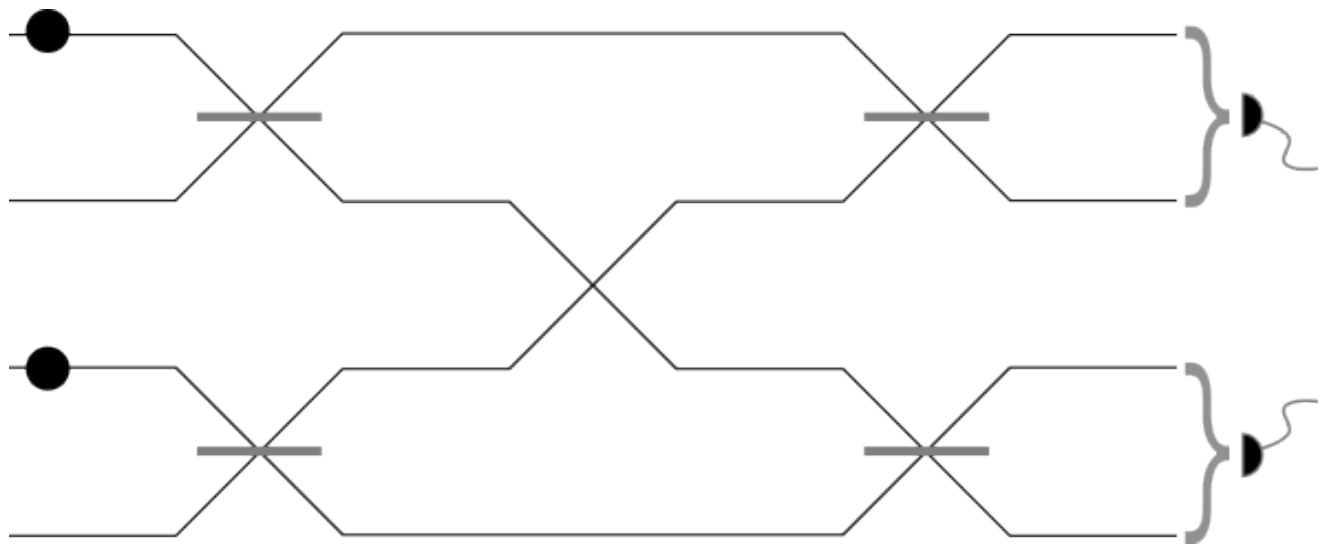

FIG. 5: Yurke-Stoler scheme for Bell state generation. Two single particles from independent sources propagate through a beamsplitter array, whose output modes form two dual-rail qubits. Each qubit is read out by a single detector that certifies the presence of one particle without resolving which of its two modes is occupied, i.e. insensitive to the logical (which-spin) value. The beamsplitters preceding it set the local measurement bases, as required, e.g., for a CHSH test.

#### 1. Basic idea: bipartite entanglement generation

The idea of generating entanglement from independent identical particles dates back to the seminal proposal of Yurke and Stoler (Yurke and Stoler, 1992a). In the scheme shown in Fig. 5, two independent sources inject single particles into separate 50:50 beamsplitters, after which one output mode from each beamsplitter is exchanged. The four output modes define two dual-rail qubits,

$$\begin{aligned} a_1^\dagger|0\rangle &\equiv |0\rangle_L, \quad a_2^\dagger|0\rangle \equiv |1\rangle_L,\\ a_3^\dagger|0\rangle &\equiv |0\rangle_R, \quad a_4^\dagger|0\rangle \equiv |1\rangle_R, \end{aligned} \quad (10)$$

where $|0\rangle$ is the vacuum state of the 4 modes. The second pair of beamsplitters acts as a local, single-qubit change of measurement basis on each dual-rail qubit. Because of that, in the following we shall consider a simplified version of the Yurke-Stoler scheme, with the final pair of beamsplitters removed, reinstating them where the measurement stage is at issue.

For the beamsplitter transformations given by Hadamard matrices, the input state evolves as

$$\begin{aligned} a_1^\dagger a_3^\dagger|0\rangle &\xrightarrow{\text{BS}} \frac{1}{2}(b_1^\dagger + b_2^\dagger)(b_3^\dagger + b_4^\dagger)|0\rangle\\ &\xrightarrow{\text{SWAP}} \frac{1}{2}(b_1^\dagger + b_3^\dagger)(b_2^\dagger + b_4^\dagger)|0\rangle\\ &\xrightarrow{\text{postselection}} \frac{1}{2}(b_1^\dagger b_4^\dagger + \eta\, b_2^\dagger b_3^\dagger)|0\rangle \\ &= \frac{1}{2}(|0\rangle_L|1\rangle_R + \eta\,|1\rangle_L|0\rangle_R), \end{aligned} \quad (11)$$

with $\eta = 1$ for bosons and $\eta = -1$ for fermions. The final state in Eq. (11) is not normalized, reflecting the fact that we discard the outcomes in which both particles end up in the same dual-rail qubit, which occurs with probability $\frac{1}{2}$.

Remarkably, independent of particle statistics, the postselected outcomes carry maximally entangled correlations. As for any postselected scheme, what is produced is not a stored entangled state but correlations between measurement outcomes, conditioned on detecting one particle per dual-rail qubit. Calling this result a *postselected state* is, however, a widely accepted convention, which we adopt.

As argued by Yurke and Stoler, these correlations violate the Clauser–Horne inequality (Clauser and Horne, 1974), demonstrating Bell nonclassicality. Although postselection can, in general, lead to spurious Bell violations, the protocol remains valid because each party discards only local events (0 or 2 detected particles), making the postselection safe (Błasiak *et al.*, 2021a). Indeed, the violation survives even without postselection by assigning the discarded events to a fixed local outcome.

This Bell violation is observed only when the measurement is performed in a basis that erases which-spin information, placing the protocol squarely within the no which-spin picture of Section II.B.3. If the four output modes are measured directly with mode-resolving detectors, only the pairs $(1, 4)$ and $(2, 3)$ are detected, yielding the logical states $|0\rangle_L\,|1\rangle_R$ and $|1\rangle_L\,|0\rangle_R$. Such measurements retain complete which-spin information and reveal only classical anticorrelations. Reinstating the second pair of beamsplitters rotates each dual-rail qubit into a complementary basis, so that detector clicks no longer identify a definite logical state. The beamsplitters therefore define the local CHSH measurement settings rather than forming part of the state preparation.

As the Yurke–Stoler protocol does not violate the Clauser–Horne inequality for distinguishable particles, its nonclassical correlations originate from particle indistinguishability. This interpretation is further illustrated by the thought experiment of Ref. (Bose and Home, 2002a). Two particles with opposite pseudospins impinge on a beamsplitter and are detected by nonabsorbing detectors that register only the presence of a particle in each output port, but not its pseudospin. Assuming the particles are identical in every degree of freedom except the pseudospin, postselecting events with one particle in each output yields

$$|\Psi^{\pm}\rangle = \frac{1}{\sqrt{2}}\left(|L{\uparrow}, R{\downarrow}\rangle + \eta\;|L{\downarrow}, R{\uparrow}\rangle\right), \qquad (12)$$

which is maximally entangled in the pseudospin degree of freedom. Although the same algebraic state could formally be obtained for distinguishable particles if the detectors were artificially made insensitive to all distinguishing degrees of freedom, identical particles guarantee this indistinguishability intrinsically: there is no additional property the detectors can exploit to identify the two contributions. Consequently, the entanglement arises from the same no-which-spin mechanism underlying the sLOCC framework of Section II.B.3, here realized through beamsplitter interference.

Consistently, any residual distinguishability degrades the entanglement. If the input particles are partially distinguishable in some additional degree of freedom $S$, with overlap $\alpha := |\langle S_1|S_2\rangle|^2 < 1$, the postselected state becomes

$$\rho = \frac{1}{2}\big(\,|L{\uparrow}, R{\downarrow}\rangle\langle L{\uparrow}, R{\downarrow}| + |L{\downarrow}, R{\uparrow}\rangle\langle L{\downarrow}, R{\uparrow}| + \eta\,\alpha\;|L{\uparrow}, R{\downarrow}\rangle\langle L{\downarrow}, R{\uparrow}| + \eta\,\alpha\;|L{\downarrow}, R{\uparrow}\rangle\langle L{\uparrow}, R{\downarrow}|\,\big). \qquad (13)$$

The entanglement of this state, measured by the concurrence $C$ (Wootters, 1998a), and the probability $D$ of discriminating $|S_1\rangle$ and $|S_2\rangle$ obey the trade-off relation

$$C + D = 1, \qquad (14)$$

reminiscent of the which-way–visibility relation known from single-particle interference (Englert, 1996). Maximal entanglement ($C = 1$) is recovered precisely when the particles are indistinguishable in $S$ ($\alpha = 1$), i.e. when no which-spin information leaks into the unprobed degree of freedom.

This line of research has been compellingly presented by Lo Franco and Compagno in (Lo Franco and Compagno, 2018) through the sLOCC operational framework introduced in Section II.A.6. As discussed there, spatial indistinguishability (or more precisely, the no which-spin information it can generate) enables the direct extraction of entanglement by observers confined to spatially separated regions. The operation of redistributing the spatial wave functions of single constituents in a many-body state over distinct spatially-separated detection regions takes the name of *spatial deformation* within this context (Nosrati *et al.*, 2020; Piccolini *et al.*, 2023b). We stress that spatial deformations are much broader than a balanced beamsplitter transformation, including unbalanced transformations with arbitrary coefficients applicable to both bosons and fermions (Piccolini *et al.*, 2023b). Applying this framework to a two-particle state $|\Psi\rangle = |\psi\uparrow, \psi'\downarrow\rangle$ with spatial wavefunctions overlapping both regions $L$ and $R$ (see Fig. 6), one defines the probability amplitudes $l = \langle L|\psi\rangle$, $l' = \langle L|\psi'\rangle$, $r = \langle R|\psi\rangle$, $r' = \langle R|\psi'\rangle$ (notice that no interference effects hinder this treatment here). The coincidence projection onto the basis $\mathcal{B}_{LR} = \{|L\uparrow, R\uparrow\rangle, |L\uparrow, R\downarrow\rangle, |L\downarrow, R\uparrow\rangle, |L\downarrow, R\downarrow\rangle\}$ yields the normalized Bell-like state

$$|\Psi_{LR}\rangle = \frac{lr'\,|L\uparrow, R\downarrow\rangle + \eta\, l'r\,|L\downarrow, R\uparrow\rangle}{\sqrt{|lr'|^2 + |l'r|^2}}, \qquad (15)$$

obtained with probability $P_{LR} = |lr'|^2 + |l'r|^2$. Overall, the whole process has firstly created spatial indistinguishability via spatial deformation, and subsequently converted the so-generated ambiguity on the internal degree of freedom into entanglement by coincidence measurement. The entanglement of this state, which depends on the spatial overlap between the original wavefunctions and the detection regions, is found to coincide with the entanglement of formation computed from the partial trace procedure within the sLOCC formalism (see Sec. II for details). This equivalence validates the sLOCC framework as a physically meaningful tool for quantifying the operational entanglement of identical particles.

Recent experiments have demonstrated that spatial indistinguishability between independent identical particles can generate observable quantum coherence and entanglement. In the experiment of Fig. 7, Sun *et al.* (Sun *et al.*, 2020) showed that two independent, orthogonally polarized photons become entangled when their spatial probability amplitudes are coherently distributed between two remote regions $L$ and $R$. By tuning the spatial overlap, they controlled the entropic measure of spatial indistinguishability of Eq. (8). Postselection of one photon in each region (sLOCC) projected the pair onto an entangled polarization state whose concurrence increased monotonically with $\mathcal{I}$. At maximal indistinguishability ($\mathcal{I} = 1$), the generated Bell state reached a fidelity of $(88.3 \pm 2.5)\%$ and violated the CHSH inequality with $S = 2.33 \pm 0.16$. Remarkably, the photons never physically interfered at a common beamsplitter: the observed entanglement originated solely from their spatial indistinguishability.

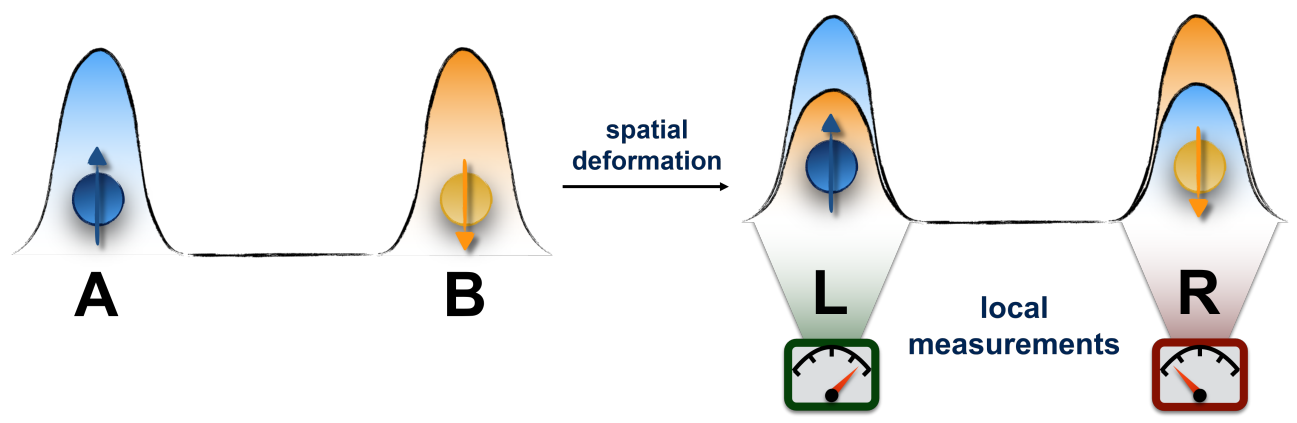

FIG. 6: Sketch of an entanglement generation protocol, exploiting deformation-generated spatial indistinguishability and spatially localized coincidence measurements.

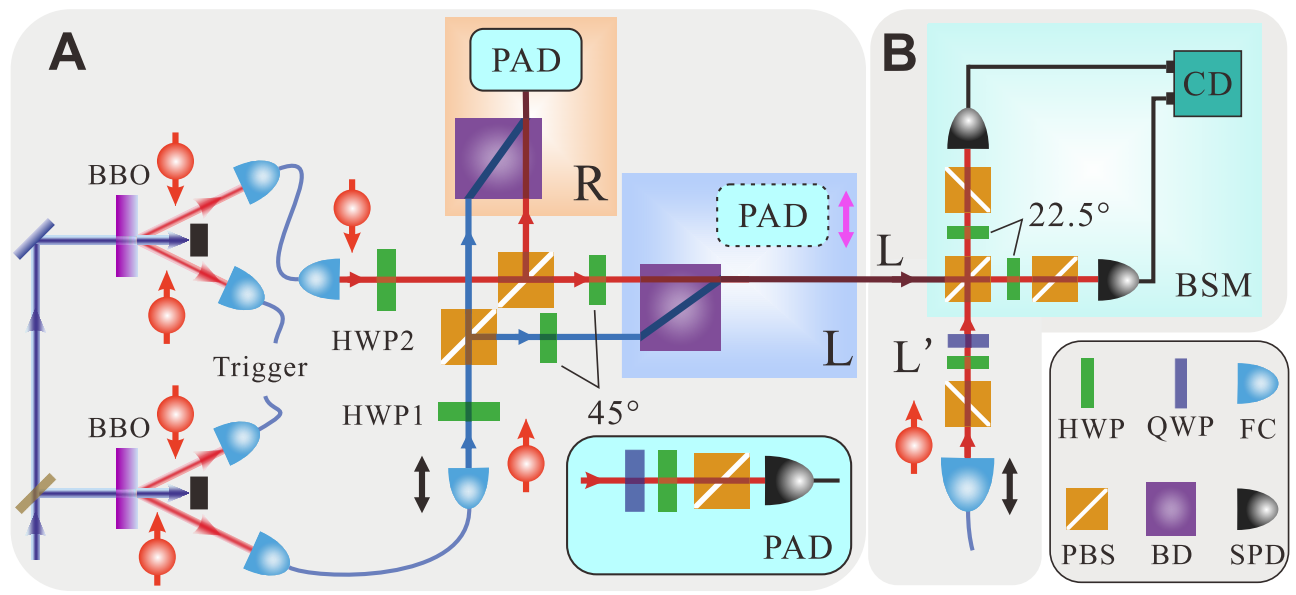

FIG. 7: Experimental setup for realizing sLOCC. Figure adapted from Ref. (Sun *et al.*, 2020).

To separate the roles of spatial overlap and temporal indistinguishability, Barros *et al.* (Barros *et al.*, 2020) introduced a unified theoretical and experimental framework in which the concurrence obeys $C(\theta, \Delta t) \propto \mathcal{S}(\theta)\Gamma(\Delta t)$ where $\mathcal{S}(\theta)$ is controlled by the spatial overlap and $\Gamma(\Delta t)$, calibrated from HOM interference, quantifies temporal indistinguishability. The measurements confirmed that entanglement vanishes whenever either resource is absent and increases monotonically with both, providing a direct quantitative link between spatial overlap, temporal indistinguishability, and entanglement.

We remark that, as in the discussed Yurke-Stoler scheme, postselection is not essential for the entanglement of these correlations. Without postselection, the state is a coherent superposition of a coincidence component, with one particle in each region, and a double-occupancy one, with both particles in the same region (Lo Franco and Compagno, 2018). Their relative weight is set by the spatial overlap of the two wavefunctions. As the double-occupancy component lies outside the one-particle-per-region subspace in which the qubits are encoded, it does not contribute to the concurrence, which is that of the Bell-like component weighted by its probability, $C = P_{LR} C(|\Psi_{LR}\rangle)$, with $C(|\Psi_{LR}\rangle)$ being the concurrence of the state of Eq. (15). Since $C > 0$ for any $P_{LR} > 0$, the state remains entangled even without postselection, whose role is only to distill the pure state $|\Psi_{LR}\rangle$ saturating Tsirelson's bound in the case of balanced overlap $P_{LR} = 1/2$. The latter condition gives $C = 1/2$, still sufficient for quantum teleportation with no classical counterpart.

The above results demonstrate that the operational entanglement originating from the indistinguishability of two particles can be exploited in quantum information processing tasks. The results in (Blasiak and Markiewicz, 2025) show that almost all states of indistinguishable particles can be used to this end in passive optical experimental setups. Specifically, any multipartite state in which not all the particles occupy the same mode can be first reduced to a two-particle one, and then used in a simplified version of the Yurke-Stoler scheme to demonstrate Bell-nonclassical correlations.

### 2. Multipartite entanglement generation

Entanglement generation schemes based on the indistinguishability of particles are not restricted to the bipartite case. In fact, a protocol for a postselected generation of the Greenberger–Horne–Zeilinger (GHZ) state $|\mathrm{GHZ}\rangle = \frac{1}{\sqrt{2}}(|000\rangle + |111\rangle)$ was proposed already in (Yurke and Stoler, 1992b), and follows the same structure as the Yurke and Stoler's Bell state generation scheme discussed in Sec. III.A.1. The only difference consists in using a third source to cast a single particle on an additional beamsplitter, and replacing the swap of modes with a cyclic permutation of length 3.

The approach of Yurke and Stoler has been generalized in (Blasiak and Markiewicz, 2019). The "no-touching" regime introduced there describes protocols where entanglement is generated between particles that never occupy the same spatial location throughout their entire evolution. In these setups, independent particles are injected into spatially separated input regions, processed by local unitaries, then have their paths permuted to different output regions where additional local processing occurs.

The "no-touching" paradigm has been used to find schemes to generate the W state $|\mathrm{W}\rangle = \frac{1}{\sqrt{3}}(|100\rangle + |010\rangle + |001\rangle)$ with success probability around 0.15 (Blasiak and Markiewicz, 2019), its $N$-partite generalization (Błasiak *et al.*, 2021b) with success probability $\frac{e^{-1}}{N^2} + O(\frac{1}{N^3})$, and arbitrary states of three qubits with success probability $\frac{1}{18}$ (Błasiak *et al.*, 2022).

The "no-touching" generation schemes can be enhanced if auxiliary particles are allowed. The $N$-partite W state generation protocol presented in (Bellomo *et al.*, 2017) involves $N+1$ initially uncorrelated, spatially separated identical particles, where $N$ particles have pseudospin $\downarrow$ and one particle has pseudospin $\uparrow$. The spatial states of each of the $N$ pseudospin-$\downarrow$ particles is placed in an equal superposition of modes $M_i$ and $C_i$ ($i = 1, 2, \ldots, N$), while the single pseudospin-$\uparrow$ particle is equally split across all $N$ modes $M_i$. Postselection is then performed to retain only events where each mode $M_i$ contains exactly one particle. Next, all particles occupying the intermediate modes $C_i$ are routed to a common ancilla mode $C$, effectively erasing the information about their origin. The final state factorizes into a tensor product of an $N$-particle $W$ state and a single-particle state in the auxiliary mode $C$. This mode acts as an "entanglement catalyst" that, through spatial indistinguishability and local operations alone, generates genuine multipartite entanglement among the remaining $N$ particles. The success probability of this scheme scales as $\frac{1}{N+1}$ for fermions and $\frac{1}{2^N}$ for bosons. An improved optical version of this protocol was proposed in (Kim *et al.*, 2020), where the use of unbalanced beamsplitters increased the success probability to $\frac{(N-1)^{N-1}}{N^{N+1}}$, or $N$ times higher with feed-forward.

A versatile scheme deviating from the "no-touching" paradigm was proposed in (Kumar *et al.*, 2023). It consists in impinging three photons with tunable polarization states on a symmetric 3-port (tritter), before postselecting outcomes where one photon emerges from each port. Specific polarization configurations of the input photons lead to the generation of the $|\mathrm{GHZ}\rangle$ and $|\mathrm{W}\rangle$ states in the same experimental setup, offering a practical advantage over traditional schemes that require rebuilding the optical circuit for each desired state type. Success probability of this scheme is $\frac{1}{9}$ for both types of states.

Another framework for considering entanglement generation schemes stems from the observation that postselection on outcomes in which all particles are found in different spatial regions can be expressed in terms of graph theory. In the *bipartite source-detector graph* (Chin *et al.*, 2021), particles and detectors are vertices while edges are single-particle transition amplitudes. By contrast, in the *pair-source graph* introduced for path identity (Gu *et al.*, 2019b; Krenn *et al.*, 2017a), vertices are output paths and edges are photon-pair sources. Concretely, let us consider a balanced bipartite source-detector graph $G_{bb} = (U \cup V, E)$, where the vertex set $U = \{1, 2, \ldots, N\}$ represents the $N$ identical particles, the vertex set $V = \{X_1, X_2, \ldots, X_N\}$ represents the $N$ detectors, and an edge $(a, X_j) \in E$ exists whenever the transformation amplitude between a specific source and detector, $T_{aj}$, is nonzero. Each edge carries two attributes: a complex weight equal to the amplitude $T_{aj}$, and a color (blue or red) corresponding to the internal state $r_j^a$ ($\uparrow$ or $\downarrow$, respectively). In this representation, the postselected states correspond precisely to *perfect matchings* of $G_{bb}$, i.e., edge subsets in which every particle vertex connects to exactly one detector vertex with no overlaps. The total state is obtained by summing over all perfect matchings, where each matching contributes a computational basis state (determined by edge colors) weighted by the product of its edge weights. This approach can be considered a special case of the graph formalism to describe experiments in quantum optics proposed in (Krenn *et al.*, 2021).

The bipartite graph can equivalently be represented as a directed graph $G_d = (W, F)$ by identifying each particle-detector pair $(a, X_a)$ with a single vertex $w_a \in W$, and mapping each edge $(a, X_j)$ to a directed edge $(w_a \to w_j \in F)$. In this representation, postselected outcomes correspond to elementary cycles. This graph-theoretic picture provides guidelines for designing interferometric schemes that generate specific classes of entangled states, including GHZ, W, Dicke and cluster states, by engineering appropriate graph topologies with suitable edge-color distributions.

The graph approach is also used in (Mahdavipour *et al.*, 2024) to describe the generation of multipartite entanglement proceeding in three steps: initialization, deformation, and activation. Starting from a product state $|\Psi_0\rangle = |\phi_1\rangle \otimes \cdots \otimes |\phi_N\rangle$ of $N$ identical particles, where each $|\phi_i\rangle = |\psi_i, \sigma_i\rangle$ comprises a spatial wavefunction and pseudospin $\sigma_i \in \{\uparrow, \downarrow\}$, a deformation operation coherently distributes the particles across $N$ detection regions. In the remote design, independent unitaries $\hat{D}_N^{(i)}$ transform each qubit as $|\phi_{iD}\rangle = \sum_{j=1}^N r_{ij} |R_j, \sigma_j\rangle$, where $r_{ij} \in \mathbb{C}$ is the probability amplitude for qubit $i$ to arrive at region $R_j$. The on-site design instead applies a single $N$-mode unitary $\hat{D}_N$ to all qubits collectively. When particles spatially overlap at detection regions, the global deformed state $|\Psi_D^{(N)}\rangle$ becomes non-factorizable, and the entanglement is activated with postselection on exactly one particle per region

$$\hat{\Pi}^{(N)} = \sum_k |R_1 \sigma_1^{(k)}, \ldots, R_N \sigma_N^{(k)}\rangle\langle R_1 \sigma_1^{(k)}, \ldots, R_N \sigma_N^{(k)}|. \tag{16}$$

The output state is computed via a weight matrix formalism that incorporates particle exchange statistics. For each detector pseudospin pattern $\{\sigma_1, \ldots, \sigma_N\}$, the weight matrix has elements $[\mathcal{R}_{\sigma_1,\ldots,\sigma_N}]_{ij} = r_{ji} \langle \sigma_i | \tau_j \rangle$, where $\tau_j$ is the original pseudospin of qubit $j$ and the inner product $\langle \sigma_i | \tau_j \rangle$ acts as a selector (unity if $\sigma_i = \tau_j$, zero otherwise). The probability amplitude for each output term is obtained through a determinant-like operation $|\mathcal{R}_{\sigma_1,\ldots,\sigma_N}|_\eta = \sum_P \eta^P \prod_{k=1}^N \langle \chi_k | \phi_{P_k D} \rangle$, where $P$ runs over all particle permutations and $\eta^P$ denotes their parities. The final $N$-qubit state is then

$$|\Psi^{(N)}\rangle = \sum_{\sigma_1,\ldots,\sigma_N \in \{\uparrow,\downarrow\}} \frac{\left|\mathcal{R}_{\sigma_1,\ldots,\sigma_N}\right|_\eta |R_1 \sigma_1, \ldots, R_N \sigma_N\rangle}{\sqrt{N_g}}, \tag{17}$$

with normalization $N_g = \sum_{\sigma_1,\ldots,\sigma_N} ||\mathcal{R}_{\sigma_1,\ldots,\sigma_N}|_\eta|^2$ and success probability $\mathcal{P}(|\Psi^{(N)}\rangle) = \mathrm{Tr}(\hat{\Pi}^{(N)} |\Psi_D^{(N)}\rangle\langle\Psi_D^{(N)}|)$.

The probabilistic nature of sLOCC measurement introduces a trade-off between fidelity $F_\psi = \langle \psi_{\mathrm{target}} | \rho | \psi_{\mathrm{target}} \rangle$ and success probability, which can be optimized by treating the edge weights $\{r_{ij}\}$ as tunable parameters within a chosen graph topology. The optimization proceeds by randomly sampling weights, and identifying the maximum probability at each fidelity level to construct a Pareto frontier. For ideal equal-weight configurations, analytical expressions emerge: $\mathcal{P}_C(|\mathrm{W}^{(N)}\rangle) = (N-1)!/N^{N-1}$ (complete digraph, bosons), $\mathcal{P}_{\mathrm{star}}(|\mathrm{W}^{(N)}\rangle) = 1/N$ (star digraph, fermions), $\mathcal{P}(|\mathrm{GHZ}^{(N)}\rangle) = 1/2^{N-1}$, and $\mathcal{P}(|C^{(N)}\rangle) = (1/9) \cdot 2^{4-N}$ for cluster states with even $N \geq 4$. Notably, fermions outperform bosons for W-state generation due to the Pauli exclusion principle eliminating unwanted interference terms.

As an experimental realization of spatial-overlap-based entanglement generation, Lee *et al.* (Lee *et al.*, 2022) engineered linear-optical transformations that convert three indistinguishable photons into genuine tripartite states of GHZ or W-class.

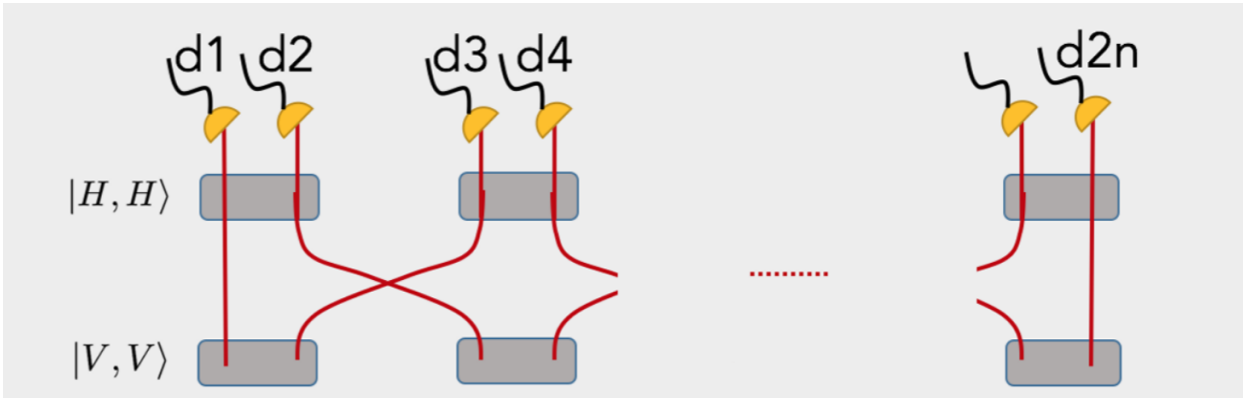

FIG. 8: Generation of GHZ state via path identity. Photon pair sources are aligned in a way that ensures that the coincident registration of particles in all detectors corresponds to the generation of the GHZ state. Adapted from Ref. (Hochrainer *et al.*, 2022).

In the GHZ configuration, a network of polarization beam displacers, half- and quarter-wave plates, and polarization beamsplitters implements a transformation that, under full indistinguishability, produces a GHZ state: making one photon distinguishable (by introducing a temporal delay much larger than the coherence time) rendered the state fully separable, as confirmed by full quantum state tomography. In the W configuration, a fiber tritter with polarization compensation distributed two horizontally ($H$) and one vertically ($V$) polarized photons into three outputs: tomography showed fidelities above the W-class witness threshold when photons were indistinguishable. By deliberately delaying one or more photons, the class degraded predictably: from genuine W to biseparable and finally separable, with fidelities dropping below the entanglement-witness lines.

Finally, let us briefly discuss entanglement generation in the path identity formalism, introduced in Section II.B.3 and covered in depth in the review (Hochrainer *et al.*, 2022). We will focus on a specific scheme generating multipartite GHZ states in an optical setup, depicted in Fig. 8. It consists of two rows of $n$ (unentangled) photon pair sources. Photons generated in the lower row have vertical polarizations, while the ones in the upper row are horizontally polarized. The sources are aligned in such a way that each of the $2n$ detectors can receive particles generated by two distinct sources without being able to distinguish between them. The alignment further ensures that a coincident detection of a single particle in each of the $2n$ detectors can only happen if all the particles were generated by the sources from the lower or upper row, resulting in a polarization-encoded GHZ state. Here, different photons are spatially overlapped on the detection regions by properly superimposing the paths of photon pairs created in different nonlinear crystals, without directly relying on beamsplitters. The resulting correlations can thus be analyzed within the sLOCC framework, although the generation mechanism is not a spatial deformation of pre-existing particles. In the pair-source graph of this setup, the GHZ state is a coherent sum of its two perfect matchings (Fig. 9).

A plethora of path identity-based schemes have been discovered resulting from the indistinguishability of particle generation processes using efficient automated search methods (Gu *et al.*, 2019b; Krenn *et al.*, 2017a). This includes schemes for the three-dimensional four-particle state $|\mathrm{GHZ}_4^3\rangle$, $N$-particle W states, and general Dicke states $|D_n^m\rangle$ (Gu *et al.*, 2019a). Path identity has also been realized experimentally for high-dimensional entangled states (Kysela *et al.*, 2020) and for entangling independent photons (Wang *et al.*, 2024).

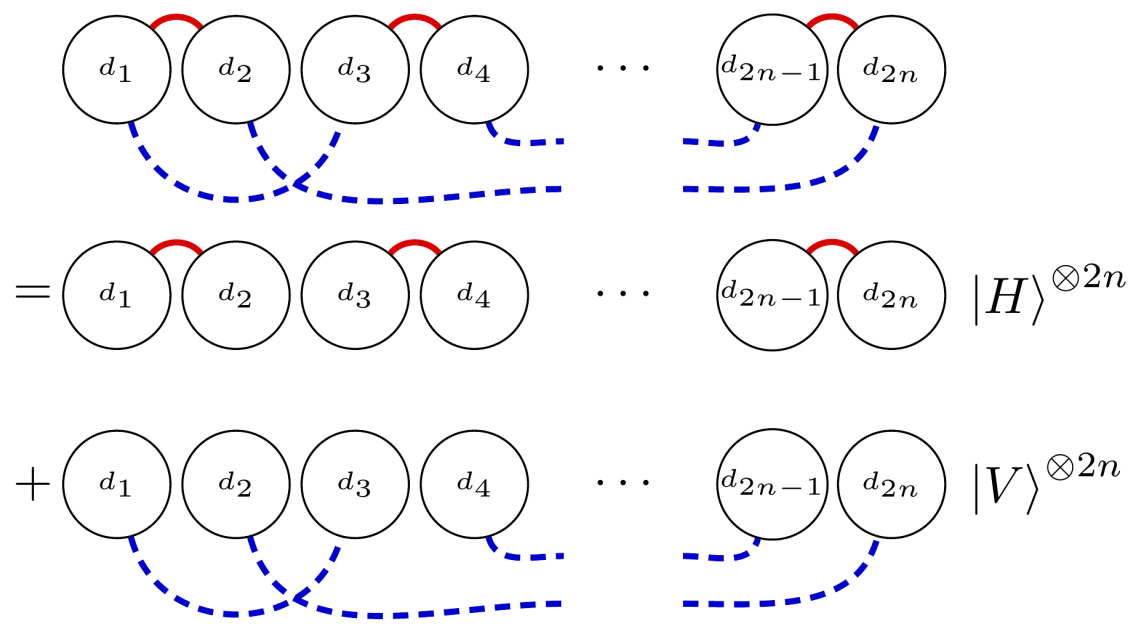

FIG. 9: Pair-source graph of the path-identity scheme of Fig. 8. Vertices are the detectors $d_1, \ldots, d_{2n}$, and each crystal is an edge connecting the two paths it emits into (solid: upper row, $|H,H\rangle$; dashed: lower row, $|V,V\rangle$). The graph is a single $2n$-cycle, whose two perfect matchings give the two terms of the GHZ state.

### 3. Protecting entanglement from noise and imperfections

An important feature of postselected entanglement extraction schemes based on single particle sources consists in their robustness to noise. Following (Bose and Home, 2002b), let us again consider two particles impinging on a beamsplitter. This time, however, the pseudospin state of each of them will be completely random. This means that the input state is an equal mixture of the states $|L\uparrow, R\uparrow\rangle$, $|L\downarrow, R\downarrow\rangle$, $|\Psi^+\rangle = (|L\uparrow, R\downarrow\rangle + |L\downarrow, R\uparrow\rangle)/\sqrt{2}$, and $|\Psi^-\rangle = (|L\uparrow, R\downarrow\rangle - |L\downarrow, R\uparrow\rangle)/\sqrt{2}$.

Note that the first three of them are symmetric, while the last one is antisymmetric. For bosons, this means that coincident detection of a particle in both output modes of a beamsplitter postselects the maximally entangled singlet state $|\Psi^-\rangle$ due to the Hong-Ou-Mandel effect (Hong *et al.*, 1987). For fermions, the protocol consists in nonabsorbing detection of both particles in the same mode (ensuring that they have opposing pseudospins due to the Pauli exclusion principle) and feeding this state back as an input. Again, a subsequent coincident detection would only stem from the singlet state.

The role of spatial indistinguishability in protecting sLOCC entanglement from noise was first demonstrated in (Nosrati *et al.*, 2020), which considered a Bell state corrupted by white noise (a Werner state) and showed that maximal spatial indistinguishability makes the extracted concurrence immune to the noise: observers always recover maximally entangled correlations, with $C_{LR} = 1$ for any noise probability $p$, whereas for distinguishable particles the concurrence decays as $1 - 3p/2$ and vanishes at $p = 2/3$. This immunity holds for both bosons and fermions. We caution, however, that the analysis of (Nosrati *et al.*, 2020) predates the refined understanding of Section II.B.3 and is framed in terms of the original measure $\mathcal{I}$ of Eq. (8), whose interpretation is problematic precisely for the interference-sensitive states involved. The maximal-indistinguishability result quoted above is unaffected, as $\mathcal{I}$ and $\tilde{I}$ coincide, but the intermediate-indistinguishability thresholds there reported should be read with this caveat in mind.

Building on this result, Nosrati *et al.* extended the analysis from noise protection to active entanglement distillation, proposing a conditional sLOCC protocol assisted by the spatial indistinguishability of identical qubits (Nosrati *et al.*, 2024). The protocol consists of a spatial deformation that creates overlap between two initially separated, possibly mixed qubits, followed by a coincidence measurement that postselects one particle in each of the remote regions $L$ and $R$. The interplay between spatial overlap and particle statistics allows entanglement to be distilled from noisy input states, including thermal Gibbs and Werner states. In particular, for bosons, maximal spatial overlap can produce a pure maximally entangled singlet whenever the initial state contains a nonzero contribution from opposite-pseudospin configurations, although with a finite success probability.

Subsequent works reinterpreted these robustness and distillation mechanisms in terms of symmetry-based filtering (Piccolini, 2024; Piccolini *et al.*, 2024a,b), where it was shown that the use of pseudospin-insensitive nonabsorbing parity-check detectors enables generation of maximally entangled N00N and Bell states for both bosons and fermions. The two are maximally entangled according to two distinct notions: the N00N state in the sense of mode entanglement (Section II.A.5), and the Bell state in the sense of detector-level entanglement, which coincides with the standard entanglement of nonidentical particles (Section II.A.6). As all these schemes can succeed even with random input pseudospin states, the states resulting from failed attempts can be reset with local, particle-number preserving noise and reused. In the asymptotic limit they can therefore be considered deterministic. Moreover, (Piccolini *et al.*, 2024b) show that this reasoning can be extended to the robust generation of bosonic $N$-partite $N$-level singlet state. The scheme consists in casting the particles on a sequence of Fourier $M$-ports ($M = N, N-1, \ldots, 2$), each one followed by nonabsorbing detectors checking whether the particles remained separate. With $N$ maximally mixed input particles, this protocol succeeds with probability $\frac{1}{N^N}$.

Finally, these results were placed on a microscopic open-quantum-system footing, where dephasing arises from explicit system–bath interactions rather than a phenomenological noise model. Ferrara *et al.* modeled the effects of dephasing on HOM interference and indistinguishability-assisted entanglement generation. Depending on the system–bath coupling regime, the dynamics can become non-Markovian, with information flowing back from the environment to the system. Counterintuitively, dephasing can stabilize coherence and entanglement instead of suppressing them (Ferrara *et al.*, 2026).

In the above schemes, spatial deformations act as the *activator* of the interference effects responsible for entanglement protection.

### 4. Physical mechanisms and role of particle statistics

The entanglement generation schemes discussed in the previous sections fall into distinctive categories, reflecting the aspects of particle indistinguishability they employ. With that, they feature starkly different dependence on the statistics of involved particles. Let us examine these connections in more detail.

*a. No-touching.* In the “no-touching” paradigm (Blasiak and Markiewicz, 2019), each particle is cast on a multiport performing a local unitary transformation. Next, all output modes of these multiports are globally permuted and divided into sets of inputs of another layer of local unitaries. The detector outcomes are postselected on events in which exactly one particle is found in the outputs of each final unitary.

Importantly, the only points where particles could potentially meet are at local multiports, but single-particle injection guarantees isolation in input regions, while postselection on events with exactly one particle detected in each target output region retroactively ensures particles never met in the output regions either. This eliminates both dynamical interactions (Hamiltonian coupling between modes) and kinematical interference effects (bunching or antibunching), meaning the resulting entanglement arises purely from the spatial overlaps of the wavefunctions at the detectors. Not surprisingly, success probabilities in the “no-touching” schemes do not depend on the particle statistics, being equal for bosons, fermions and even anyons (Blasiak and Markiewicz, 2019; Błasiak *et al.*, 2022, 2021b).

*b. Interference-based.* On the other hand, we have also discussed schemes such as (Bose and Home, 2002b), where the input particles can meet during the evolution. Because of that, their final state depends on interference effects governed by particle statistics or, potentially, particle interactions (Bose and Home, 2013). This approach enables the design of robust schemes based on suppressed interferometric outcomes (Dittel *et al.*, 2018b), at the cost of being tailored to only one type of particles.

There are limits to entanglement that can be generated through multipartite interference (Tichy *et al.*, 2013b). A combinatorial bound shows that the generalized Schmidt rank of any generated state satisfies $R \leq N!$, reflecting that at most $N!$ distinct many-particle paths can contribute to the final state for bosons and fermions alike. Moreover, there are dimensional bounds on the manifold of accessible states: for fermions $\dim(\Xi_F) \leq (d-1)N^2 - N + 2$, while for bosons $\dim(\Xi_B) \leq dN^2 - 2N + 2$, with the bosonic manifold always being larger. This difference originates from the algebraic properties of the determinant (fermions) versus the permanent (bosons) that appear in many-particle scattering amplitudes. However, this bosonic advantage comes at a cost: the Pauli exclusion principle assists fermionic entanglement generation by forbidding multiply-occupied modes, whereas bosonic bunching reduces success probability.

*c. Spatial deformations.* The spatial deformations formalism, proposed in (Lo Franco and Compagno, 2018), bears many

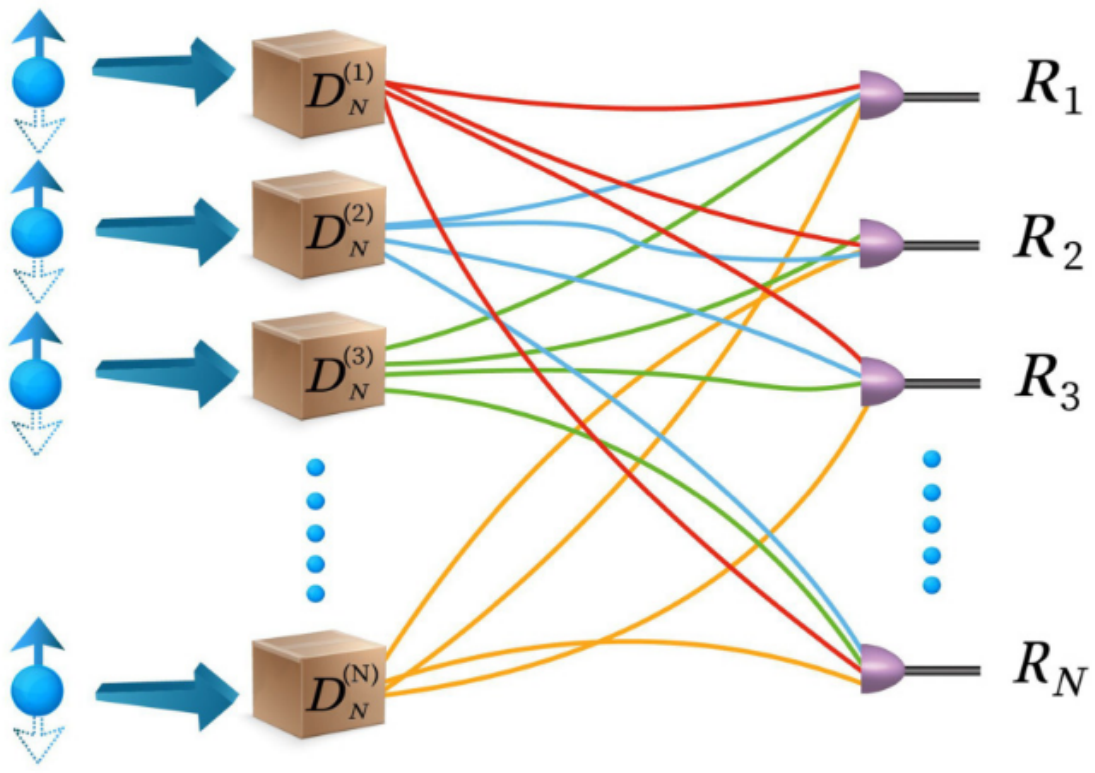

FIG. 10: Entanglement generation protocol based on spatial deformations. $N$ qubits undergo deformations, which send them into $N$ spatial regions. A sLOCC projection measurement postselects the final state. Adapted from Ref. (Mahdavipour *et al.*, 2024).

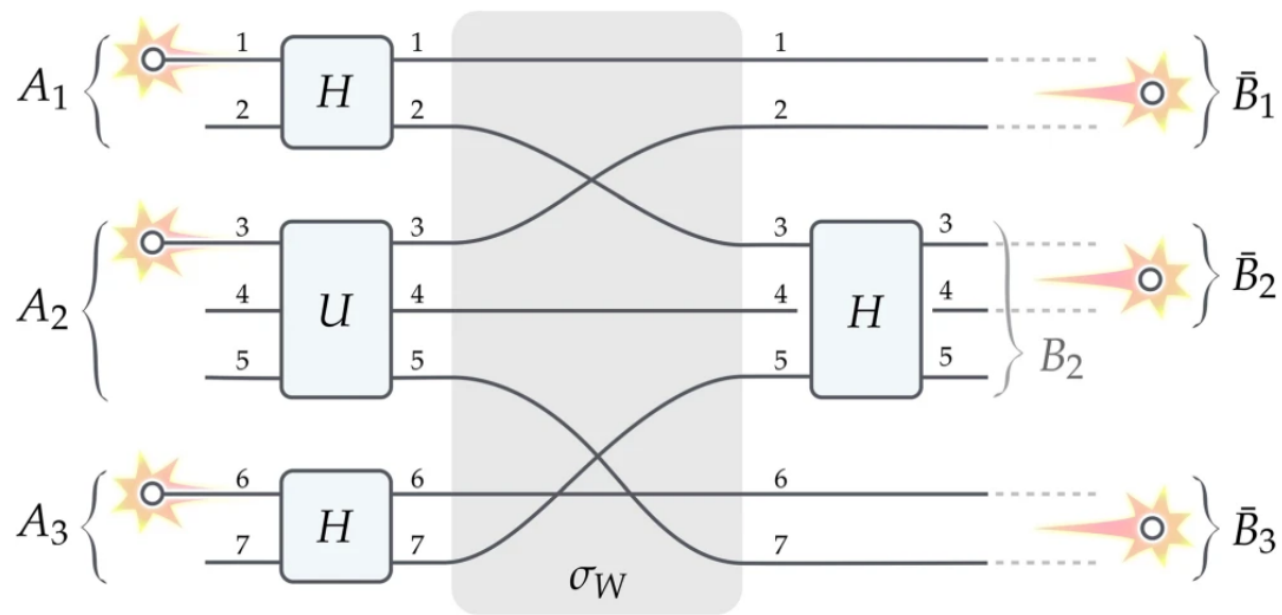

FIG. 11: No-touching scheme for W state generation. Independent single particles impinge on a first layer of multiports. The output ports are then permuted and directed into a second layer, whose output modes define dual-rail qubits. Success is conditioned on detecting exactly one particle per qubit. Adapted from Ref. (Blasiak and Markiewicz, 2019)

similarities with the "no-touching" approach. The central idea is to manipulate the spatial wavefunctions of input particles such that they overlap only at the detectors' locations, where postselection takes place. Each single-particle wavefunction undergoes a spatial deformation. The effective transformation of the multipartite state of indistinguishable particles is, in general, non-unitary. To illustrate this, consider a specific instance of the general setup depicted in Fig. 10, involving three particles distributed across three spatial regions. Each particle carries a two-level internal degree of freedom (pseudospin), with creation operators $a^\dagger$ and $b^\dagger$ corresponding to the two internal states, while the subscript denotes the spatial region. The initial state $b_1^\dagger a_2^\dagger a_3^\dagger|0\rangle$ describes one qubit in each region, with the qubit in region 1 in internal state $b$ and the qubits in regions 2 and 3 in internal state $a$. The spatial deformations spread each particle's wavefunction across multiple regions

$$b_1^\dagger a_2^\dagger a_3^\dagger|0\rangle \rightarrow \frac{1}{\sqrt{N_\eta}} \frac{b_1^\dagger + b_2^\dagger + b_3^\dagger}{\sqrt{3}} \frac{a_1^\dagger + a_2^\dagger}{\sqrt{2}} \frac{a_1^\dagger + a_3^\dagger}{\sqrt{2}}|0\rangle, \quad (18)$$

where the factor $1/\sqrt{N_\eta}$ ensures proper normalization of the resulting state. Crucially, this normalization factor differs for bosons and fermions due to their distinct exchange statistics. As a result, the probability of successful postselection, conditioned on detecting exactly one particle in each spatial region, depends on whether the particles are bosons or fermions. The successfully postselected state is the W state, which in an analogous "no-touching" scheme depicted in Fig. 11 is generated with a statistics-independent success probability. The statistics-dependent normalization factor $1/\sqrt{N_\eta}$ in the spatial deformations approach reflects the role of many-body interference in this framework, consistent with the discussion at the beginning of this section: when the deformation causes particles to spatially overlap, it activates interference effects that depend on the exchange symmetry of the many-body state, distinguishing the spatial deformations approach from the interference-free "no-touching" paradigm.

*d. Path identity.* The path identity approach generates entanglement through two key ingredients: a superposition of particles from different origins, and the alignment of particles so that their paths become identical. Here the relevant ambiguity is of a different character than in the schemes discussed so far: rather than overlapping pre-existing particles at the detectors, path identity withholds which-source information already at the particle-generation stage, so there is no path information to erase. The correlations revealed by the final coincidence postselection can nonetheless be analyzed within the same sLOCC framework (see Section II.B.3).

In pair-source implementations of this mechanism, such as the construction in Fig. 8, a distinctive requirement is that the individual sources must *not* emit a fixed number of particles: only then can the final state be a coherent superposition of contributions from different origins. In the optical realizations based on spontaneous parametric down-conversion, this indefiniteness is precisely the per-source superposition of vacuum and a photon pair. The entanglement then resides in the postselected sector containing a definite total number of photons, generated in unidentified crystals. This requirement is, however, specific to pair-sources rather than to path identity itself, as a broader graph framework describes path identity for deterministic single-photon (Krenn *et al.*, 2021; Ruiz-Gonzalez *et al.*, 2023), and multiphoton (Gu *et al.*, 2020) sources.

### 5. Coherence generation

The spatial overlap of identical particles can generate quantum resources beyond entanglement. In Ref. (Castellini *et al.*, 2019), particle indistinguishability is established as a source of quantum coherence, even for independently prepared systems. Consider two identical particles with spatial wavefunctions $|\psi\rangle$ and $|\psi'\rangle$ and internal states (pseudospins) $\sigma, \tau \in \{\uparrow, \downarrow\}$, initially prepared in the mixed state

$$\rho^I = \sum_{\sigma,\tau=\downarrow,\uparrow} p_{\sigma\tau} |\psi\sigma, \psi'\tau\rangle \langle\psi\sigma, \psi'\tau|, \quad (19)$$

where the coefficients $p_{\sigma\tau}$ satisfy $\mathrm{Tr}[\rho^I] = 1$. As already reported, in the no-label formalism each term $|\psi\sigma, \psi'\tau\rangle$ is a global state describing two identical particles which cannot be written as a tensor product $|\psi\sigma\rangle \otimes |\psi'\tau\rangle$. When this state is projected onto the subspace spanned by the basis $\mathcal{B}^I = \{|L\sigma, R\tau\rangle ; \sigma, \tau = \downarrow, \uparrow\}$, where $|L\rangle$ and $|R\rangle$ are states spatially localized in separate regions $L$ and $R$, it reads

$$\rho^I_{LR} = \frac{1}{N} \sum_{\sigma,\tau=\downarrow,\uparrow} p_{\sigma\tau} \Big( |l|^2 |r'|^2 \, |L\sigma, R\tau\rangle \langle L\sigma, R\tau| + \eta l r' l'^* r^* \, |L\sigma, R\tau\rangle \langle L\tau, R\sigma| + \eta l' r l^* r'^* \, |L\tau, R\sigma\rangle \langle L\sigma, R\tau| + |l'|^2 |r|^2 \, |L\tau, R\sigma\rangle \langle L\tau, R\sigma| \Big), \tag{20}$$

where $N$ is a normalization factor, $l = \langle L|\psi\rangle$, $r' = \langle R|\psi'\rangle$, $l' = \langle L|\psi'\rangle$, and $r = \langle R|\psi\rangle$ are probability amplitudes for finding a particle in the localized states, and $\eta = +1$ for bosons and $\eta = -1$ for fermions. The off-diagonal terms proportional to $lr'$ and $l'r$ represent quantum coherence. This "indistinguishability-based coherence" is quantified using the $l_1$-norm of coherence (Streltsov *et al.*, 2017),

$$C_{l_1}(\rho) = \sum_{i \neq j} |\rho_{ij}|, \tag{21}$$

applied to the reduced single-particle density matrix obtained under sLOCC projection. It vanishes entirely when the particles do not spatially overlap ($l' = r = 0$), reducing indistinguishable particles to effectively distinguishable ones. This sharply contrasts with distinguishable particles, for which the analogous state $\rho^{NI}$ remains diagonal (incoherent) in the basis $\mathcal{B} = \{|L\sigma\rangle_A \otimes |R\tau\rangle_B\}$ regardless of spatial overlap.

The results of (Castellini *et al.*, 2019) closely parallel those of (Lo Franco and Compagno, 2018). Just as Lo Franco and Compagno show that spatial overlap of identical particles creates entanglement absent for distinguishable particles, Castellini et al. demonstrate that the same overlap creates coherence. In both cases, the operational basis $\{|L\sigma, R\tau\rangle\}$ serves as the preferred reference, reflecting the physical constraint that identical particles can only be accessed through spatially separated measurement regions rather than individual addressing. Together, these works establish indistinguishability as a unified source of quantum resources, where spatial overlap enables extraction of both coherence and entanglement from independently prepared identical particles.

Sun *et al.* (Sun *et al.*, 2022) then showed that two independent photons, when coherently distributed into two distant regions, exhibit a novel form of coherence that arises purely from spatial overlap and particle identity, rather than from single-particle superposition. In the experimental realization, the two-photon resource state $|\Psi_{LR}(\theta)\rangle = \cos\theta \, |L\uparrow, R\downarrow\rangle + \sin\theta \, |L\downarrow, R\uparrow\rangle$ is engineered by splitting the spatial amplitudes with half-wave plates and polarizing beamsplitters. The resulting coherence is given analytically by $C_{l_1}(\Psi_{LR}(\theta)) = |\sin 2\theta|$, which increases monotonically with the degree of spatial indistinguishability $\mathcal{I}(\theta) = -\cos^2\theta \, \log_2(\cos^2\theta) - \sin^2\theta \, \log_2(\sin^2\theta)$.

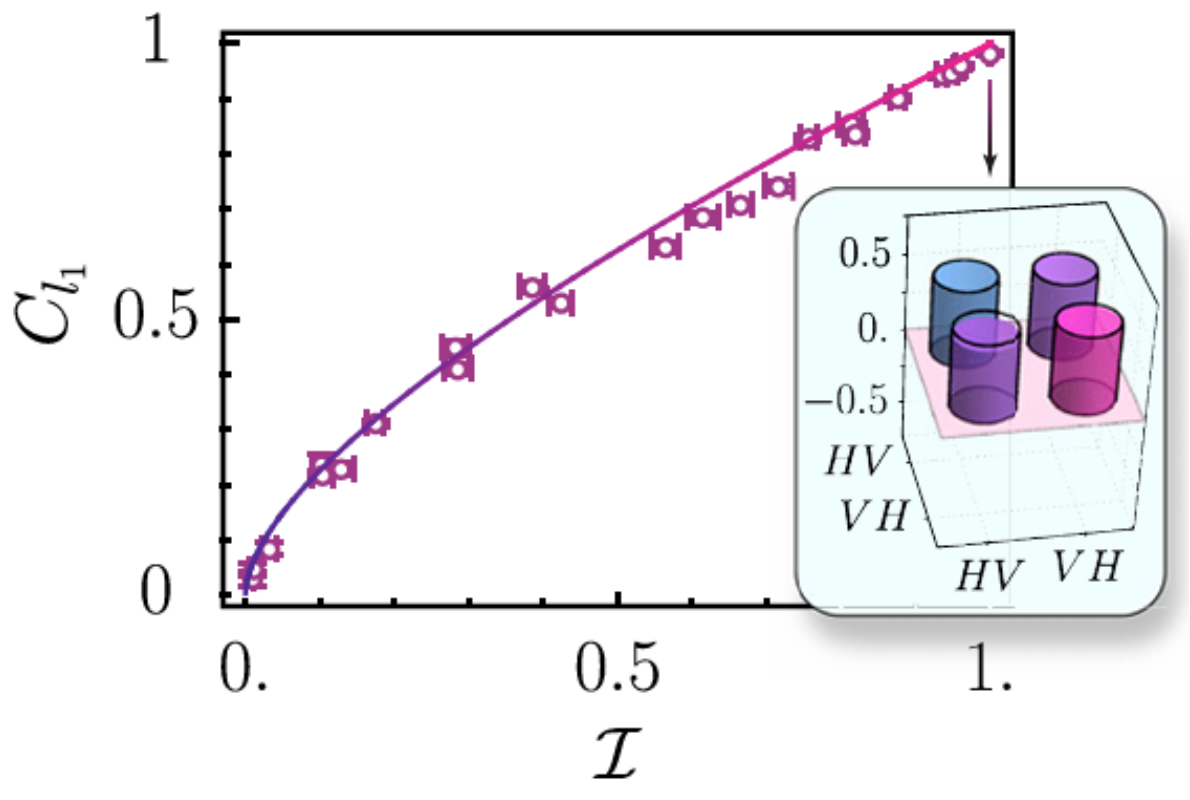

FIG. 12: The coherence $C_{l_1}$ increases monotonically with the spatial indistinguishability $\mathcal{I}(\theta)$. Figure adapted from Ref. (Sun *et al.*, 2022).

Experimentally, when the photons are distinguishable, the reduced density matrix was diagonal and $C_{l_1} = 0$. As the spatial indistinguishability was tuned toward $\mathcal{I}(\theta) \to 1$, off-diagonal elements emerge and $C_{l_1}$ increase accordingly, in excellent agreement with theoretical predictions. At maximal indistinguishability, the measured state fidelity reach 0.988, confirming the high quality of the generated coherence. The monotonic relation between $C_{l_1}$ and $\mathcal{I}$ is illustrated in Fig. 12.

Beyond generating indistinguishability-based coherence, Perez-Leija *et al.* show that two-particle exchange coherence is more robust to noise compared to single-particle coherence (Perez-Leija *et al.*, 2018). They find that both bosonic and fermionic systems retain coherence associated with particle indistinguishability, even when initialized in classically correlated states. This behavior was experimentally investigated using dynamically disordered photonic waveguide networks, where indistinguishable photon pairs converge toward characteristic steady-state correlation patterns that remain distinct from those of distinguishable photons.

### B. Heralded schemes

In this section, we briefly present the key ideas underlying heralded entanglement generation with single-particle sources, focusing on the role of indistinguishability in the process. For a more thorough review of the topic, we refer the reader to (Forbes *et al.*, 2025).

#### 1. Basic idea

To explain the idea of heralded entanglement generation, let us present a conceptually simple and versatile method proposed in (Paesani *et al.*, 2021). It is based on a zero-transmission law, specifying which output patterns are allowed when photons propagate through an $m$-mode discrete Fourier transform (DFT) interferometer (Tichy *et al.*, 2010). In such an interferometer, the creation operator $\hat{a}_j^\dagger$ for a photon in input mode $j$ transforms according to

$$\hat{a}_j^\dagger \to \frac{1}{\sqrt{m}} \sum_{k=0}^{m-1} e^{2\pi \mathrm{i} jk/m} \hat{b}_k^\dagger, \tag{22}$$

where $\hat{b}_k^\dagger$ creates a photon in output mode $k$. This transformation is characterized by the unitary matrix $U_{j,k} = \exp[2\pi \mathrm{i} jk/m]/\sqrt{m}$. The zero transmission law then states that if $m$ single photons are injected into the $m$ input modes of this interferometer, only those output configurations $\vec{c} = (c_1, c_2, \ldots, c_m)$ satisfying

$$\sum_{k=0}^{m-1} kc_k \mod m = 0, \tag{23}$$

can have nonzero probability amplitude. For instance, we can see that for two photons impinging on a Hadamard beamsplitter, the antibunched output configuration $(1, 1)$ is ruled out, which is the well known HOM effect (Hong *et al.*, 1987).

To see how this tool can be useful in designing entanglement generation schemes, consider a Fourier 5-port, in which mode 0 will contain a heralding photon number resolving detector. If it registers 3 photons, the only output configurations satisfying Eq. (23) are (3,1,0,0,1) and (3,0,1,1,0). Direct calculation confirms that these two outcomes have equal nonvanishing weights. Thus, detecting 3 photons in mode 0 certifies that a Bell state was created in the two dual-rail qubits formed by the remaining modes. The same technique was used to design first proof-of-principle schemes for heralded generation of arbitrary $d$-level $N$-partite GHZ states, an important resource in universal high-dimensional quantum computation, with $(N^{dN} - 1)/(N^d - 1)$ particles (Paesani *et al.*, 2021). Exponentially more efficient schemes targeting these states, requiring $dN$ particles, were later discovered using graph techniques similar to those described in previous section (Chin *et al.*, 2024a, 2021, 2024b).

This example illustrates key features of heralded entanglement generation. In contrast to the postselected schemes, heralded ones provide immediate feedback on their success. If a specific pattern of particle detections is registered in selected modes, it guarantees that the target entangled state was produced in the remaining ones. This characteristic allows probabilistic heralded sources of entangled states to, in principle, be combined to form a quasi-deterministic source. However, this method requires more modes and initial particles compared to postselection schemes, and, as we shall see below, is limited to bosons and prone to particle losses and distinguishability.

### 2. Role of particle statistics

Heralded entanglement generation schemes based on single-particle sources and linear-optical transformations are limited to bosonic systems. Following (Karczewski *et al.*, 2019), consider a state of $N$ fermions $|\mathrm{asym}_N\rangle = f_1^\dagger \cdots f_N^\dagger |0\rangle$, where $f_i^\dagger$ represents the fermionic creation operator for mode $i = 1, \ldots, N$. Suppose now that a single fermion is detected in a heralding mode

$$f = \sum_{i=1}^{N} \alpha_i f_i + \beta f_\perp = \alpha f_{\mathrm{sup}} + \beta f_\perp, \tag{24}$$

where $f_\perp$ acts on modes not in the support of $|\mathrm{asym}_N\rangle$. This situation can be described as

$$f|\mathrm{asym}_N\rangle = f f_1^\dagger \cdots f_N^\dagger |0\rangle = \alpha f_{\mathrm{sup}} f_1^\dagger \cdots f_N^\dagger |0\rangle. \tag{25}$$

However, as $|\mathrm{asym}_N\rangle$ is invariant under single-particle unitary transformations $U f_i = \tilde{f}_i$, we have

$$f|\mathrm{asym}_N\rangle = \alpha f_{\mathrm{sup}} f_{\mathrm{sup}}^\dagger \tilde{f}_2^\dagger \cdots \tilde{f}_N^\dagger |0\rangle \equiv |\mathrm{asym}_{N-1}\rangle. \tag{26}$$

This state $|\mathrm{asym}_{N-1}\rangle$ corresponds to a single Slater determinant, and therefore lacks correlations (Eckert *et al.*, 2002). The same argument holds for the subtraction of multiple fermions.

From the practical viewpoint, heralded entanglement generation schemes are most often considered in optical systems. On the one hand, quantum optics provides a mature experimental platform: single photons can be reliably generated using spontaneous parametric down-conversion or single-photon emitters such as quantum dots, their quantum states can be precisely manipulated using linear optical elements, and they can be detected with high efficiency using superconducting nanowire or avalanche photodiode detectors. Furthermore, photons are naturally robust carriers of quantum information because they interact only weakly with their environment, which suppresses decoherence and makes them ideal for transmitting entanglement over long distances in quantum communication networks.

On the other hand, this same weak interaction presents a fundamental obstacle: photons do not directly interact with one another under ordinary conditions, which means that deterministic two-qubit gates are difficult to implement. It is precisely this limitation that motivates the heralded approach: rather than forcing photons to interact directly, one exploits their quantum indistinguishability and the coherent superposition of many-particle paths to generate entanglement probabilistically, with successful events identified through specific detection patterns at the output (Forbes *et al.*, 2025).

### 3. Effects of noise and imperfections

*a. Particle loss* Realistic implementations of heralded entanglement generation schemes are inevitably affected by particle loss, which can occur at various stages of the protocol: during propagation through the optical network, at inefficient detectors, or due to imperfect particle sources. In contrast to postselected schemes, where particle loss reduces only the rate of successful events without compromising the quality of the output state, heralded schemes suffer also from reduced fidelity of the generated entangled state.

The impact of particle loss depends on where in the setup it occurs. If a particle is lost from a heralding mode that was supposed to register a detection, a genuinely successful entanglement generation event may be incorrectly discarded. Conversely, and more problematically, if the heralding mode initially contained more particles than intended (for instance, due to higher-order emission events from the source), losing the excess particles can cause an unsuccessful event to produce the correct heralding pattern, leading to false acceptance. In this case, the output state in the target modes does not possess the desired entanglement structure, yet it is erroneously certified as valid. Losses in the target modes, on the other hand, directly degrade the fidelity of the output state by removing particles that should carry the entanglement, resulting in a state with reduced particle number.

*b. Particle distinguishability* The generation of entanglement through interference fundamentally relies on the indistinguishability of the input particles. When they become partially distinguishable, for instance due to mismatches in their spectral, temporal, or spatial profiles, which-way information is introduced into the system. This degrades the interference underlying entanglement generation. In postselected schemes, partial distinguishability reduces the visibility of interference and thus the quality of entanglement, but the postselection condition (detecting exactly one particle in each output region) ensures that the output state remains within the logical subspace of valid $N$-qudit states. This built-in robustness arises because postselection filters out events that would otherwise populate invalid configurations.

Heralded schemes, however, do not benefit from this protection. Even when the heralding detectors register the correct pattern, partial distinguishability can cause the state in the target modes to fall outside the intended logical subspace. This occurs because the heralding measurement does not fully project onto the desired many-particle output pattern when the particles carry distinguishing information in unmonitored degrees of freedom. The result is an output state that, despite passing the heralding check, may contain components with incorrect mode occupations or unwanted correlations with the distinguishing degrees of freedom. Such errors are particularly problematic because they are not flagged by the heralding signal, leading to a reduction in fidelity that cannot be diagnosed without costly procedures, like full state tomography, on the target modes.

A promising method of mitigating these problems by increasing the indistinguishability of input particles is proposed in (Sparrow, 2018). It consists in arranging the interference of multiple partially distinguishable particles in such a way that a specific heralding pattern is more likely to occur when particles are indistinguishable. This way, its detection offers a probabilistic method of preparing fewer, more indistinguishable particles from the input ones. The simplest indistinguishability distillation scheme is presented in Fig. 13a. Two photons, each in a mixed internal state $\rho = \sum_i p_i |\xi_i\rangle\langle\xi_i|$, are sent to a balanced beamsplitter. The key observation is that identical photons bunch with unit probability due to HOM interference, while orthogonal photons bunch with probability $1/2$. Thus observing a bunching event, heralded by detecting vacuum at one output port, increases the likelihood that the photons were in the same internal mode. A second beamsplitter then extracts a single photon via heralded subtraction, conditioned on detecting exactly one photon in the auxiliary mode. The resulting single-photon state has modified weights $p_i' \propto p_i(p_i + 1)$, meaning that dominant terms are amplified relative to smaller ones. For nearly pure initial states with $p_0 = 1 - \epsilon$, the filtering approximately halves the error probability, yielding $p_0' \approx 1 - \epsilon/2$, with success probability of $(P + 1)/8$, where $P = \sum_i p_i^2$ is the initial purity. A multi-copy generalization of this scheme, depicted in Fig. 13b, achieves $n$-fold decrease of error probability with a success probability vanishing faster than exponentially with $n$.

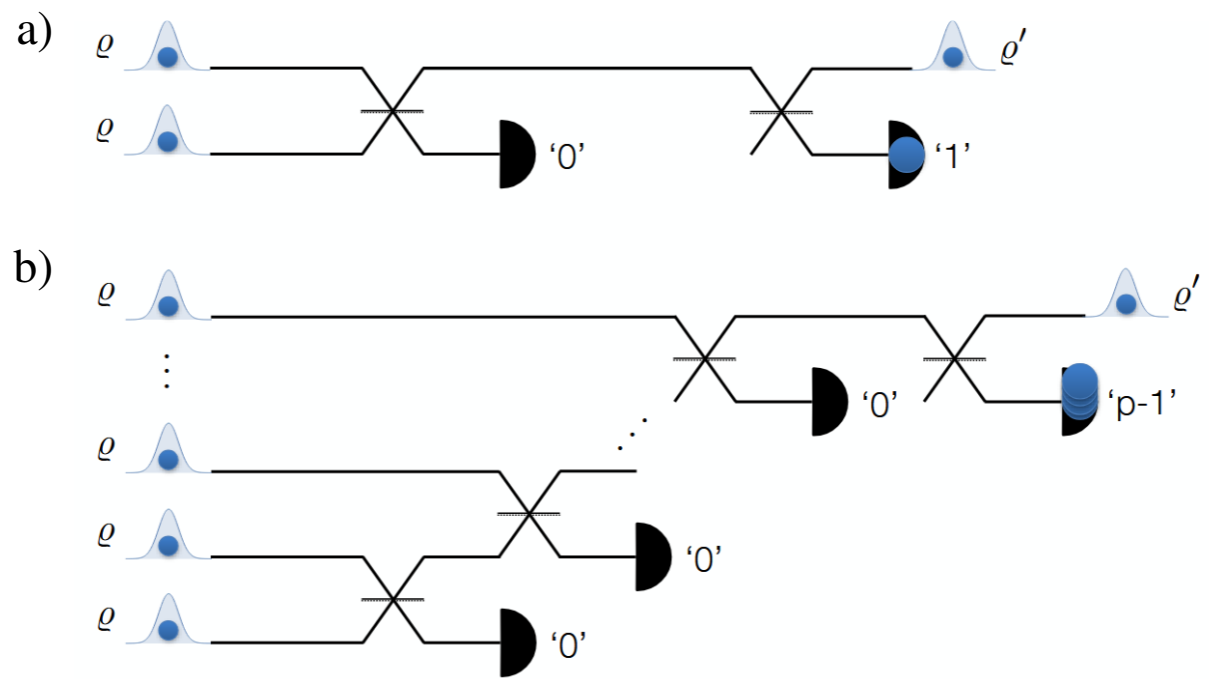

FIG. 13: Simple indistinguishability distillation schemes. A pair (a) or multiple (b) photons impinge on a layer of beamsplitters. Bunching of all photons into a single mode, confirmed by null detections in the remaining modes, is followed by a final beamsplitter. Detection of all but one photon in one output port heralds the distillation of a single photon with increased indistinguishability in the other. Adapted from Ref. (Sparrow, 2018).

The idea behind these schemes, dubbed "indistinguishability distillation", was subsequently developed in (Marshall, 2022). The scheme proposed there requires $O((\epsilon/\epsilon')^2)$ photons to reduce the distinguishability error from $\epsilon$ to $\epsilon'$, or equivalently $O(n^2)$ photons to achieve an $n$-fold error reduction through concatenated rounds of the protocol. These results were improved in (Saied *et al.*, 2025), by recognizing that Fourier transforms on finite abelian groups provide a natural framework for distillation, enabling error reduction by a factor of $n$ using only $O(n)$ photons in a single round. Independently, similar conclusions were reached in (Somhorst *et al.*, 2025) by studying multiphoton interference in discrete Fourier transform multiports, demonstrating that the success probability tends to a constant for arbitrary error reductions and that multiple valid herald measurements exist, providing different paths toward improved indistinguishability. In the case of three photons, an optimized distillation protocol that maximizes visibility gain by accounting for multiphoton effects such as collective photonic phases was presented and experimentally validated in (Hoch *et al.*, 2025).

## IV. APPLICATIONS IN QUANTUM METROLOGY

Quantum-enhanced metrology seeks to estimate physical parameters more precisely than is possible classically with comparable resources (Albarelli *et al.*, 2020; Braun *et al.*, 2018; Giovannetti *et al.*, 2011; Helstrom, 1969; Holevo, 1976;

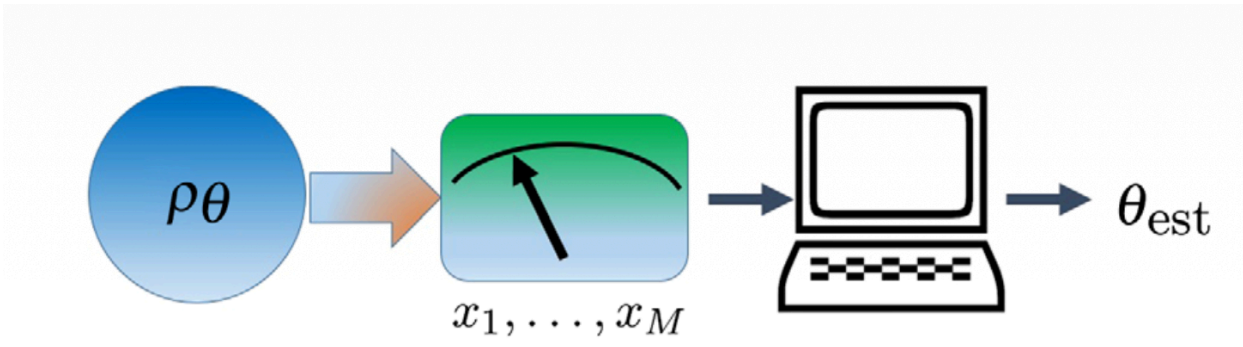

FIG. 14: General framework of quantum parameter estimation. A probe state $\rho_\theta$, carrying information about an unknown parameter $\theta$, is measured by a POVM, producing outcomes $x_1, \ldots, x_M$. An estimator constructed from these data yields an estimate $\theta_{\rm est}$. Adapted from Ref. (Braun *et al.*, 2018).

Paris and Rehacek, 2004; Paris, 2009; Pezze and Smerzi, 2014; Pezze *et al.*, 2018; Polino *et al.*, 2020; Sidhu and Kok, 2020; Tóth and Apellaniz, 2014; Zhang and Zhuang, 2021). Its applications span gravitational-wave detection, precision spectroscopy, atomic clocks, and quantum imaging. The ultimate precision is determined jointly by the probe state and the measurement strategy. While most studies focus on resources such as entanglement and squeezing, here we examine the distinct operational role of indistinguishability.

### A. Brief review of quantum parameter estimation

Rather than reviewing quantum metrology in detail, we recall only the concepts needed to identify where indistinguishability enters as a metrological resource. In classical parameter estimation, an unknown parameter $\theta$ is inferred from measurement outcomes distributed according to $p(x|\theta)$ (Fig. 14). For $M$ independent measurements and an unbiased estimator, the estimation uncertainty is bounded by the Cramér–Rao inequality (Cramér, 1999; Rao *et al.*, 1945), $\delta\theta \geq 1/\sqrt{MF_\theta}$, where

$$F_\theta = \int \mathrm{d}x\, p(x|\theta) \left[ \frac{\partial}{\partial\theta} \ln p(x|\theta) \right]^2, \tag{27}$$

is the Fisher information.

In the quantum setting, the parameter is encoded in a family of probe states $\rho_\theta$, typically through a unitary transformation, $\rho_\theta = U_\theta \rho U_\theta^\dagger$. A measurement described by a positive operator-valued measure (POVM) $\{M_o\}$ produces the outcome distribution (Helstrom, 1969) $p(o|\theta) = \mathrm{Tr}[M_o \rho_\theta]$. Quantum parameter estimation therefore involves two optimizations: the preparation of the probe state and the choice of measurement. For a fixed probe state, the Fisher information associated with any POVM is bounded by the quantum Fisher information (QFI), $F_\theta \leq F_Q$, with $F_Q[\rho_\theta] = \mathrm{Tr}\left[\rho_\theta L_\theta^2\right]$, where the symmetric logarithmic derivative $L_\theta$ is defined implicitly by

$$\frac{\partial \rho_\theta}{\partial \theta} = \frac{1}{2} \left( L_\theta \rho_\theta + \rho_\theta L_\theta \right). \tag{28}$$

For a pure state undergoing unitary encoding, $|\psi_\theta\rangle = e^{-i\theta G}|\psi\rangle$, the QFI simplifies to $F_Q = 4(\Delta G)^2$, where $(\Delta G)^2$ is the variance of the generator $G$ in the input state. This expression shows that metrological enhancement depends on the structure of the probe state relative to the generator of the parameter transformation.

In principle, the measurement that saturates the quantum Cramér–Rao bound can be constructed from the symmetric logarithmic derivative $L_\theta$: projectors onto its eigenstates define an optimal POVM (Braunstein and Caves, 1994). However, this construction generally depends on the true value of $\theta$, which is precisely the unknown parameter (Paris, 2009; Pezze and Smerzi, 2014). In practice, one therefore treats the quantum Cramér–Rao bound as a local benchmark and asks whether physically accessible measurements can approach it in the asymptotic limit of many repetitions.

Several approaches address the absence of prior knowledge, including the van Trees inequality (Van Trees, 2004), Bayesian methods (Kramer and Sorenson, 2002; Lane *et al.*, 1993; Macieszczak *et al.*, 2014), and adaptive measurement strategies (Armen *et al.*, 2002; Fujiwara, 2006; Wheatley *et al.*, 2010). Also, the quantum Cramér–Rao bound is generally attainable only asymptotically. For finite data sets, which are most relevant experimentally, the bound does not need to be saturable, and finite-sample corrections become important (Meyer *et al.*, 2025).

As two paradigmatic examples, consider first $N$ independent distinguishable probes, each undergoing the local unitary encoding $|\psi_\theta\rangle = e^{-i\theta G}|\psi\rangle$. Optimizing over all separable probe states and measurements still yields the bound $F_Q \leq N$ (under the usual single-probe normalization). Consequently, the quantum Cramér–Rao bound is limited by

$$\Delta\theta_{\min} \propto \frac{1}{\sqrt{N}}, \tag{29}$$

known as the standard quantum limit. Thus, Surpassing this bound requires nonclassical correlations between the probes.

A paradigmatic entangled probe is the GHZ state (Greenberger *et al.*, 1989)

$$|\mathrm{GHZ}\rangle = \frac{1}{\sqrt{2}} \Big( |0, \ldots, 0\rangle + e^{-iN\theta} |1, \ldots, 1\rangle \Big), \tag{30}$$

where the parameter is collectively encoded on all $N$ particles. In this case, $F_Q = N^2$, yielding the Heisenberg limit,

$$\Delta\theta_{\min} \propto \frac{1}{N}. \tag{31}$$

Both the standard quantum limit and the Heisenberg limit are conventionally formulated for individually addressable particles, such as trapped atoms or ions, whose internal states define effective spin-$\frac{1}{2}$ systems (Giovannetti *et al.*, 2006, 2011; Pezze *et al.*, 2018). In the following, we discuss how this framework is modified when the probes are identical and cannot be individually addressed.

### B. Indistinguishability in quantum parameter estimation

The concept of indistinguishability becomes more nuanced in interferometric platforms. A broad class of phase-estimation protocols shares the same mathematical structure (Lee *et al.*, 2002b; Wineland *et al.*, 1994; Yurke *et al.*, 1986). The essential ingredient is not the specific experimental realization, such as Ramsey interferometry (Ramsey, 1950) or the Mach–Zehnder interferometer shown in Fig. 15, but the existence of two modes, denoted by $A$ and $B$, between which a relative phase is encoded. The metrological task is then to estimate this relative phase.

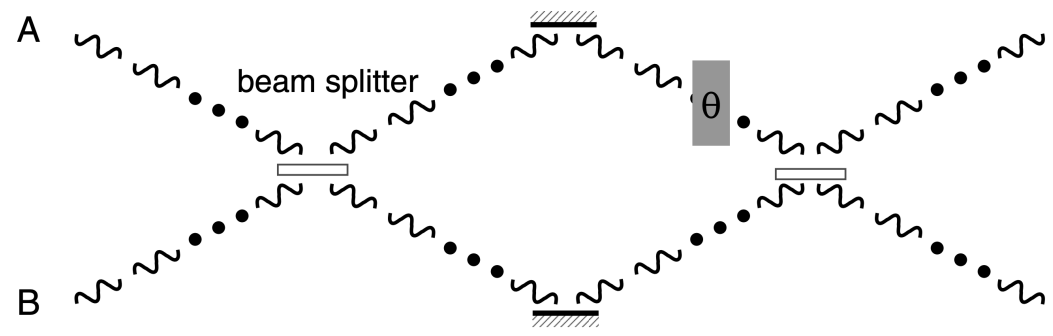

FIG. 15: An optical Mach–Zehnder interferometer for estimating a relative phase $\theta$ between two modes. The input modes $A$ and $B$ are mixed, phase shifted, recombined, and detected. Adapted from Ref. (Lee *et al.*, 2002b).

A useful baseline is the input state $|N,0\rangle$, in which $N$ indistinguishable photons (or, more generally, bosons) occupy a single input mode. After the first balanced beamsplitter and phase encoding, the state becomes

$$|\psi(\theta)\rangle = \sum_{n=0}^{N} \sqrt{\frac{\binom{N}{n}}{2^N}}\, e^{in\theta}\, |N-n,n\rangle\,, \tag{32}$$

which yields the standard quantum-limit scaling, $\Delta\theta_{\min} \propto \frac{1}{\sqrt{N}}$, as dictated by the quantum Cramér–Rao bound (Braunstein and Caves, 1994; Helstrom, 1969; Holevo, 1976; Kok *et al.*, 2002; Lang and Caves, 2013; Lee *et al.*, 2002b). The origin of this scaling is evident from the detection statistics. The natural observable is the photon-number difference, $\hat{M} = \hat{n}_A - \hat{n}_B$ for which $\langle\hat{M}\rangle = N\cos\theta$, while the fluctuations scale as $\sqrt{N}$ (Yurke, 1986). Thus, the signal grows linearly with $N$, whereas the noise grows only as $\sqrt{N}$. Physically, the phase is accumulated independently by each photon, so the measurement probes the sum of $N$ single-particle interference processes rather than genuine $N$-particle interference (Kok *et al.*, 2002; Mandel, 1991). This behavior is analogous to coherent laser light, for which $\Delta\theta_{\min} \propto 1/\sqrt{\bar{n}}$, with $\bar{n}$ the mean photon number (Gerry and Knight, 2023).

A different situation arises when both input modes are populated. Holland and Burnett proposed using the two-mode Fock state $|N-n,n\rangle$ as a probe for phase estimation in a Mach–Zehnder interferometer (Holland and Burnett, 1993), later analyzed in detail by Lang and Caves (Lang and Caves, 2013). After the first beamsplitter, the output is no longer described by the single binomial distribution of Eq. (32). Instead, the amplitudes result from coherent superpositions of many-particle interference pathways, modifying the output photon-number statistics and enabling asymptotic Heisenberg scaling, $\Delta\theta_{\min} \propto 1/N$.

Another paradigmatic resource state for quantum-enhanced metrology is the NOON state (Lee *et al.*, 2002b)

$$|\mathrm{NOON}_\theta\rangle = \frac{1}{\sqrt{2}}\left(|N,0\rangle + e^{iN\theta}\,|0,N\rangle\right), \tag{33}$$

which corresponds to the coherent superposition in which all $N$ indistinguishable photons occupy either mode $A$ or mode $B$. The encoded phase is therefore amplified by a factor $N$, producing interference fringes proportional to $\cos(N\theta)$ that oscillate $N$ times faster than those of the single-input state in Eq. (32) (Kok *et al.*, 2002; Lee *et al.*, 2002b). Consequently, the phase sensitivity reaches the Heisenberg limit.

The operational distinction between the binomial input state and the NOON state therefore does not lie in the presence of indistinguishable particles alone, as both contain $N$ identical photons. Rather, it lies in how indistinguishability manifests itself through quantum interference. In the binomial state, the phase is accumulated independently by each photon, producing first-order interference fringes proportional to $\cos\theta$. The NOON state of Eq. (33), by contrast, yields fringes proportional to $\cos(N\theta)$. The metrological enhancement therefore originates from genuine multiphoton interference, where the phase information is encoded collectively. This operational role of indistinguishability is closely connected to the multiphoton interference witnesses discussed in Sec. II.B.2, where the visibility of collective interference directly quantifies genuine $N$-photon indistinguishability.

This observation has motivated extensive efforts to realize NOON-like states across different physical platforms, since their metrological advantage relies on genuine many-particle interference of indistinguishable particles (Polino *et al.*, 2020). The minimal realization is the two-photon HOM effect (Hong *et al.*, 1987), which represents the simplest example of genuine multiphoton interference (see Sec. II.B.1 and Ref. (Bouchard *et al.*, 2021)). In this regime, ideal path-entangled states achieve a factor-of-two phase enhancement (Slussarenko *et al.*, 2017). Scaling to larger photon numbers remains challenging because NOON states are difficult to generate and become increasingly fragile to loss and decoherence. Photonic NOON states have been demonstrated up to $N = 5$ (Afek *et al.*, 2010), spin-NOON states up to $N = 10$ (Jones *et al.*, 2009), while even weak Markovian decoherence restores the standard quantum limit (Demkowicz-Dobrzański *et al.*, 2012; Huelga *et al.*, 1997).

In realistic interferometers, achieving a quantum advantage requires not only high transmission and detection efficiency but also a high degree of photon indistinguishability. For two-photon interferometry, this condition can be expressed as $\eta^N \nu^2 N > 1$, where $\eta$ is the single-photon efficiency and $\nu$ quantifies the photon indistinguishability (Slussarenko *et al.*, 2017). A simple model of partial indistinguishability takes the input state as $|\Psi_{\mathrm{in}}\rangle = \sqrt{\nu}\,|1,1\rangle + \sqrt{1-nu}\,|1,\bar{1}\rangle$, where the first term represents perfectly overlapping photons and the second photons occupying orthogonal modes. The corresponding QFI is $F_Q = 2(1+\nu)$ showing that metrological performance increases continuously with indistinguishability, reaching its maximum for perfectly indistinguishable photons ($\nu = 1$) and reducing to the distinguishable limit for $\nu = 0$. This supplies a direct operational link between photon indistinguishability and quantum-enhanced phase estimation.

### C. Indistinguishability and quantum imaging

Quantum imaging provides another important application of indistinguishability-enabled quantum metrology; for recent reviews, see Refs. (Defienne *et al.*, 2024; D'errico and Karimi, 2026). An early milestone was the proposal of quantum lithography by Boto *et al.* (Boto *et al.*, 2000), which employed path-entangled $N$-photon (NOON) states to generate sub-Rayleigh interference fringes. For these states, the effective de Broglie wavelength scales as $\lambda/N$, producing interference fringes that oscillate $N$ times faster than in classical interferometry. As in quantum metrology, this enhancement originates from genuine multiphoton interference of indistinguishable particles, giving rise to Heisenberg-limited phase sensitivity and a $\sqrt{N}$-fold improvement over the standard quantum limit.

More generally, quantum estimation theory has shown that the diffraction limit is not a fundamental bound on spatial resolution but rather a limitation of conventional measurements, particularly direct intensity imaging (Helstrom, 1969; Lupo and Pirandola, 2016; Nair and Tsang, 2016; Napoli *et al.*, 2019). By exploiting quantum interference, optimal measurement strategies can extract spatial information inaccessible to direct detection (Defienne *et al.*, 2024; Tsang, 2019). Among these, two-photon interference imaging is especially attractive because it realizes the advantages of indistinguishability through HOM-type interference without requiring large-$N$ NOON states. Therefore, it is considerably more robust to loss and decoherence while remaining experimentally accessible (Demkowicz-Dobrzański *et al.*, 2012; Huelga *et al.*, 1997).

Practical implementations remain limited by photon loss, noise, and the difficulty of preparing high-$N$ entangled states. Consequently, two-photon interference has emerged as a practical resource for quantum-enhanced imaging. Parniak *et al.* (Parniak *et al.*, 2018) demonstrated that HOM interference with spatially resolved detection can surpass the Rayleigh limit for estimating the separation and centroid of two point emitters. More recently, HOM interference has been formulated as a quantum sensing primitive for estimating temporal delays and transverse displacements (Triggiani *et al.*, 2023; Triggiani and Tamma, 2024). In the latter scheme (see Fig. 16), the relative displacement $\Delta x$ between two input wave packets is encoded in momentum-space interference according to

$$P_\nu(\Delta k; X) = \frac{1}{2} C(\Delta k) \left[1 + \alpha(X)\nu \cos(\Delta k\, \Delta x)\right]. \tag{34}$$

where $C(\Delta k)$ is the momentum envelope, $\nu$ denotes the photon indistinguishability, and $\alpha = \pm 1$ distinguishes bunching and coincidence events. Thus, the displacement is encoded as HOM interference fringes whose visibility is directly controlled by the degree of indistinguishability. For Gaussian wave packets, the corresponding Fisher information satisfies

$$F_\nu(\Delta x) \gtrsim \left(1 - \sqrt{1-\nu^2}\right) H, \tag{35}$$

where $H$ is the ideal QFI (Triggiani and Tamma, 2024). So, imperfect indistinguishability continuously degrades the achievable precision rather than eliminating the quantum advantage, further illustrating that indistinguishability acts as an operational metrological resource.

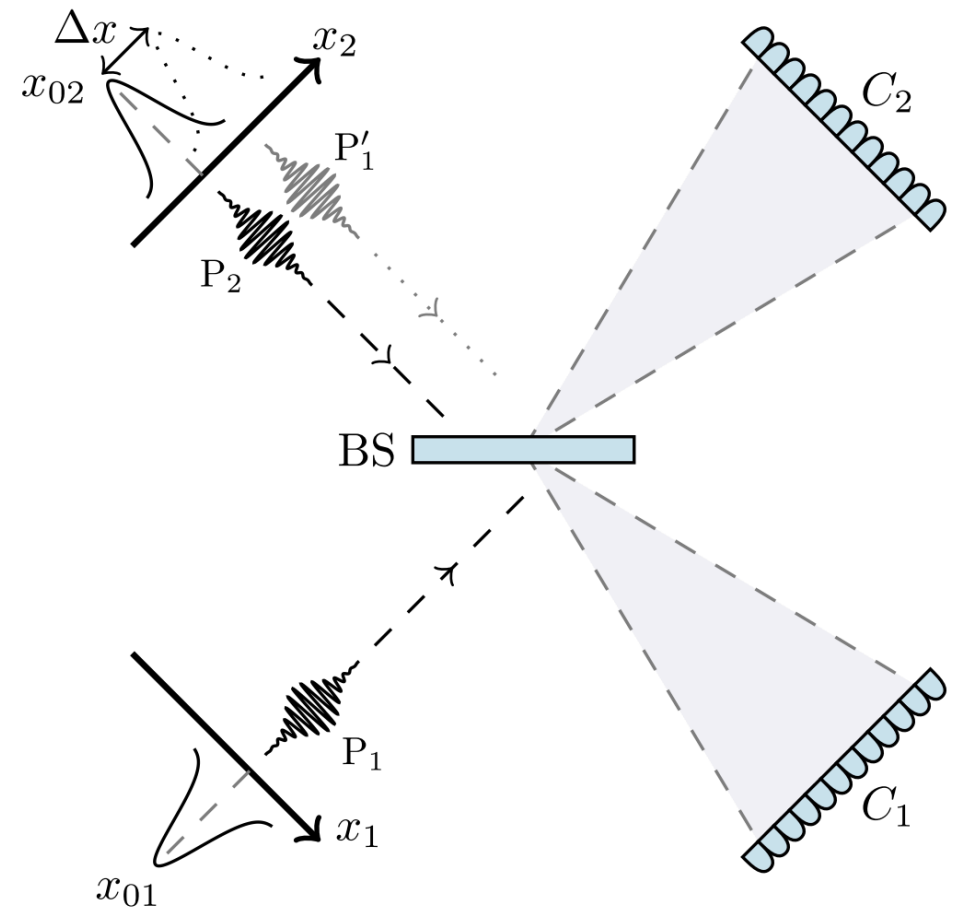

FIG. 16: Spatially resolved two-photon interference for transverse-displacement estimation. Two single photons, centered at transverse positions $x_{01}$ and $x_{02}$, impinge on the two input ports of a balanced beamsplitter. The parameter of interest is the displacement $\Delta x = x_{01} - x_{02}$ between the probe photon $P_2$ and the mirror image $P'_1$ of the reference photon $P_1$. The output photons are detected by cameras $C_1$ and $C_2$ in the far-field regime. Adapted from Ref. (Triggiani and Tamma, 2024).

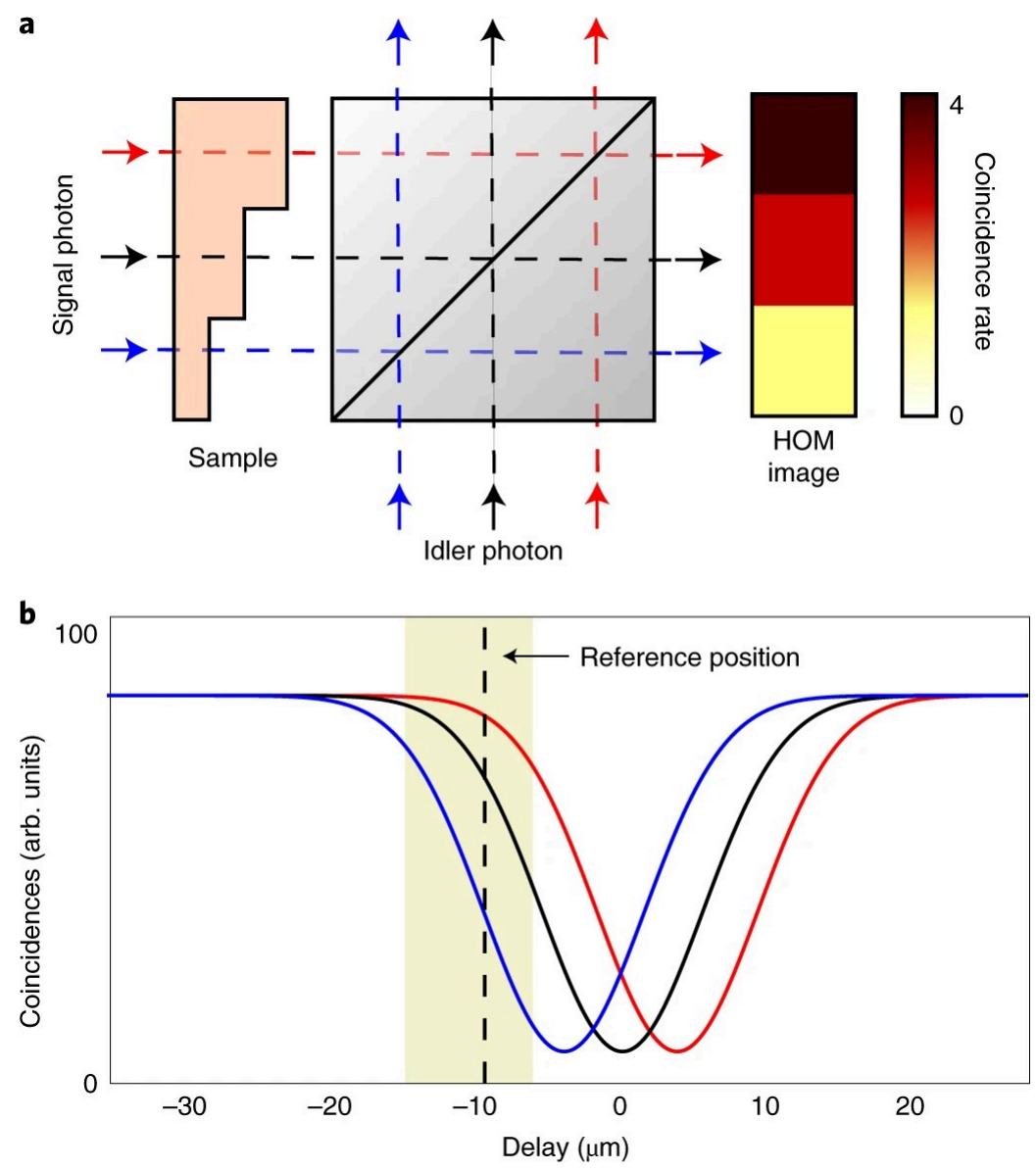

FIG. 17: (a) A signal photon probes a sample with spatially varying thickness and interferes with a reference idler photon at a 50:50 beamsplitter. Coincidence measurements at the outputs probe the local two-photon indistinguishability.(b) With the idler delay fixed near the reference position, sample-induced delays produce different coincidence probabilities, yielding a contrast image from which the optical-thickness profile can be reconstructed. Adapted from Ref. (Ndagano *et al.*, 2022).

In the same direction, Ndagano *et al.* (Ndagano *et al.*, 2022) implemented a full-field HOM microscope, reconstructing the depth profile of transparent samples from spatially resolved coincidence measurements (Fig. 17). One photon probes the sample, while the other serves as a reference before the pair interferes at a beamsplitter. The sample introduces a spatially varying optical delay, or equivalently an optical-thickness profile, which shifts the local HOM dip across the image. By mapping these shifts, the sample depth profile can be reconstructed. HOM microscopy thus converts spatially varying photon distinguishability into measurable coincidence visibility, providing another example where photon indistinguishability is a resource for quantum-enhanced imaging.

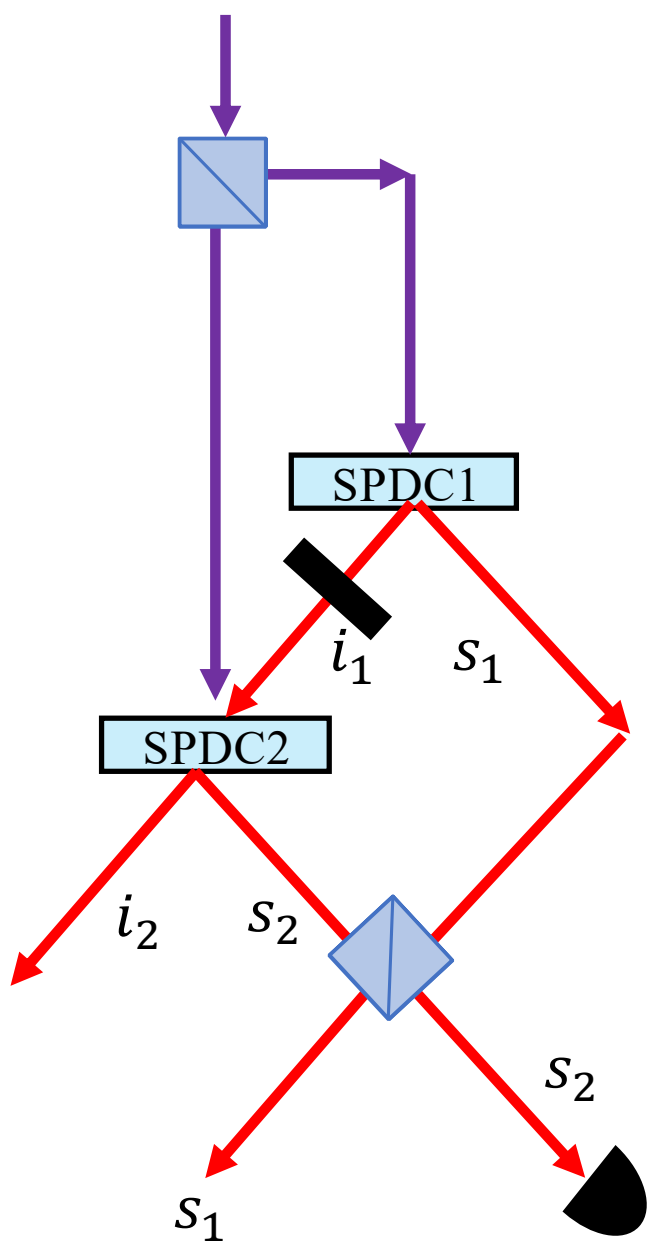

FIG. 18: Zou–Wang–Mandel interferometer. Two coherently pumped SPDC crystals generate indistinguishable signal–idler pair-creation alternatives. Adapted from Ref. (D'errico and Karimi, 2026).

The Zou–Wang–Mandel (ZWM) interferometer, depicted in Fig. 18, represents one of the most striking demonstrations of indistinguishability in quantum imaging and detection (Wang *et al.*, 1991). Unlike the HOM effect, where interference originates from overlapping identical photons at a beamsplitter, the ZWM interferometer relies on the indistinguishability of two photon-pair generation alternatives. Thus, interference arises even though the detected photons never overlap, a phenomenon now understood in terms of path identity, or path indistinguishability (see (Hochrainer *et al.*, 2022)).

In the low-gain regime, and for equal pair-generation amplitudes in the two nonlinear crystals, the relevant state is

$$|\psi_{\text{in}}\rangle = \frac{1}{\sqrt{2}} a_i^\dagger \left( a_{s_1}^\dagger + e^{i\delta} a_{s_2}^\dagger \right) |0\rangle \,, \tag{36}$$

where the common idler creation operator $a_i^\dagger$ expresses path identity: no information exists about which crystal produced the photon pair. Introducing a sample into the idler path modifies this identity through its complex transmission coefficient, $a_i^\dagger \to t\, a_i^\dagger + r\, a_0^\dagger$, where $t = |t|e^{i\phi}$. After interference of the signal photons at a balanced beamsplitter, the detected count rate is

$$R_s(\delta) = R_0 \left[1 - |t| \cos(\delta + \phi)\right] , \tag{37}$$

so that the interference visibility is $\mathcal{V} = |t|$, while the fringe shift directly measures the sample phase $\phi = \arg(t)$ (Wang *et al.*, 1991). Loss reduces the visibility by introducing which-source information, whereas a transparent phase object shifts the interference fringes without reducing their contrast.

This principle underlies quantum imaging with undetected photons (Lemos *et al.*, 2014). Remarkably, the object is placed in the idler arm, although no detected photon ever interacts with it. When the transmission becomes position dependent, $t \to t(\mathbf{x})$, the detected signal reconstructs both the amplitude and phase of the object through changes in interference visibility and phase. The imaging information is therefore encoded entirely through the preservation or loss of path indistinguishability. This paradigm has since developed into a versatile imaging platform; for recent reviews, see Refs. (Barreto Lemos *et al.*, 2022; Defienne *et al.*, 2024; D'errico and Karimi, 2026).

### D. Indistinguishability and state discrimination

Quantum state discrimination provides a natural operational framework where quantum resources acquire a direct information-theoretic meaning. Given two quantum states, $\rho_0$ and $\rho_1$, prepared with equal prior probabilities, the minimum achievable error probability is determined by their trace distance through the Helstrom bound (Helstrom, 1969),

$$P_{\text{err}}^{\text{opt}} = \frac{1}{2}\left(1 - \frac{1}{2}\left\|\rho_0 - \rho_1\right\|_1\right) , \tag{38}$$

Consequently, orthogonal states can be discriminated perfectly, whereas nonorthogonal states inevitably incur a finite error probability.

Within quantum resource theory, coherence acquires an operational meaning when the allowed measurements or operations are restricted with respect to a fixed reference basis (Streltsov *et al.*, 2017). Under such restrictions, coherence becomes a resource that enhances quantum state discrimination by reducing the minimum achievable error probability below that attainable using incoherent strategies. This connection is particularly transparent in phase-discrimination tasks, where the robustness of coherence directly quantifies the operational advantage provided by coherent superpositions (Napoli *et al.*, 2016; Piani *et al.*, 2016).

As discussed in Sec. III.A.5, spatial overlap between independently prepared identical particles generates coherence in the operational basis defined by the localized detection regions. Castellini *et al.* (Castellini *et al.*, 2019) showed that this indistinguishability-induced coherence can be directly exploited for quantum phase discrimination. Within this framework, one prepares the state $|\Psi\rangle = |\psi\downarrow, \psi'\uparrow\rangle$. After projection onto the subspace with one particle in region $L$ and one in region $R$, the state becomes $|\Psi_{LR}\rangle$ of Eq. (15). The two coherent amplitudes of this state constitute the resource that enables the discrimination protocol.

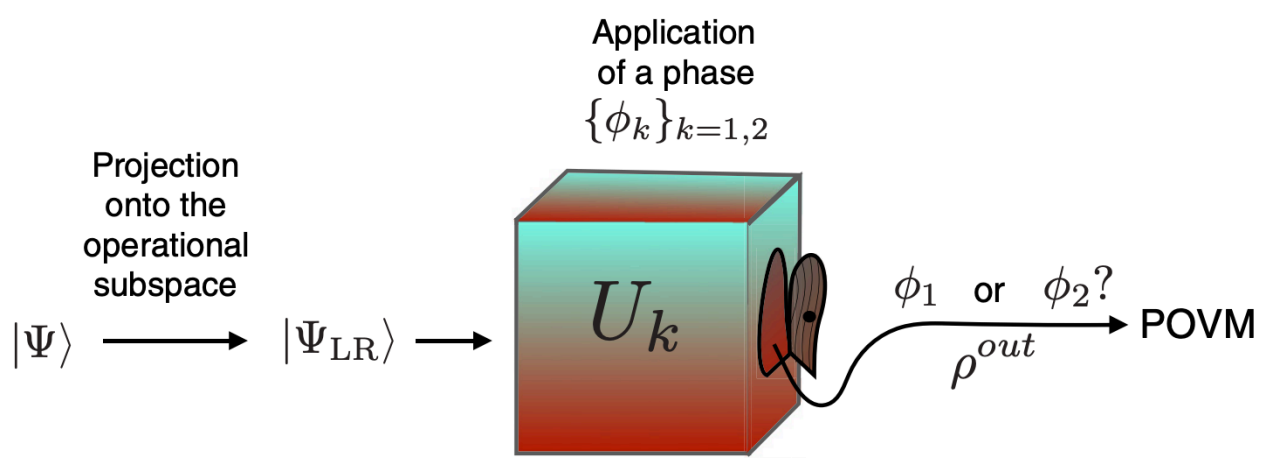

FIG. 19: Sketch of the phase discrimination game, from Ref. (Castellini *et al.*, 2019).

Suppose that one of two possible phases $\{\phi_k\}_{k=1,2}$ is encoded through the unitary $U_k = e^{i\hat{G}\phi_k}$, as illustrated in Fig. 19. The resulting state is

$$|\Psi_{LR}^k\rangle = \frac{lr'e^{i\omega_{\downarrow\uparrow}\phi_k}\,|L\downarrow, R\uparrow\rangle + \eta l'r e^{i\omega_{\uparrow\downarrow}\phi_k}\,|L\uparrow, R\downarrow\rangle}{\sqrt{|lr'|^2+|l'r|^2}}, \tag{39}$$

leading to the ensemble $\{p_k, |\Psi_{LR}^k\rangle\}_{k=1,2}$. The optimal minimum-error discrimination probability is then

$$P_{\rm err}^{\rm opt} = \frac{1}{2}\left(1-\sqrt{1-4p_1p_2|\langle\Psi_{LR}^1|\Psi_{LR}^2\rangle|^2}\right), \tag{40}$$

which depends explicitly on the overlap between the encoded states. If only one detection alternative is present, the phase contributes merely a global phase and no discrimination advantage is obtained. When both indistinguishable alternatives contribute, however, the phase is encoded as a relative phase between coherent components, reducing the minimum discrimination error. Thus, indistinguishability-induced coherence acquires a direct operational meaning by enhancing the distinguishability of the encoded quantum states.

This protocol mirrors the entanglement activation mechanism of Lo Franco and Compagno (Lo Franco and Compagno, 2018): in both cases, spatial overlap transforms particle indistinguishability into an operational quantum resource. Here, however, the resource is coherence, whose advantage is revealed through quantum state discrimination rather than entanglement generation.

This mechanism was experimentally demonstrated with photons by Sun *et al.* (Sun *et al.*, 2022). Independent photon pairs generated by spontaneous parametric down-conversion were spectrally filtered, spatially overlapped, and subjected to phase encoding in an unbalanced Mach–Zehnder interferometer, while sLOCC measurements were implemented through projective detection in the localized regions $L$ and $R$. State tomography confirmed the preparation of a highly coherent state with a fidelity of 0.988. As shown in Fig. 20, the measured error probability $P_{\rm err}$ closely followed the Helstrom–Holevo bound and outperformed the optimal distinguishable-particle strategy. Furthermore, by simulating fermionic exchange statistics through basis transformations, the experiment demonstrated that both bosonic and fermionic indistinguishability enhance phase discrimination, with fermions yielding an even lower $P_{\rm err}$ in selected parameter regimes.

Unambiguous quantum state discrimination (UQSD), introduced independently by Ivanovic, Dieks, and Peres (Dieks,

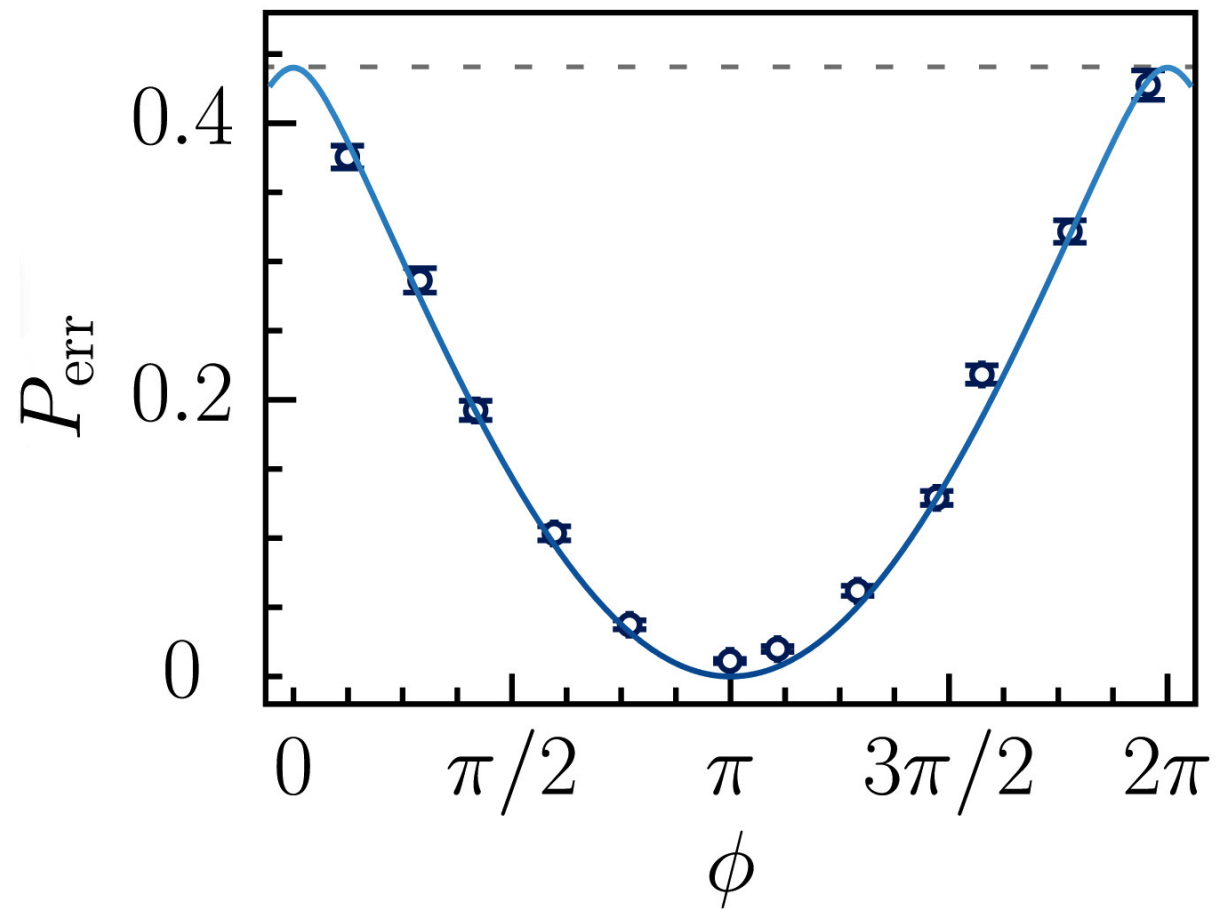

FIG. 20: Error probability $P_{\rm err}$ for binary phase discrimination as a function of the phase parameter $\phi$. The dashed curve denotes the corresponding Helstrom–Holevo benchmark for the best strategy without indistinguishability-induced coherence. Adapted from Ref. (Sun *et al.*, 2022).

1988; Ivanovic, 1987; Peres and Terno, 1998), complements minimum-error discrimination by allowing inconclusive outcomes while forbidding incorrect ones. For two equiprobable pure states $|d_A\rangle$ and $|d_B\rangle$, the optimal success probability is

$$P_{\rm succ}^{\rm UQSD} = 1 - |\langle d_A|d_B\rangle|\,. \tag{41}$$

so perfect discrimination is possible only for orthogonal states, whereas identical states cannot be distinguished. For identical particles, this result makes the overlap of their accessible single-particle states an operational measure of indistinguishability, which can therefore be exploited as a quantum resource.

As a further example, two bosons impinge on a balanced beamsplitter carrying internal states $|d_A\rangle$ and $|d_B\rangle$, which encode the available which-source information (Pathania and Qureshi, 2022; Stanisic and Turner, 2018). Their overlap determines the coincidence probability, $P_C = \frac{1}{2}(1-|\langle d_A|d_B\rangle|^2)$: perfectly indistinguishable particles suppress coincidences completely, while orthogonal states recover the classical coincidence rate. The corresponding HOM visibility is $V = |\langle d_A|d_B\rangle|^2$. Conversely, identifying the particle source amounts to discriminating the states $|d_A\rangle$, $|d_B\rangle$. The associated path distinguishability is $D = 1 - |\langle d_A|d_B\rangle|^2$, leading to the complementarity relation $D+V = 1$ (Pathania and Qureshi, 2022). Thus, HOM interference can be interpreted as an operational path-discrimination protocol in which particle indistinguishability limits the available which-source information.

## V. QUANTUM COMMUNICATION WITH IDENTICAL PARTICLES

Rather than surveying all communication protocols involving identical particles, we focus on three paradigmatic instances: quantum teleportation, entanglement swapping, and quantum cloning. These protocols illustrate how indistinguishability can be exploited to generate, transfer, and process quantum information, revealing operational advantages that have no analog for distinguishable particles.

### A. Quantum teleportation with identical particles

Quantum teleportation is a cornerstone of quantum communication, enabling the faithful transfer of an unknown quantum state between distant nodes using shared entanglement and classical communication. In conventional protocols, the carriers (photons, atoms, or electrons) are identical at the microscopic level but are rendered operationally distinguishable through spatial separation or orthogonal encoding, so that standard teleportation theory for distinguishable qubits applies.

However, for identical particles quantum teleportation is fundamentally constrained by superselection rules (SSRs), most notably particle-number conservation. Since local operations and measurements must preserve the local particle number, the operational notion of entanglement differs from the conventional one (Wiseman and Vaccaro, 2003). Consequently, some states that are separable in the usual tensor-product description cannot be prepared locally under SSR constraints, enabling phenomena such as perfect quantum data hiding (Verstraete and Cirac, 2003). More generally, SSRs break the operational equivalence between entanglement-assisted classical communication and a quantum channel, while the accessible entanglement is determined by the superselection-induced variance across particle-number sectors (Schuch *et al.*, 2004a,b).

Quantum teleportation with identical massive particles under particle-number SSRs has been investigated from several perspectives. In the single-particle regime, Heaney *et al.* showed that the mode entanglement of a single massive particle can, in principle, enable quantum communication tasks such as teleportation and dense coding, provided that the communicating parties share an external phase reference or have access to a particle reservoir (Heaney and Vedral, 2009).

For finite many-particle systems, Marzolino and Buchleitner formulated teleportation entirely in terms of modes within a second-quantized framework, enforcing particle-number conservation throughout the protocol (Marzolino and Buchleitner, 2015). They showed that deterministic unit-fidelity teleportation is impossible for any finite-particle entangled resource: although local particle-number–conserving operations can compensate phase correlations, amplitude distortions associated with the particle-number distribution of the shared resource cannot be removed. This limitation persists for noisy and imperfect resources (Marzolino and Buchleitner, 2016). Benatti *et al.* further argued that mode entanglement, defined with respect to physically accessible mode partitions, provides the appropriate operational resource for quantum information with identical particles (Benatti *et al.*, 2021). Within this framework, mode-entangled states can achieve teleportation fidelities beyond the classical limit, but particle-number conservation continues to preclude deterministic perfect teleportation for finite systems.

A different approach takes advantage of *spatial indistinguishability* as the teleportation resource (Lo Franco and Compagno, 2018). Three independently prepared identical particles are distributed over three localized regions $L'$, $L$, and $R$. The initial state is fully separable, $|\Psi_{\text{in}}\rangle = |\psi\rangle_{L'} \otimes |\uparrow\rangle_L \otimes |\downarrow\rangle_R$, so no entanglement is prepared in advance. The particle in region $L'$ encodes the unknown state $|\psi\rangle = \alpha\,|\uparrow\rangle + \beta\,|\downarrow\rangle$ to be teleported, while the other two particles are then made spatially overlapping in both $L$, $R$ via balanced spatial deformation, rendering them operationally indistinguishable.

After a sLOCC projection onto the sector containing two particles in the sender's region ($L' \cup L$) and one particle in the receiver's region $R$, the post-selected state admits the Bell decomposition

$$|\Psi\rangle = \frac{1}{2}\sum_{\mu=0}^{3} |\Phi_\mu\rangle_{L'L} \otimes \sigma_\mu\,|\psi\rangle_R\,, \tag{42}$$

where $\{|\Phi_\mu\rangle\}$ are the Bell states and $\sigma_\mu \in \{\mathbb{I}, \sigma_x, \sigma_y, \sigma_z\}$. A Bell-state measurement in the sender's region, followed by classical communication, therefore allows the receiver to reconstruct the unknown state exactly. In contrast to conventional teleportation, the Bell resource is not distributed beforehand but is activated operationally through spatial overlap and sLOCC projection, demonstrating that spatial indistinguishability itself can serve as the communication resource.

This mechanism was demonstrated experimentally with photonic qubits by Sun *et al.* (Sun *et al.*, 2020). As shown in Fig. 7, controlled spatial overlap and post-selection generated the required Bell resource without pre-shared entanglement. Teleportation of six independent input states yielded fidelities between 78% and 93%, exceeding the classical limit and confirming the operational role of spatial indistinguishability in quantum communication.

### B. Entanglement swapping with identical particles

Entanglement swapping is a central primitive for long-distance quantum communication and quantum repeater architectures(Sangouard *et al.*, 2011)) and distributed quantum computation (Eisert *et al.*, 2000; Main *et al.*, 2025). Conventionally, two independent entangled pairs are prepared, and a Bell-state measurement (BSM) projects the two remote particles—initially uncorrelated—onto an entangled state (Briegel *et al.*, 1998; Kimble, 2008; Żukowski *et al.*, 1993).

For identical particles, entanglement swapping is instead enabled by two-photon interference. Consider two distant emitters (Maunz *et al.*, 2007b; Moehring *et al.*, 2007), each prepared in an atom–photon entangled state,

$$\begin{aligned} |\psi\rangle_{AL} &= \frac{1}{\sqrt{2}}\left(|A\,g_1\rangle\,|L\uparrow\rangle + |A\,g_2\rangle\,|L\downarrow\rangle\right), \\ |\psi\rangle_{BR} &= \frac{1}{\sqrt{2}}\left(|B\,g_1\rangle\,|R\uparrow\rangle + |B\,g_2\rangle\,|R\downarrow\rangle\right). \end{aligned} \tag{43}$$

Having the photons interfere in modes $L$ and $R$ on a balanced beamsplitter and postselecting one photon in each output project the photonic state onto

$$|\Psi^-\rangle_{LR} = \frac{1}{\sqrt{2}} (|\mathrm{L} \uparrow, \mathrm{R} \downarrow\rangle - |\mathrm{L} \downarrow, \mathrm{R} \uparrow\rangle), \tag{44}$$

thereby projecting the remote emitters onto the Bell state

$$|\Psi^-\rangle_{AB} = (|\mathrm{A}\, g_1, \mathrm{B}\, g_2\rangle - |\mathrm{A}\, g_2 \mathrm{B}\, g_1\rangle) / \sqrt{2}. \tag{45}$$

Consequently, the swapping fidelity is directly governed by the two-photon interference visibility, making photon indistinguishability the key operational resource.

The same interference-based mechanism extends to superconducting circuit QED, where microwave-photon interference entangles remote transmon qubits (Narla *et al.*, 2016), and to quantum repeater architectures, where repeated interference-assisted swapping distributes entanglement over long distances (Sangouard *et al.*, 2011), enabling applications ranging from quantum key distribution to distributed quantum computation (Eisert *et al.*, 2000; Main *et al.*, 2025).

Within the sLOCC framework (Lo Franco and Compagno, 2018), Castellini *et al.* showed that independently prepared identical particles can realize entanglement swapping through many-particle interference without Bell-state measurements (Castellini *et al.*, 2022). Entanglement is activated by postselecting local particle-number measurements, while the operational implementation is determined by particle statistics. For fermions, antisymmetrization and the Pauli exclusion principle automatically suppress unwanted amplitudes. Postselecting one particle in each remote mode projects the remaining particles onto

$$|\Psi_f^{(4)}\rangle = |\Psi_M\rangle \otimes |\Psi_{AB}^-\rangle, \tag{46}$$

where

$$|\Psi_M\rangle = |M \uparrow, M \downarrow\rangle, \tag{47}$$

and

$$|\Psi_{AB}^-\rangle = \frac{1}{\sqrt{2}} (|A \downarrow, B \uparrow\rangle - |A \uparrow, B \downarrow\rangle). \tag{48}$$

Thus, local particle-number measurements alone suffice to swap entanglement, with a success probability of $2/9$. For bosons, bunching introduces additional interference pathways that require postselection or interferometric filtering. The corresponding four-particle projection succeeds with probability $6/25$, and a Bell-state measurement at the intermediate node is required to distinguish the generated remote states. These interference-based protocols naturally extend to multi-node quantum networks through concatenated swapping operations.

This protocol was experimentally demonstrated by Wang *et al.* (Wang *et al.*, 2022b) in a three-node photonic network (Fig. 21). High-visibility interference between photons from independent SPDC sources, combined with localized product-state measurements, generated entanglement between remote nodes without initially shared entangled pairs. State tomography yielded fidelities exceeding 0.82, while concurrence values above 0.72 confirmed successful entanglement distribution, establishing indistinguishability-enabled swapping as a viable alternative to conventional Bell-measurement protocols.

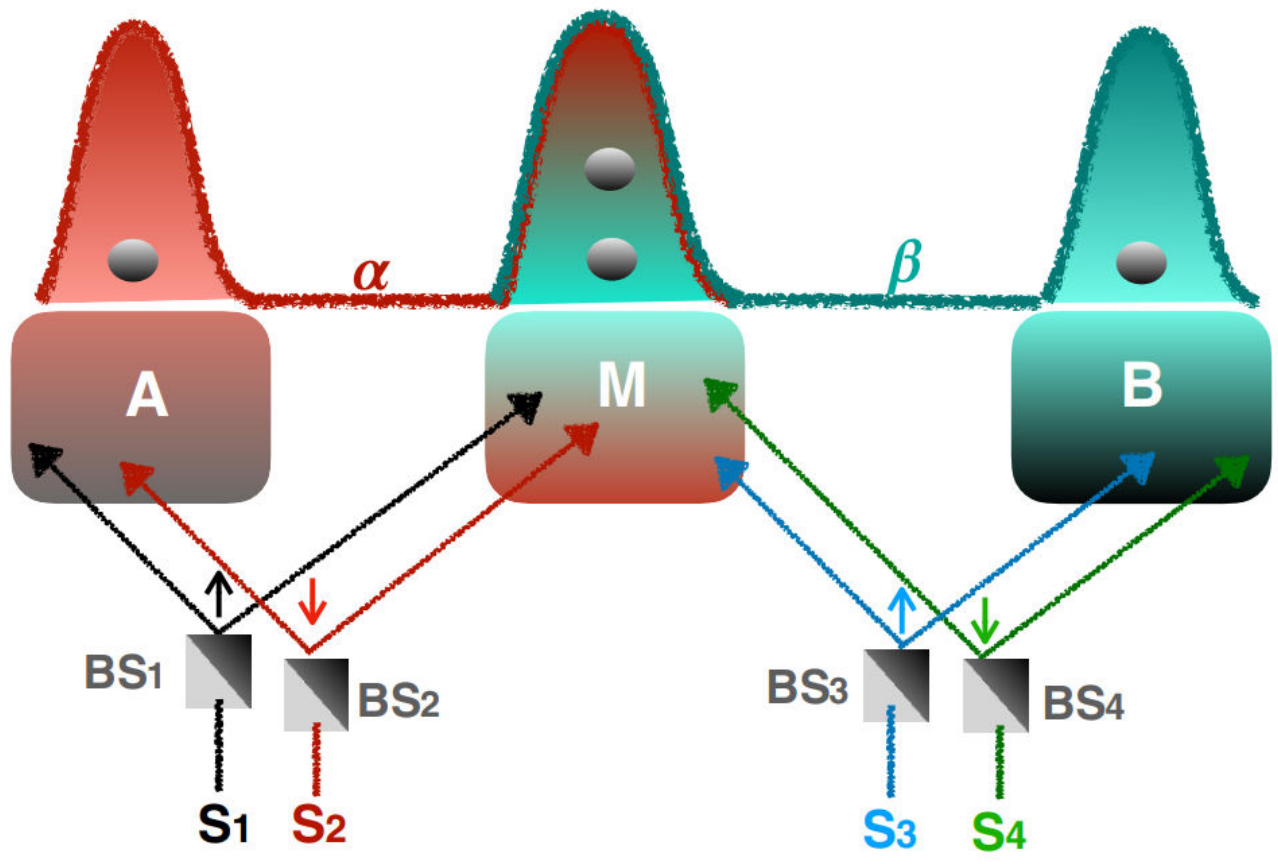

FIG. 21: Scheme for entanglement swapping for fermions and bosons. Figure adapted from Ref. (Castellini *et al.*, 2022).

The concept has also been extended to hyper-hybrid entanglement, where swapping exploits both spatial and internal degrees of freedom. In this setting, quantum swapping has been shown to be feasible with only two indistinguishable bosons (Li *et al.*, 2018) and also with indistinguishable fermions (Das *et al.*, 2020), whereas no analogous hyper-hybrid resource exists for distinguishable particles (Das *et al.*, 2020).

### C. Quantum cloning with identical particles

Quantum cloning provides one of the earliest and clearest demonstrations that particle indistinguishability can be exploited as a quantum information resource (Scarani *et al.*, 2005). Although the no-cloning theorem forbids perfect copying of an arbitrary unknown quantum state (Wootters and Zurek, 1982), optimal approximate cloning is both possible and operationally important, with applications ranging from quantum communication to quantum cryptography.

The physical realization of optimal cloning is strictly connected to particle statistics. Bosonic indistinguishability enables stimulated emission, which preferentially amplifies the occupation of an already populated mode and realizes the optimal universal quantum cloning transformation (Simon *et al.*, 2000). Differently, fermionic systems do not exhibit an analogous mechanism because the Pauli exclusion principle forbids multiple particles from occupying the same quantum state.

In second quantization, an arbitrary input photonic qubit is written as $|\psi\rangle_{\mathrm{in}} = \alpha\hat{a}_H^\dagger|0\rangle + \beta\hat{a}_V^\dagger|0\rangle$, where $\hat{a}_H^\dagger$ and $\hat{a}_V^\dagger$ create horizontally and vertically polarized photons, respectively. The cloning process is realized by stimulated parametric emission in a nonlinear crystal, described by the interaction Hamiltonian

$$\hat{H}_{\mathrm{int}} = i\hbar\chi(\hat{a}_H^\dagger\hat{b}_H^\dagger + \hat{a}_V^\dagger\hat{b}_V^\dagger) + \mathrm{h.c.}, \tag{49}$$

where $\chi$ is the nonlinear coupling strength and $\hat{b}_{H,V}^\dagger$ create photons in the idler modes. The output state is

$$|\Psi_{\mathrm{out}}\rangle = e^{-i\hat{H}_{\mathrm{int}}t/\hbar}|\psi\rangle_{\mathrm{in}}|0\rangle_b. \tag{50}$$

Expanding the evolution to first order in the parametric gain, corresponding to the low-gain regime, one gets

$$|\Psi_{\rm out}\rangle \approx |\psi\rangle_{\rm in}|0\rangle_b + \chi t(\alpha|2_H,0_V\rangle + \beta|0_H,2_V\rangle) + \dots, \quad (51)$$

where the second term describes stimulated emission into the polarization mode already occupied by the input photon. This bosonic enhancement realizes the physical cloning process via mode occupation rather than engineered quantum gates.

The resulting single-copy fidelity of the cloned output is

$$F = \langle\psi|\rho_{\rm clone}|\psi\rangle = \frac{5}{6} \quad \text{for the } 1 \to 2 \text{ qubit case}, \quad (52)$$

which coincides with the optimal universal $1 \to 2$ cloning fidelity (Bruß *et al.*, 1998; Bužek and Hillery, 1996; Werner, 1998). Stimulated emission provides a natural implementation of optimal bosonic cloning, with the parametric gain controlling the number of generated copies (Simon *et al.*, 2000).

These predictions were confirmed experimentally using quantum-injected optical parametric amplifiers (Lamas-Linares *et al.*, 2002; Martini *et al.*, 2002, 2004; Pelliccia *et al.*, 2003). Near-optimal cloning fidelities were achieved through stimulated emission of indistinguishable photons, supplying direct evidence that bosonic symmetry underlies optimal quantum cloning.

This result generalizes optimal universal cloning from qubits to arbitrary $d$-dimensional systems, with the achievable fidelity decreasing monotonically as the Hilbert-space dimension increases. The optimal universal $1 \to 2$ cloning fidelity for a $d$-dimensional system is (Nagali *et al.*, 2010, 2009)

$$F = \frac{1}{2} + \frac{1}{d+1}. \quad (53)$$

The same principle has since been extended to continuous-variable systems. Liu *et al.* (Liu *et al.*, 2021) realized an all-optical optimal $N \to M$ cloning protocol for coherent states using four-wave mixing followed by a beam-splitter network. The optimal fidelity is

$$F = \frac{MN}{MN + M - N}. \quad (54)$$

As predicted theoretically, the cloning fidelity increases with the number of input replicas $N$ and decreases with the number of output clones $M$. The experiment closely followed this behavior, achieving fidelities of about 93.3% for the $4 \to 5$ cloning protocol.

## VI. THERMAL MACHINES OF IDENTICAL PARTICLE ENSEMBLES

Identical particles (bosons, fermions, and, in low-dimensional settings, anyons) are the elementary constituents of quantum many-body systems, including quantum thermodynamic devices. Quantum thermodynamics concerns the interconversion of heat, work, and information in such systems. In this *Colloquium*, we review historical and recent developments that clarify when and how indistinguishability becomes operationally relevant in thermodynamic processes and thermal machines.

### A. Indistinguishability and Gibbs paradox

#### 1. History and early resolution

The Gibbs' thought experiment (Gibbs, 1878) illustrates the first notable consequence of particle (in)distinguishability. Historically, it considers two ideal gases, each containing $N$ particles in a volume $V$, separated by a removable partition (Darrigol, 2018; Gibbs, 1878). For distinguishable gases, removing the partition increases the entropy by $\Delta S = 2N \ln 2$, reflecting the larger number of accessible microstates (Tolman, 1938). By contrast, no entropy change occurs when the gases are identical, $\Delta S = 0$. This discontinuous change in the entropy of mixing as the gases become indistinguishable constitutes the Gibbs paradox (Ehrenfest and Ehrenfest, 1912; von Neumann, 1927). Although numerous resolutions have been proposed (Darrigol, 2018; Skilling, 2013), here we focus on what this paradox reveals about distinguishability and its implications for quantum thermodynamics.

#### 2. Indistinguishability and early resolution to Gibbs paradox

In a quantum formulation of Gibbs mixing, two gases initially occupy opposite halves of a box, each containing $N$ particles in thermal equilibrium at inverse temperature $\beta = (k_B T)^{-1}$. Each particle possesses a spatial degree of freedom and a two-level internal degree of freedom (pseudospin), prepared in the states $|\theta_L\rangle = \cos\theta\, |\uparrow\rangle + \sin\theta\, |\downarrow\rangle$, and $|\theta_R\rangle = \cos\theta\, |\uparrow\rangle - \sin\theta\, |\downarrow\rangle$. The initial density matrix of each gas is (Allahverdyan and Nieuwenhuizen, 2006; Klein, 1958; Landé, 1965)

$$\rho_X = Z^{-1} \sum_r e^{-\beta E_r} |r\rangle\langle r| \otimes |\theta_X\rangle^{\otimes N} \langle\theta_X|^{\otimes N}, \quad X = \{L, R\}, \quad (55)$$

where $\{|r\rangle\}$ are the spatial eigenstates in volume $V$ and $Z = \sum_r e^{-\beta E_r}$ is the partition function. Assuming that the mixing process leaves the internal states unchanged, the final state is

$$\rho_m = Z^{-1} \sum_r e^{-\beta E_r} |r\rangle\langle r| \otimes \left( \frac{1+\alpha}{2} |+\rangle\langle +| + \frac{1-\alpha}{2} |-\rangle\langle -| \right), \quad (56)$$

where $|\pm\rangle = \left(|\theta_L\rangle^{\otimes N} \pm |\theta_R\rangle^{\otimes N}\right) / \left\| |\theta_L\rangle^{\otimes N} \pm |\theta_R\rangle^{\otimes N} \right\|$, and $\alpha = |\langle\theta_L|\theta_R\rangle|^N$ quantifies the distinguishability of the two gases, ranging from $\alpha = 1$ for identical preparations to $\alpha = 0$ for orthogonal ones. The corresponding entropy of mixing per particle is

$$\frac{\Delta S}{2N} = -\frac{1+\alpha}{2} \ln\left(\frac{1+\alpha}{2}\right) - \frac{1-\alpha}{2} \ln\left(\frac{1-\alpha}{2}\right). \quad (57)$$

Unlike the classical result, the entropy now varies continuously from 0 for identical gases ($\alpha = 1$) to $\ln 2$ for orthogonal preparations ($\alpha = 0$), thereby resolving the Gibbs paradox (Klein, 1958; Landé, 1965).

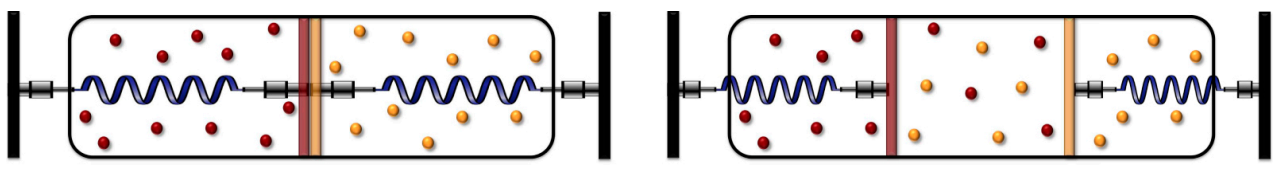

(a) Initial setup. (b) Drawing work from mixing.

FIG. 22: Optomechanical analogue of Gibbs' mixing process. (a) Two photon gases, labeled *red* and *yellow* occupy the left and right halves of a box of total volume $2B$. Mixing is controlled by two semipermeable membranes: red photons pass through the red membrane but are reflected by the yellow one, and vice versa for yellow photons. (b) Each membrane is mounted on a spring and moves in response to the radiation pressure exerted by the confined gas, allowing mechanical work to be extracted during the expansion from $V$ to $2V$. Adapted from Ref. (Holmes *et al.*, 2020b).

Eq. (57) removes the discontinuity of the Gibbs paradox by replacing the classical entropy of mixing with the von Neumann entropy of nonorthogonal quantum states. However, this resolution has been questioned by Dieks (Dieks and van Dijk, 1988) and later by Allahverdyan *et al.* (Allahverdyan and Nieuwenhuizen, 2006), who argued that it implicitly identifies the von Neumann entropy with the thermodynamic entropy. While the latter counts the number of operationally distinguishable microstates, nonorthogonal quantum states cannot be perfectly discriminated by any measurement (Peres, 2002), making the notion of distinct microstates fundamentally ambiguous. This observation has motivated a variety of operational and information-theoretic approaches to the Gibbs paradox, which we recall below.

### 3. Work extraction in mixing indistinguishable ensembles

A schematic illustration of the classical Gibbs mixing protocol is shown in Fig. 22. Two gases initially occupy the left and right halves of a box and expand against semipermeable membranes connected to springs, storing the work extracted during mixing. This mechanical picture provides the basis for the quantum thermodynamic treatment.

In quantum thermodynamics, one defines work and heat through changes in the system's internal energy under a (possibly) time-dependent Hamiltonian. The first law reads $\mathrm{d}U = \delta Q + \delta W$ where the internal energy is $U = \mathrm{Tr}[\rho H]$. For a general state $\rho$ and Hamiltonian $H$, the differential of $U$ is $\mathrm{d}U = \mathrm{Tr}[H\,\mathrm{d}\rho] + \mathrm{Tr}[\rho\,\mathrm{d}H]$ . This decomposition identifies the heat and work as $\delta Q \equiv \mathrm{Tr}[H\,\mathrm{d}\rho]$ and $\delta W \equiv \mathrm{Tr}[\rho\,\mathrm{d}H]$ (Deffner and Campbell, 2019; Vinjanampathy and Anders, 2016).

Rather than interpreting the Gibbs paradox through the von Neumann entropy, Allahverdyan *et al.* (Allahverdyan and Nieuwenhuizen, 2006) reformulated it in terms of the work extractable during mixing. They considered two-level quantum gases, each with energy gap $\varepsilon$, prepared in the pure states $|\theta_{\mathrm{L}}\rangle$ and $|\theta_{\mathrm{R}}\rangle$, and quantified the maximum extractable work using the ergotropy (Allahverdyan *et al.*, 2004),

$$\mathcal{W} = \mathrm{Tr}[\rho H] - \min_{U} \mathrm{Tr}\left[U\rho U^{\dagger} H\right], \tag{58}$$

where the minimization is taken over all unitary transformations. The work associated with mixing is then defined as the change in ergotropy, $\Delta\mathcal{W} = \mathcal{W}_i - \mathcal{W}_f$, yielding

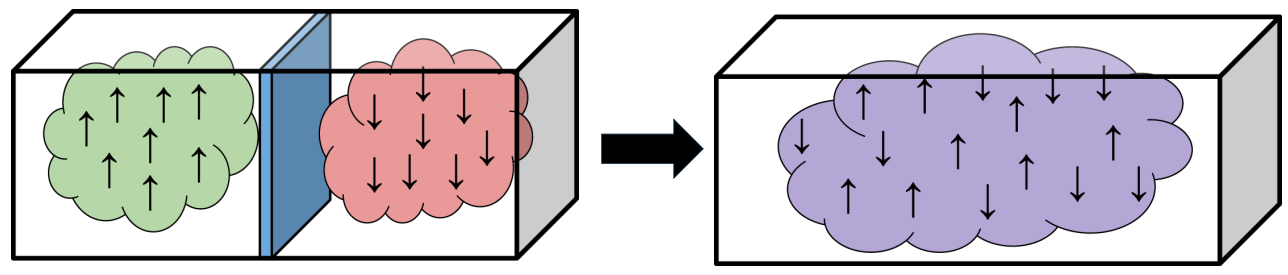

FIG. 23: Two distinct gases of $n$ particles, with opposite spin orientations and at the same temperature and pressure, are initially separated by a partition. When the partition is removed, the gases mix and reach equilibrium. Adapted from Ref. (Yadin *et al.*, 2021).

$$\Delta\mathcal{W} = \frac{N\varepsilon}{2}\left(1 - |\langle\theta_{\mathrm{L}}|\theta_{\mathrm{R}}\rangle|\right). \tag{59}$$

Unlike the classical entropy of mixing, the extractable work varies continuously with the overlap parameter, vanishing for identical gases ($|\langle\theta_{\mathrm{L}}|\theta_{\mathrm{R}}\rangle| = 1$) and being maximum for perfectly distinguishable gases ($|\langle\theta_{\mathrm{L}}|\theta_{\mathrm{R}}\rangle| = 0$).

A complementary resolution adopts an information-theoretic perspective (Maruyama *et al.*, 2005), emphasizing that the thermodynamic cost of mixing is ultimately limited by the information that can be extracted from the quantum states. After mixing, one may attempt to restore the initial configuration using semipermeable membranes or, equivalently, an information-powered Maxwell demon (see Ref. (Maruyama *et al.*, 2009) for a review). For orthogonal preparations, an optimal measurement perfectly identifies the gas and enables reversible separation. More generally, the accessible information is bounded by the Holevo quantity (Holevo, 1973),

$$I \le \chi \equiv S(\rho) - \sum_i p_i S(\rho_i), \tag{60}$$

where $\rho = \sum_i p_i \rho_i$. For nonorthogonal states, perfect discrimination is impossible, so only part of the encoded information can be recovered. Consequently, no physical device can completely reverse the mixing process, and any attempt to do so necessarily incurs an irreversible entropy increase, preserving the second law (Maruyama *et al.*, 2009; Peres, 2002).

Although complete reversibility is therefore unattainable, the thermodynamic cost of mixing can still be quantified within a nonequilibrium free-energy framework. Defining the free energy as $F(\rho) = \mathrm{Tr}[\rho H] - TS(\rho)$, the free-energy difference $\Delta F$ bounds the extractable work as $\Delta W \le -\Delta F$, providing a consistent thermodynamic description of Gibbs mixing for partially distinguishable quantum states (Goold *et al.*, 2016a; Parrondo *et al.*, 2015).

Yadin *et al.* (Yadin *et al.*, 2021) further developed the nonequilibrium free-energy approach by explicitly incorporating particle statistics through the symmetrization postulate. They introduced a toy model in which bosons or fermions occupy $d$ energetically degenerate spatial cells, with $d/2$ cells assigned to each half of the box (Fig. 23). They showed that the entropy of mixing depends not only on the distinguishability of the gases but also on the information accessible to the observer. An informed observer, who has access to the spin degree of freedom, recovers the classical result $\Delta S_{\mathrm{info}} \approx 2n \ln 2$, whereas an ignorant observer, who accesses only the spatial degrees of freedom, would classically assign $\Delta S_{\text{igno}} \approx 0$. For partially distinguishable spin preparations, the mixing entropy varies continuously with the state overlap, providing a smooth resolution of the Gibbs paradox. Crucially, this classical picture changes once exchange symmetry is imposed. Even for the ignorant observer, bosonic and fermionic statistics yield a finite mixing entropy, $\Delta S_{\text{igno}} \approx 2n \ln 2$, because symmetrization itself contributes to the number of accessible many-body states. Thus, the entropy of mixing vanishes only for classical identical particles treated without the symmetrization postulate, highlighting that quantum statistics constitute an intrinsic thermodynamic resource beyond the observer's knowledge.

Holmes *et al.* (Holmes *et al.*, 2020b) proposed an optomechanical realization of the Gibbs mixing protocol by replacing the semipermeable membranes with polarization-dependent beamsplitters (PBSs) acting on two photon gases confined in the two halves of a cavity. The two gases are prepared in the polarization states $|\theta_L\rangle = |\uparrow\rangle$ and $|\theta_R\rangle = \cos\theta\,|\uparrow\rangle + \sin\theta\,|\downarrow\rangle$, respectively. Radiation pressure displaces the membrane, and the extracted mechanical work is defined as $W_M = \frac{1}{2} m\omega_M^2 \langle X_M\rangle^2$, where $m$, $\omega_M$, and $\langle X_M\rangle$ are, respectively, mass, mechanical frequency, and average displacement of the membrane. For the PBS-in-the-middle configuration, they obtain

$$W_M = \frac{\hbar^2 g_\uparrow^2 \langle N\rangle^2}{2m\omega_M} \sin^4\theta, \tag{61}$$

where $g_\uparrow$ is a photon-membrane coupling strength and $\langle N\rangle$ the total number of photons per gas; $W_M$ varies continuously with distinguishability, vanishing for identical photon gases ($\theta = 0$) and reaching its maximum for orthogonal polarizations ($\theta = \pi/2$). In the high-temperature regime, the extracted work approaches the classical linear dependence on temperature. At low temperatures, the thermal photon population is strongly suppressed; since $W_M \propto \langle N\rangle^2$, the work per photon scales as $W_M/\langle N\rangle \propto \langle N\rangle$ and vanishes faster than linearly.

The vanishing work for indistinguishable photons appears to contrast with the finite work predicted by Yadin *et al.* (Yadin *et al.*, 2021), but the two results are complementary rather than contradictory. Holmes *et al.* calculate the work realized by an optomechanical protocol, whereas Yadin *et al.* impose bosonic symmetrization kinematically and derive a nonequilibrium free-energy bound, $\Delta W \leq 2nT \ln 2$, on the maximum extractable work. Consequently, the latter represents the maximum thermodynamically available work, while the former is the work extracted by a particular physical implementation.

Besides the extractable work output, Holmes et al. (Holmes *et al.*, 2020a) considered the optomechanical interpretation of classical Gibbs with a movable beamsplitter membrane. They demonstrated that the total energy transferred to the membrane is *quadratically enhanced* for indistinguishable photons.

### B. Quantum thermal machines of identical particles

Quantum thermal machines employ quantum systems as working media to convert heat into useful work (Alicki, 1979; Deffner and Campbell, 2019; Scovil and Schulz-DuBois, 1959). It is therefore natural to consider working media composed of identical particles. A heat engine operating between two reservoirs at temperatures $T_h > T_c$ absorbs heat $Q_h$, releases heat $Q_c$, and extracts work $W = Q_h - Q_c$, with efficiency $\eta = W/Q_h$. The second law implies the Carnot bound $\eta \leq \eta_C \equiv 1 - T_c/T_h$, which represents the maximum efficiency attainable by an engine operating between equilibrium thermal reservoirs.

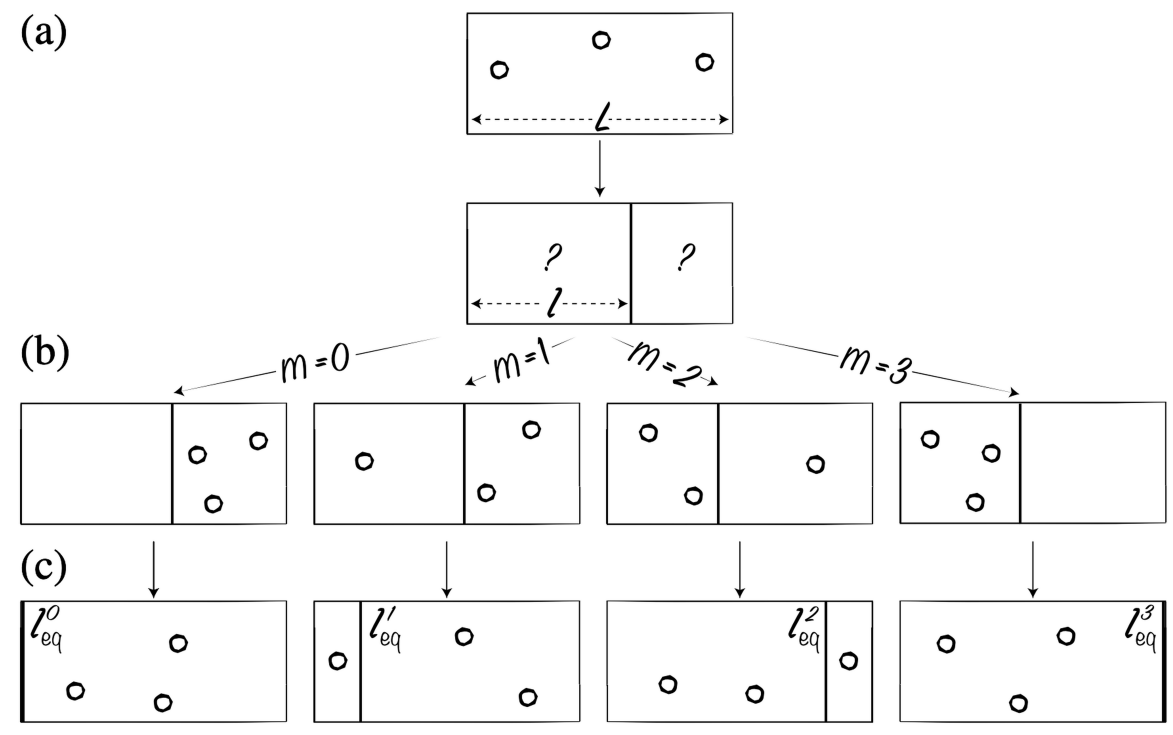

FIG. 24: Three quantum particles are confined in a one-dimensional box of length $L$. (a) A partition is inserted at position $l$. (b) A measurement is performed, yielding the number $m$ of particles on the left side. (c) The system then undergoes an isothermal expansion to the equilibrium position $l^m_{\text{eq}}$, after which the partition is removed. Adapted from Ref. (de Oliveira Junior *et al.*, 2025).

Motivated by the computational advantages offered by quantum resources, considerable effort has been devoted to understanding whether coherence, entanglement, correlations, squeezing, or indistinguishability can enhance thermodynamic performance. Although nonequilibrium resources may apparently allow efficiencies beyond the standard Carnot limit (Niedenzu *et al.*, 2018; Roßnagel *et al.*, 2014), it is also debated that such enhancements originate from the additional free energy carried by these resources. Once the energetic cost of preparing and maintaining non-Gibbsian states is properly included, Carnot-type bounds remain valid (Cangemi *et al.*, 2024; Gardas and Deffner, 2015; Strasberg *et al.*, 2017).

Within this framework, indistinguishability does not increase the maximum thermodynamic efficiency, but it can substantially modify the finite-time performance of quantum heat engines. Exchange statistics reshape the many-body spectrum and alter the work distribution, particularly at low temperatures (Gong *et al.*, 2014), while finite-time operation introduces nonadiabatic excitations and incomplete thermalization that reduce the efficiency below the Carnot limit (Abah and Lutz, 2014, 2016, 2017; del Campo *et al.*, 2014; Guéry-Odelin *et al.*, 2019; Rezek and Kosloff, 2006). As a result, ensembles of identical particles provide a natural platform for investigating how indistinguishability influences the power, work output, and efficiency of microscopic different thermal machines, which is the focus of the following sections.

### 1. Quantum Szilard engine and indistinguishability

The quantum Szilárd engine (Kim *et al.*, 2011) provides one of the earliest explicit demonstrations of the thermodynamic role of indistinguishability. The engine operates quasistatically under isothermal conditions, so that the working medium remains in thermal equilibrium throughout the cycle. As illustrated in Fig. 24, the protocol consists of four steps. The cycle consists of four quasistatic strokes: (i) an isothermal insertion of a partition; (ii) a measurement of the number $m$ of particles on the left side by a Maxwell demon; (iii) an isothermal expansion of the partition to its equilibrium position $l^m_{\rm eq}$; and (iv) the isothermal removal of the partition. Here, Maxwell's demon is a hypothetical information-gathering agent that converts information into extractable work from a single thermal reservoir (Maruyama *et al.*, 2009). Since each stroke is quasistatic, the work is determined by the corresponding free-energy changes, yielding (Kim *et al.*, 2011)

$$W_{\rm tot} = -k_{\rm B}T \sum_{m=0}^{N} f_m \ln \frac{f_m}{f_m^*}, \tag{62}$$

where $f_m$ and $f_m^*$ are the probabilities of finding $m$ particles on the left before and after the isothermal expansion, respectively.

Although the engine apparently extracts work from a single thermal reservoir, the second law is preserved because the information acquired by the Maxwell demon has a thermodynamic cost (Bennett, 1982; Brillouin, 1951; Landauer, 1961; Sagawa and Ueda, 2009). Accordingly, the extractable work is bounded by the mutual information obtained during the measurement (Sagawa and Ueda, 2008). Sagawa *et al.* formulate this balance through a generalized second law,

$$\langle W\rangle - \langle \Delta F\rangle \geq -k_{\rm B}T\, I(X, M), \tag{63}$$

where $I(X, M)$ is the mutual information between the microscopic state $X$ and the measurement outcome $M$. For a cyclic engine with $\langle \Delta F\rangle = 0$, this bound implies $\langle W\rangle \geq -k_{\rm B}T\, I(X, M)$. Equivalently, the extracted work satisfies $-\langle W\rangle \leq k_{\rm B}T\, I(X, M)$ and reduces to $-\langle W\rangle \leq k_{\rm B}T \ln 2$ for one bit of information.

Indistinguishability enters through the many-body partition function, which factorizes for distinguishable particles but must satisfy the symmetrization postulate for bosons and fermions. Hence, the measurement probabilities $f_m$ become statistics dependent. For the typical case of two particles with the partition inserted at the center of the box, the three bosonic outcomes (left-left, left-right, and right-right) occur with equal probability at low temperatures, yielding $W_{\rm tot} = \frac{2}{3} k_{\rm B}T \ln 2$. By contrast, Pauli exclusion forces two identical fermions to occupy opposite sides of the partition, making the outcome deterministic and preventing any work extraction.

The extractable work depends sensitively on the partition-insertion position, which determines the degeneracy of the many-body ground state. In the limit $T \to 0$, a nondegenerate ground state leads to a unique measurement outcome, so that the information gain and the extractable work vanish for bosons, fermions, and distinguishable particles (Plesch *et al.*, 2014). Cai *et al.* (Cai *et al.*, 2012) showed that the ground-state degeneracy depends strongly on the insertion position. For bosons, degeneracy occurs only for the symmetric insertion $l = L/2$, whereas for fermions it appears at multiple positions, approximately $l/L \simeq i/(N+1)$, with $i = 1, \ldots, N$. Consequently, as shown in Fig. 25, $W/T_1$ vanishes at low temperatures for nondegenerate configurations but approaches a finite value at degenerate insertion points. At high temperatures, both bosonic and fermionic engines recover the classical multiparticle Szilárd limit. Similar behavior has recently been observed in a photonic implementation, where bosonic bunching enhances the measured extractable work compared with a distinguishable-particle baseline (Hloušek *et al.*, 2024).

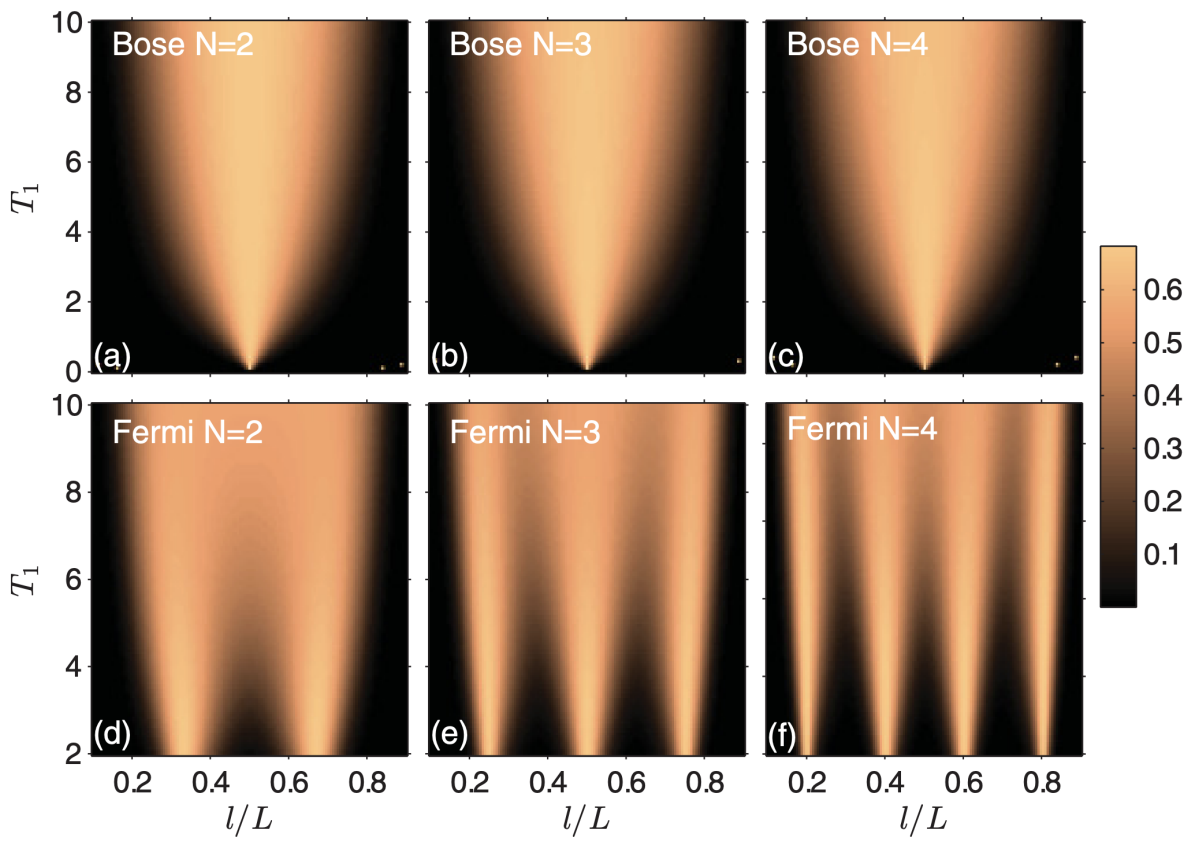

FIG. 25: Extracted work per temperature, $W/T_1$, as a function of the insertion position $l/L$ and the working temperature $T_1$. Panels (a)–(c) show Bose systems with $N = 2, 3, 4$, respectively, while panels (d)–(f) show Fermi systems with $N = 2, 3, 4$. Adapted from Ref. (Cai *et al.*, 2012).

Indistinguishability can appear to enhance the extracted work in Szilárd engines. However, a detailed analysis shows that the enhancement depends on the engine design and, especially, on the measurement and feedback protocol. This observation motivated a discussion on the thermodynamic role of particle statistics (see, e.g., Refs. (Kim *et al.*, 2013; Plesch *et al.*, 2013)). Plesch *et al.* (Plesch *et al.*, 2014) reformulated the problem in information-theoretic terms by showing that the maximum extractable work is determined solely by the number $M$ of operationally distinguishable outcomes available to the demon,

$$W_{\rm max} = k_{\rm B}T \ln M, \tag{64}$$

consistent with Landauer's principle. Accordingly, $W_{\rm d} = N k_{\rm B}T \ln 2$ for $N$ distinguishable particles, $W_{\rm b} = k_{\rm B}T \ln(N+1)$ for bosons, and $W_{\rm f} = k_{\rm B}T \ln 2$ for fermions.

This information-theoretic viewpoint was further generalized by Thomas *et al.* (Thomas *et al.*, 2019), who considered inserting $g$ partitions into a box containing $n$ particles. The corresponding work is determined by the number of distinguishable particle configurations within the degenerate manifold, yielding $(g+1)^n$ configurations for distinguishable particles, $\frac{(n+g)!}{n!g!}$ for bosons, and $\frac{(g+1)!}{n!(g+1-n)!}$ for fermions.

The information gain, and hence the extractable work, can be further enhanced by engineering the accessible many-body configurations. Zhuang *et al.* (Zhuang and Liang, 2014) showed that allowing the demon to access internal degrees of freedom, in addition to particle positions, increases the number of distinguishable outcomes and therefore the maximum extractable work. Bengtsson *et al.* (Bengtsson *et al.*, 2018b) demonstrated that attractive interactions between bosons provide another route to enhancement by favoring particle clustering on one side of the partition, thereby increasing the average force imbalance and the extracted work. In contrast, repulsive interactions drive the engine toward fermionic behavior (Cai *et al.*, 2012). Remarkably, in the attractive regime the extracted work increases with particle number, whereas it saturates for noninteracting bosons and classical particles (Bengtsson *et al.*, 2018a,b).

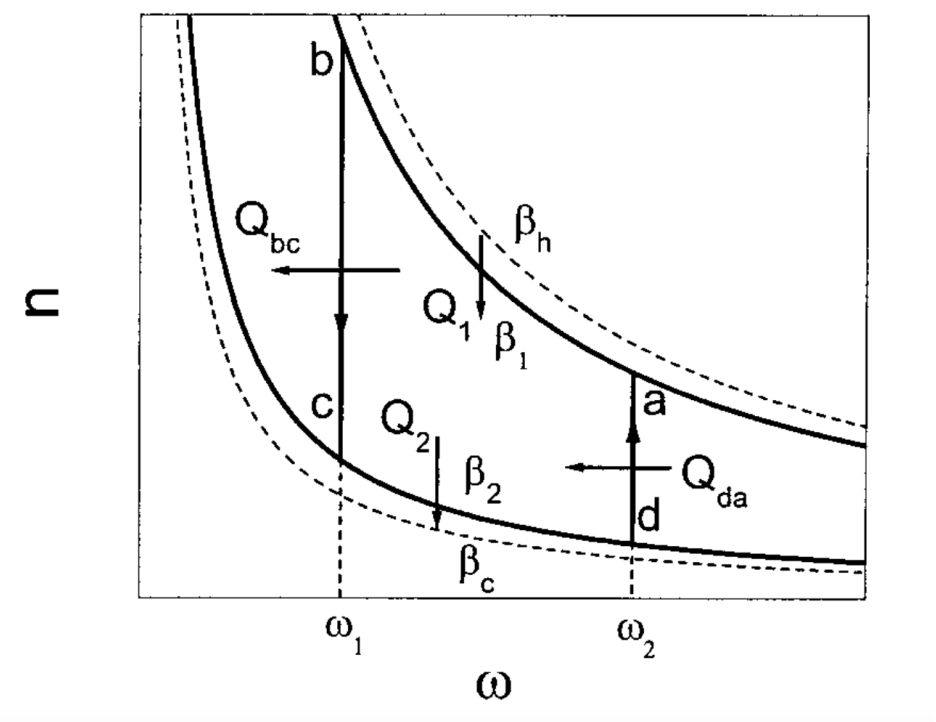

FIG. 26: The $n - \omega$ diagram of an irreversible quantum heat engine using harmonic oscillators as the working substance. Adapted from Ref. (Lin and Chen, 2003).

### 2. Quantum Stirling engine and indistinguishability

The Stirling cycle comprises two isothermal strokes at $T_h$ and $T_c$, connected by two isochoric (constant-volume) strokes. A quantum Stirling engine retains this four-stroke structure but employs a quantum working medium, such as a confined quantum gas or harmonic-oscillator modes (Lin and Chen, 2003; Wu *et al.*, 1998). The Stirling cycle, illustrated in Fig. 26, consists of two isothermal ($a \to b$ and $c \to d$)) and two isochoric strokes ($b \to c$ and $d \to a$). The oscillator frequency $\omega$ is varied quasistatically during the isothermal strokes and held fixed during the isochoric strokes, where the working medium thermalizes with the hot or cold reservoir with $\beta_h$ and $\beta_c$ inverse temperature.

The Hamiltonian of a single-mode harmonic oscillator is $\hat{H}_s = \omega(\hat{a}^\dagger \hat{a} \pm \frac{1}{2})$, where the + (−) sign corresponds to bosons (fermions), and $n = \langle \hat{a}^\dagger \hat{a} \rangle$ denotes the mean occupation number. The total work over one cycle by (Lin and Chen, 2003; Wu *et al.*, 1998)

$$W_{\rm tot} = \oint dW = -\oint (n \pm \frac{1}{2}) d\omega, \tag{65}$$

where the equilibrium occupation is $n = [e^{\beta\omega} \mp 1]^{-1}$ (Gupt *et al.*, 2021; Lin and Chen, 2003). Indistinguishability therefore enters through the quantum statistics governing the thermal occupation of the working medium. In the low-temperature quantum-degenerate regime ($\beta\omega \gg 1$), bosonic and fermionic engines become thermodynamically equivalent. By contrast, in the classical limit ($\beta\omega \ll 1$), the bosonic engine recovers the Carnot efficiency under perfect regeneration, whereas Pauli exclusion can prevent the fermionic engine from reaching the same limit (Gupt *et al.*, 2021).

Thomas *et al.* (Thomas *et al.*, 2019) embedded the quantum Szilárd engine into a Stirling cycle, providing a direct connection between indistinguishability and heat-engine performance. The cycle consists of two isothermal strokes, corresponding to the insertion and removal of the partition, connected by two isochoric thermalization strokes. In the low-temperature limit, it approaches a reversible engine with $W \approx k_B(T_h - T_c)\ln 2$ and $\eta \approx 1 - T_c/T_h$. Exchange statistics then determine the work output: for two particles, $W = 2k_B(T_h - T_c)\ln 2$ for distinguishable particles, $W = k_B(T_h - T_c)\ln 3$ for bosons, and $W = 0$ for fermions.

Realistic quantum engines operate in finite time and therefore at finite power, inevitably generating entropy that lowers the efficiency below the Carnot limit. Indeed, the Carnot bound is achievable only under reversible operation with vanishing entropy production, $\Sigma \to 0$, and is approached in practice only in the quasistatic limit, $\tau_{\rm cyc} \to \infty$ (Kosloff, 2013). Away from the quasistatic limit, the working medium is generally not in equilibrium throughout the cycle. During each stroke, heat exchange is limited by finite relaxation rates and is accompanied by nonzero entropy production. Since quantum statistics constrain the accessible occupations and transition pathways, indistinguishability can significantly modify these relaxation processes and, then, the finite-time performance of the engine.

This behavior has been analyzed for Stirling engines with harmonic-oscillator working media (Gupt *et al.*, 2021; Lin and Chen, 2003). In the quantum-degenerate regime, the effective heat transfer becomes nonlinear and model dependent, causing the efficiency at maximum power (EMP) to deviate from the Curzon–Ahlborn value, $\eta_{\rm EMP} = 1 - \sqrt{T_c/T_h}$. Differently, in the classical limit ($\beta\omega \ll 1$), heat transfer becomes effectively Newtonian and the Curzon–Ahlborn efficiency is recovered (Curzon and Ahlborn, 1975; Geva and Kosloff, 1992; Gupt *et al.*, 2021; Lin and Chen, 2003). More generally, whether energy is exchanged through single-particle or collective many-body dissipation determines the relaxation pathways and heat currents. For example, Eglinton *et al.* (Eglinton *et al.*, 2023) showed that Bose–Einstein condensation enhances collective heat transport, enabling Bose-gas engines to deliver substantially higher power at a given efficiency than an equivalent collection of independent single-particle engines.

### 3. Quantum Otto engine and indistinguishability

In the quantum Otto cycle (Kosloff, 1984), a control parameter, such as the energy-level spacing or trap frequency $\omega$, plays the role of the volume in the classical engine. As illustrated in Fig. 27, the cycle consists of two isentropic strokes, during which $\omega$ is varied while the working medium is isolated from the reservoirs, and two isochoric strokes, during which $\omega$ is held fixed while the working medium thermalizes with the hot and cold baths. The internal energy is $E = \mathrm{Tr}(\rho H)$, and the first law, $\mathrm{d}E = \delta Q + \delta W$ identifies work with changes in the Hamiltonian, $\delta W = \mathrm{Tr}(\rho\, \mathrm{d}H)$, and heat with changes in the quantum state, $\delta Q = \mathrm{Tr}(H\, \mathrm{d}\rho)$. Consequently, work is exchanged only during the isentropic strokes, whereas heat is exchanged only during the isochoric strokes.

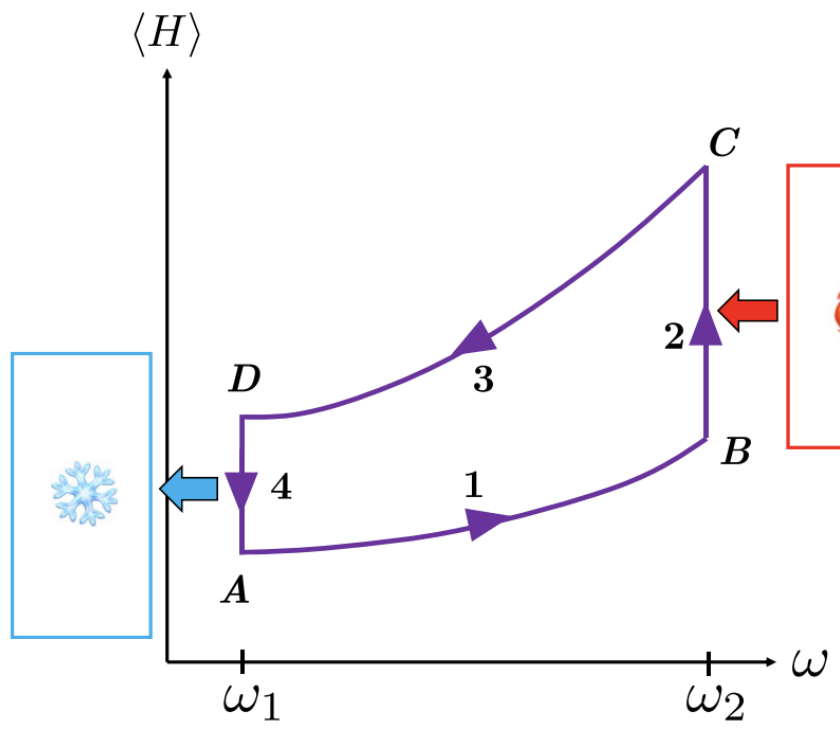

FIG. 27: The $E - \omega$ diagram of an quantum Otto engine using harmonic oscillators as the working substance. Adapted from Ref. (Myers and Deffner, 2020).

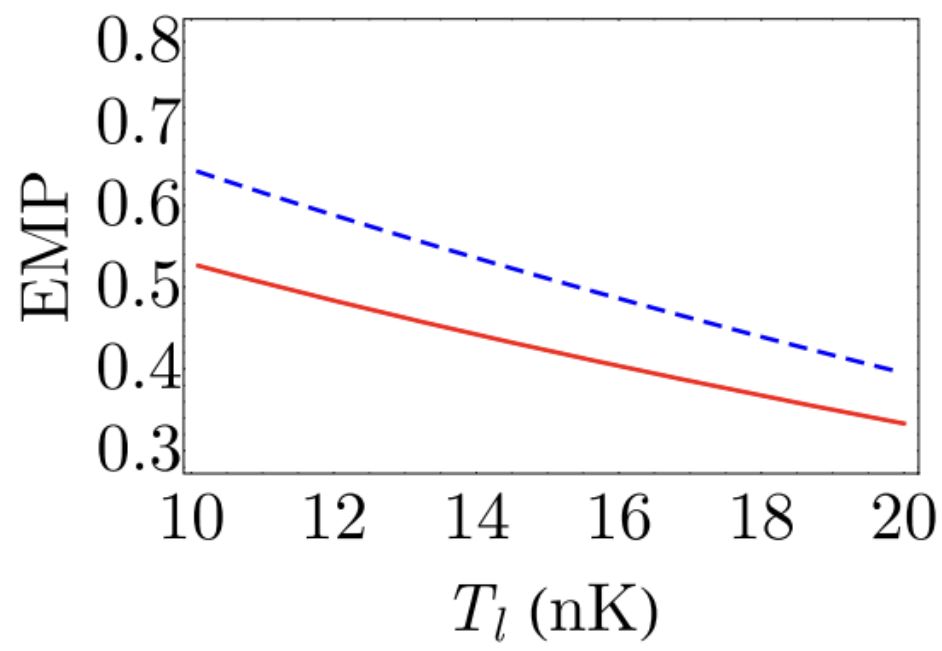

FIG. 28: EMP as a function of the cold thermal reservoir temperature for a BEC working medium (blue dashed line). In each plot the Curzon–Ahlborn efficiency (red solid line) is reported for comparison. Adapted from Ref. (Myers *et al.*, 2022).

The working medium of a quantum Otto engine can consist of $N$ noninteracting particles confined in a power-law trap,

$$H = \sum_{i=1}^{N} \frac{p_i^2}{2m} + \frac{1}{2} m\, \omega^a\, x_i^a, \tag{66}$$

where $\omega$ is the control parameter and $a = 2$ corresponds to the harmonic trap. Particle statistics enters through the $N$-particle partition function $Z^{(N)}$, from which the mean energy follows as $E = -\partial_\beta \ln Z^{(N)}$. Consequently, the work output depends on the occupation statistics of the working medium.

For $N$ distinguishable harmonic oscillators, the work is additive,

$$W_{\mathrm{dis}} = N(\omega_2 - \omega_1) \left[ \frac{1}{e^{\beta_h \omega_2} - 1} - \frac{1}{e^{\beta_c \omega_1} - 1} \right], \tag{67}$$

whereas for $N$ indistinguishable particles in the canonical ensemble,

$$W = (\omega_2 - \omega_1) \sum_{k=1}^{N} k \left[ \frac{1}{e^{\beta_h k \omega_2} - 1} - \frac{1}{e^{\beta_c k \omega_1} - 1} \right]. \tag{68}$$

For harmonic confinement, this expression is identical for bosons and fermions, but its dependence on $N$ differs from the distinguishable case: at low temperatures the work saturates because most particles occupy the lowest-energy modes, so adding further particles changes the population imbalance between the hot and cold strokes only marginally (Boubakour *et al.*, 2023). For nonharmonic power-law traps, the nonuniform level spacing allows bosonic and fermionic statistics to affect work output differently (Zheng and Poletti, 2015).

Myers and Deffner (Myers and Deffner, 2020) explicitly demonstrated that indistinguishability can enhance the finite-time performance of a quantum Otto engine with two noninteracting particles. By incorporating exchange symmetry through the (anti)symmetrization of both thermal state and driven harmonic-oscillator dynamics (Husimi, 1953; Myers and Deffner, 2020), they found that a bosonic working medium achieves a higher EMP than distinguishable or fermionic working media. Nevertheless, the efficiency remains below the Curzon–Ahlborn bound, $\eta_{\mathrm{EMP}} = 1 - \sqrt{T_c/T_h}$.

A natural question is whether indistinguishability can enhance the finite-time efficiency beyond the Curzon–Ahlborn value. Myers *et al.* (Myers *et al.*, 2022) addressed this question using an endoreversible Otto engine with an ideal Bose gas as the working medium. Endoreversible operation means that the trapped Bose gas remains in internal equilibrium, while heat exchange on the isochoric strokes is incomplete due to finite thermal conductance. Within the trap thermodynamic limit, the internal energy across the Bose–Einstein is

$$E(T, \omega) = \begin{cases} 3k_B T \left( \frac{k_B T}{\hbar \omega} \right)^3 g_4(1), & T \le T_c, \\ 3k_B T \left( \frac{k_B T}{\hbar \omega} \right)^3 g_4(z), & T > T_c, \end{cases} \tag{69}$$

where the critical temperature is $T_c = \left( \frac{\hbar \omega}{k_B} \right) \left[ \frac{N}{\zeta(3)} \right]^{1/3}$. Here $g_\nu(z) = \sum_{n=1}^{\infty} z^n / n^\nu$ is the Bose function, $z = \exp(\mu / k_B T)$ is the fugacity, and $\zeta(\nu)$ is the Riemann zeta function. Myers *et al.* (Myers *et al.*, 2022) found that the EMP recovers the Curzon–Ahlborn value above the condensation temperature, whereas operation in the condensed phase can exceed this benchmark (see Fig. 28). These results demonstrate that many-body quantum degeneracy alone, even in the absence of interactions, can enhance the finite-time performance of a quantum heat engine.

A natural question is whether this enhancement persists for finite systems with interactions. Jaramillo *et al.* (Jaramillo *et al.*, 2016) addressed this problem using $N$ identical bosons confined in a harmonic trap with inverse-square Calogero–

Sutherland interactions,

$$H = \sum_{i=1}^{N} \frac{p_i^2}{2m} + \frac{1}{2}m\omega^2 x_i^2 + \frac{1}{m}\sum_{i<j}\frac{\lambda(\lambda-1)}{(x_i - x_j)^2}, \qquad (70)$$

where the interaction strength $\lambda$ interpolates between an ideal Bose gas ($\lambda = 0$) and the fermionized regime ($\lambda = 1$). Jaramillo *et al.* (Jaramillo *et al.*, 2016) showed that, under highly nonadiabatic driving, many-body engines with weak interactions ($\lambda < 1/2$) can outperform an ensemble of $N$ independent single-particle engines, with the largest enhancement occurring for ideal bosons. This advantage gradually disappears as $\lambda$ approaches the fermionized regime.

Subsequent studies have shown that interaction-assisted enhancement is not universal. For harmonically trapped few-body Otto engines with tunable contact interactions, Boubakour *et al.* (Boubakour *et al.*, 2023) found that distinguishable particles can outperform indistinguishable ones in efficiency at maximum power and approach the Curzon–Ahlborn benchmark. Related Otto engines based on one-dimensional Bose gases have explored alternative interaction and coupling mechanisms to optimize finite-time performance (Boubakour *et al.*, 2023; Chen *et al.*, 2018; Estrada *et al.*, 2024; Nautiyal, 2025; Nautiyal *et al.*, 2024). In most of these works, the primary emphasis is on interaction and degeneracy engineering rather than on isolating indistinguishability as an independent thermodynamic resource. Collectively, these studies indicate that the thermodynamic advantage of indistinguishability depends sensitively on the interplay between particle statistics, interactions, and driving protocol.

Apart from trapped Bose gases, spin-based working media provide an alternative platform for exploring the thermodynamic consequences of indistinguishability. Watanabe *et al.* (Watanabe *et al.*, 2020) studied an Otto engine composed of $N$ identical two-level systems collectively coupled to hot and cold reservoirs and to a harmonic-oscillator load. The total Hamiltonian $H(t) = H_E(t) + H_S + H_B + H_{EB}(t) + H_{ES}(t)$ describes the collective working medium. The first term gives the controllable energy splitting $\omega(t)$ between the two internal states, while the second term, with coupling strength $\Delta$, coherently mixes them.

Watanabe *et al.* found that the energy transferred to the load is enhanced for indistinguishable bosonic resources, with the enhancement increasing with particle number before saturating. As shown in Fig. 29(a), the enhancement factor $\mathcal{E} = \Delta U_{\rm ind}/\Delta U_{\rm dis}$ increases with $N$, depends on the driving parameter $\theta_{t_1} = \tan^{-1}[\omega(t_1)/\Delta]$, and decreases with temperature. Importantly, the enhancement persists for $\theta_{t_1} = \pi/2$ ($\Delta = 0$), demonstrating that it originates from indistinguishability rather than interactions. By contrast, no comparable enhancement is found for fermions because Pauli exclusion suppresses the relevant collective transitions. The corresponding increase in ergotropy further confirms that indistinguishability enhances the useful work delivered to the load rather than merely increasing its internal energy.

A complementary perspective considers indistinguishability from the viewpoint of the thermal reservoir. When identical particles couple collectively to a common bath, the environment cannot distinguish which particle exchanges an excitation, enabling collective dissipation and bath-induced coherences (Funo and Tajima, 2025; Jaseem *et al.*, 2023; Latune *et al.*, 2019a,b). Such collective thermalization can increase the ergotropy transferred to a load, establishing a direct connection between indistinguishable working media and quantum batteries (Pokhrel and Gea-Banacloche, 2025; Watanabe *et al.*, 2020).

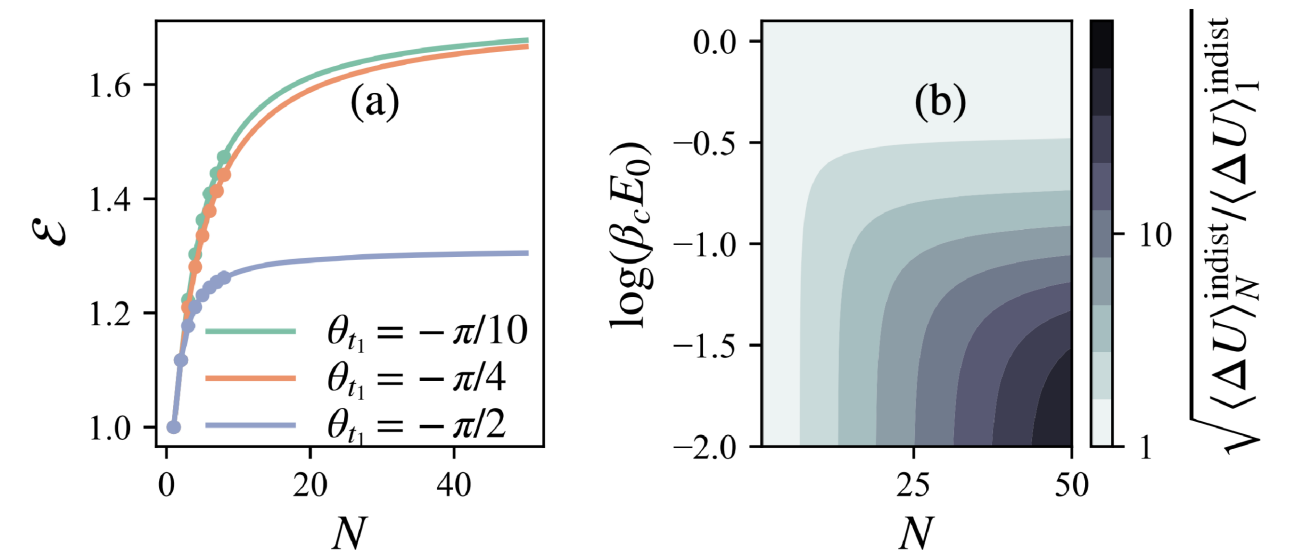

FIG. 29: (a) Enhancement factor $\mathcal{E} = \Delta U_{\rm ind}/\Delta U_{\rm dis}$ as a function of the number of resources $N$ for different driving parameters $\theta_{t_1} = \tan^{-1}[\omega(t_1)/\Delta] \in \{-\pi/4, -\pi/10, -\pi/2\}$. (b) $\sqrt{\Delta U_{\rm ind}/\Delta U_{\rm dis}}$ as a function of $N$ and the cold inverse temperature on a logarithmic scale, $\log(E_0\beta_c)$, with $E_0\beta_h = 1/4$. Adapted from Ref. (Watanabe *et al.*, 2020).

Beyond the Bose–Fermi dichotomy, one may also consider working media obeying generalized exchange statistics. Myers *et al.* (Myers and Deffner, 2021) investigated Otto engines with anyonic working media. For statistical anyons, the efficiency at maximum power is independent of the exclusion parameter, although indistinguishable working media outperform distinguishable ones for sufficiently small temperature ratios. Differently, for topological anyons confined in a two-dimensional harmonic trap, the spectrum depends explicitly on the anyonic phase, leading to a phase-dependent efficiency at maximum power that can exceed the bosonic value in appropriate regimes.

### 4. Particle-statistics-driven engines

Inspired by the quantum Otto cycle, Koch *et al.* (Koch *et al.*, 2023) experimentally realized a Pauli engine using a quantum-degenerate two-component $^6$Li gas confined in an opto-magnetic trap. As illustrated in Fig. 30, work is performed by adiabatically varying the trap frequency, while the interaction strength is tuned across a broad Feshbach resonance to drive the BCS–BEC crossover. Rather than exchanging heat with thermal reservoirs, the interaction stroke changes the internal energy of the working medium through its quantum statistics. Accordingly, the efficiency is defined as $\eta = W_{\rm ext}/\Delta E_{\rm stat}$, where $\Delta E_{\rm stat}$ is the interaction-induced increase in internal energy. They observed statistics-driven free-energy conversion over repeated cycles, with both the extracted work and the efficiency increasing with the compression ratio, and the work scaling as $W \propto N^{1.73(18)}$, distinctly different from the classical noninteracting prediction.

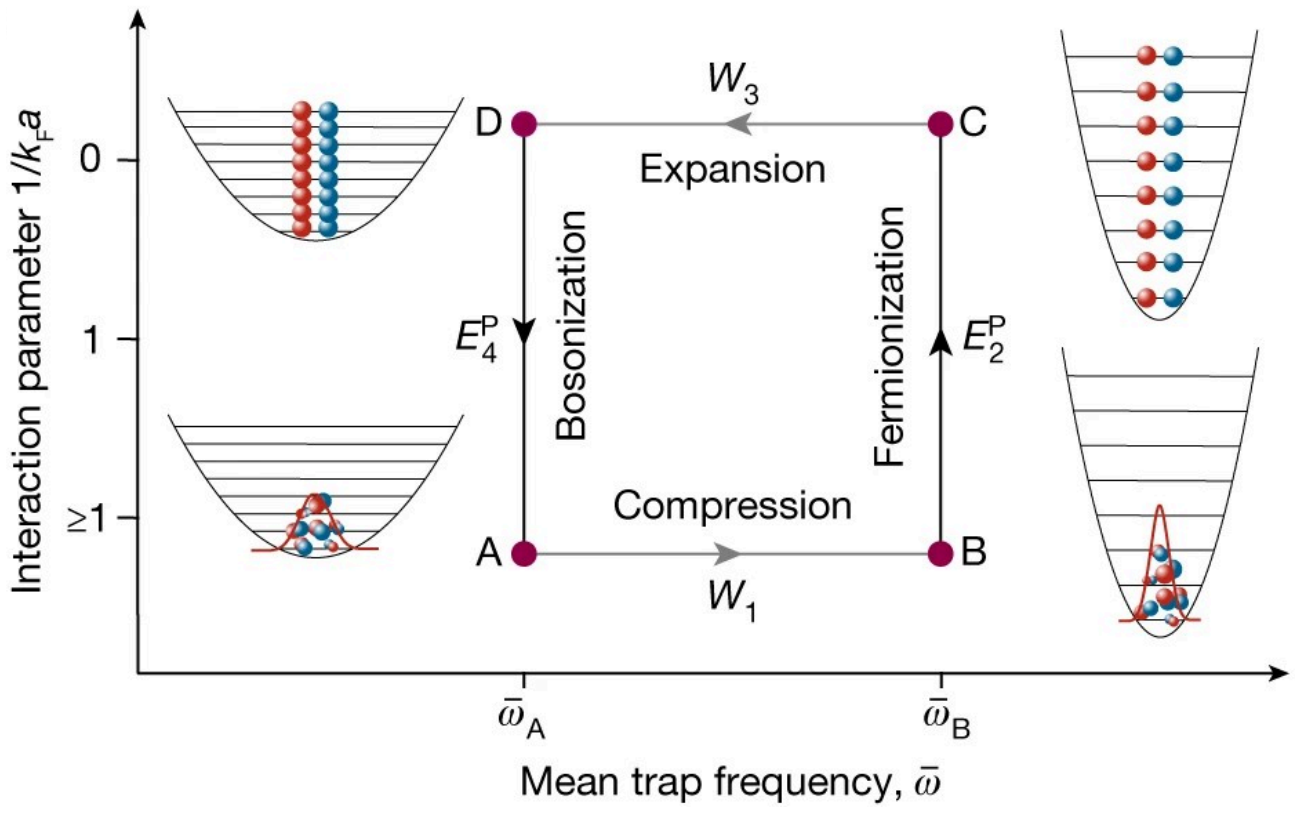

FIG. 30: Cycle of the Pauli engine. Adiabatic modulation of the trap frequency provides the work strokes ($A \to B$ and $C \to D$), while tuning the interaction parameter $(k_F a)^{-1}$ (with $k_F$ the Fermi wavevector magnitude and $a$ the $s$-wave scattering length) drives the BCS–BEC crossover and replaces heat exchange through statistics-control strokes ($B \to C$ and $D \to A$). Adapted from Ref. (Koch *et al.*, 2023).

The absence of thermal reservoirs in the Pauli engine has motivated alternative engine paradigms driven purely by particle statistics. Menon *et al.* (Menon *et al.*, 2025) investigated this idea using the one-dimensional Lieb–Liniger model,

$$H = \sum_{i=1}^{N} \frac{p_i^2}{2m} + c \sum_{i<j} \delta(x_i - x_j), \tag{71}$$

where $c$ denotes the contact-interaction strength. The model interpolates continuously between an ideal Bose gas ($c = 0$) and the Tonks–Girardeau limit ($c \to \infty$), where strongly interacting bosons acquire fermionic thermodynamic properties. Exploiting this crossover, they identified operating regimes with enhanced finite-time performance approaching the Curzon–Ahlborn benchmark.

Going beyond the Bose–Fermi dichotomy, Dunlop *et al.* (Dunlop *et al.*, 2025) proposed heat-engine cycles based on anyonic working media, where controlled bosonization and fermionization replace the conventional compression and expansion strokes. Their results suggest that continuously tunable exchange statistics provide an additional resource for optimizing quantum-engine performance.

## VII. CONCLUSIONS AND OUTLOOK

In this *Colloquium*, we have reviewed how particle identity, indistinguishability, exchange statistics, and many-particle interference can be harnessed as quantum resources. Building on the theoretical foundations of entanglement and indistinguishability for identical particles (Sec. II), we argued that indistinguishability should not be viewed merely as a kinematic consequence of the symmetrization postulate, but as an operational resource whose role depends on the state preparation, interference, measurements, and information accessible within a given protocol. A central message is that there is no unique notion of indistinguishability. Historically, the Hong–Ou–Mandel (HOM) experiment established it through two-particle interference and its visibility (Bouchard *et al.*, 2021; Hong *et al.*, 1987), while later developments extended this concept to genuine many-body interference and indistinguishability witnesses (Brod *et al.*, 2019b; Englbrecht *et al.*, 2024; Pont *et al.*, 2022). In parallel, the sLOCC framework associated indistinguishability with the absence of accessible which-way (or, more generally, which-pseudospin) information under localized measurements (Lo Franco and Compagno, 2018; Nosrati *et al.*, 2020; Piccolini, 2024), whereas path-identity protocols engineer indistinguishability directly at the source by eliminating which-path information before interference occurs (Hochrainer *et al.*, 2022; Krenn *et al.*, 2017b). More broadly, indistinguishability has also been formulated within contextuality-based approaches (Asadian and Cabello, 2022; Srikumar *et al.*, 2026). These notions are not competing definitions but complementary operational characterizations of the same physical phenomenon.

As discussed in Sec. III, these operational notions of indistinguishability provide complementary mechanisms for generating and distilling coherence and entanglement through both postselected and heralded protocols. HOM interference and its multiphoton generalizations generate entanglement through two- and many-particle interference at mode-mixing elements and, more recently, have enabled the distillation of photon indistinguishability itself. By contrast, the sLOCC framework activates and distill coherence and entanglement from spatial overlap through localized measurements, whereas path-identity protocols generate coherence directly at the source by eliminating which-path information before interference occurs. Postselected protocols offer favorable resource scaling, robustness to loss and noise, and natural compatibility with multi-rail photonic encodings, but the generated resource is consumed during verification. Heralded protocols overcome this limitation by producing verified entangled states that can be directly incorporated into larger quantum architectures, albeit at the cost of increased experimental complexity and greater sensitivity to loss and partial distinguishability.

Continued advances in deterministic single-photon sources, integrated photonics, high-efficiency detectors, and coherent control of matter platforms are expected to make indistinguishability-based protocols increasingly practical. An important challenge is to develop precise scalable architectures that combine robustness of postselected schemes with on-demand capability of heralded protocols. Such advances would enable large-scale quantum networks, fault-tolerant photonic quantum computing, and hybrid architectures using indistinguishability as an active quantum resource.

Sections IV and V were devoted to the operational applications of identical particles in quantum metrology and quantum communication. HOM interference and its multiphoton generalizations provide the canonical mechanism for quantum-enhanced metrology, enabling phase estimation, imaging, sensing, state discrimination, quantum cloning, and boson sampling through two- and many-particle interference. The sLOCC framework exploits spatial indistinguishability to activate coherence and entanglement, providing the resources required for phase discrimination, quantum teleportation, entanglement swapping, and related communication protocols. The path-identity approach instead engineers indistinguishability directly at the source by eliminating which-path information, enabling quantum imaging and spectroscopy using undetected photons. Across these applications, the achievable advantage depends on the available state preparation, measurements, and superselection constraints, proving that indistinguishability is fundamentally a task-dependent operational resource rather than an intrinsic property of the quantum state.

TABLE I: Compact comparison of three representative entanglement-generation methods relevant to identical particle quantum information processing: sLOCC framework, HOM two-particle interference, and path identity approach.

| Aspect | sLOCC framework | HOM interference | Path identity |
|---|---|---|---|
| **Mechanism** | Generates operational entanglement from identical particle states through spatial overlap combined with spatially localized operations and classical communication (sLOCC). | Generates two-particle correlations by interfering indistinguishable particles on a mode-mixing element, producing bunching or anti-bunching. | Generates coherent superpositions by engineering multiple sources to share identical output paths, such that no which-path information is available. |
| **Platforms** | Applicable to massive identical particles and photons with controllable spatial overlap; suitable for both bosons and fermions. | Applicable to photons and a variety of quantum carriers (neutral atoms, phonons, electrons, plasmons, etc.) where two-particle interference has been demonstrated. | Primarily implemented with photonic SPDC and multi-source interferometry; extendable to architectures requiring structured multi-path coherence. |
| **Key resources** | Precise spatial-mode engineering, local measurement modules, and classical communication. | A mode-mixing element together with control of all degrees of freedom required to render the particles indistinguishable. | Phase-stable multi-source interferometry that yields identical output modes, supported by high-quality source engineering. |
| **Scalability** | Generally applicable to identical-particle systems, though scaling requires increasingly complex spatial control. | Widely usable as a building block, with scalability limited by loss and probabilistic multi-particle interference. | Supports high-dimensional and multi-photon entanglement, with scalability set by source number and phase stability. |
| **Role of indistinguishability** | Spatial indistinguishability acts as the fundamental resource; partial overlap can still produce extractable entanglement. | Complete indistinguishability across relevant degrees of freedom is essential for interference visibility. | Indistinguishability is imposed at the source level by design, preventing the emergence of which-path information. |
| **Postselection** | Not intrinsically required; sLOCC procedures enable entanglement extraction and verification without destructive post-selection. | Often relies on conditional detection in photonic systems; other platforms may use analogous heralding strategies. | Can avoid post-selection when path identity is fully enforced, though some implementations still employ conditional detection. |
| **Primary advantages** | Provides a direct route to harness entanglement rooted in the symmetrization of identical particles; readily suited to massive-particle platforms where beam-splitter interference is challenging. | Conceptually simple, experimentally mature, and a standard tool for probing indistinguishability and generating entanglement. | Enables complex, high-dimensional, and multi-photon entanglement using minimal resources, with indistinguishability intrinsically incorporated. |
| **Primary limitations** | Requires accurate control of spatial-mode overlap; scaling to many particles is experimentally demanding. | Scalability is hindered by probabilistic photon generation and detector limitations. | Sensitive to phase fluctuations between multiple sources; deterministic operation depends on stringent stability. |
| **Representative applications** | Entanglement extraction from identical particle states; entanglement swapping with massive particles; entanglement distillation; robust coherence generation and phase discrimination; quantum teleportation; sLOCC-based quantum information protocols. | Entanglement generation and Bell-state projection; linear-optical quantum computing; quantum-enhanced imaging and metrology; optimal quantum cloning; entanglement swapping; enhanced performance of finite-time quantum thermal machines; characterization and distillation of single-photon sources; boson sampling. | High-dimensional entanglement generation; quantum imaging and spectroscopy with undetected photons; induced-coherence interferometry; indistinguishability distillation; multi-source quantum interference; quantum information based on localized operations and classical communication. |

In Sec. VI, we recalled how indistinguishability of identical particles underlies the modern resolution of the Gibbs mixing paradox. Classical treatments invoke an idealized semipermeable membrane that perfectly discriminates particle species and therefore always predicts a nonzero entropy of mixing. However, quantum mechanics precludes perfect discrimination once the constituent quantum states overlap in all degrees of freedom, causing the entropy of mixing to decrease continuously and ultimately vanish. In this sense, indistinguishability is an operational notion determined by the information accessible to physically realizable measurements rather than by particle identity alone.

We discussed in Sec. VI that the thermodynamic consequences of indistinguishability are operational and strongly design-dependent. Observable effects arise only when indistinguishability becomes experimentally accessible, typically for bosonic working media operating at low temperatures, where occupation is restricted to the ground state or a small manifold of low-lying states together with (near-)degenerate energy spectra. Under these conditions, indistinguishability manifests through exchange statistics, including bosonic bunching and fermionic antibunching, as well as through collective many-body state of the working medium, whose partition function is determined by the particle statistics. Indistinguishability effects can further reshape thermalization pathways, fluctuations, and transport, thereby affecting work extraction, power, and efficiency without altering the fundamental Carnot bound. Appropriately engineered bosonic working media can, in some cases, surpass the Curzon–Ahlborn benchmark under finite-time operation. Beyond performance enhancement, indistinguishability also enables qualitatively new engine architectures, such as particle-statistics–driven engines, where tuning exchange statistics replaces conventional heat-exchange strokes and can reduce bath-induced dissipation and quantum friction during finite-time driving.

Through Secs. III–VI, we have shown that the operational role of indistinguishability extends well beyond particle symmetrization. Depending on the physical platform and task, indistinguishability arises through spatial overlap, multiparticle interference, path identity, or exchange statistics, and is exploited through appropriate measurements within a given protocol. These operational paradigms enable a broad spectrum of applications in entanglement generation, quantum metrology, communication, computation, and thermodynamics. The three operational paradigms summarized in Table I— sLOCC framework, HOM interference, and path identity approach—provide complementary realizations of indistinguishability as a quantum resource.

In prospect, an important open question is whether indistinguishability provides a genuine resource-theoretic advantage or its benefits remain restricted to carefully engineered architectures that convert indistinguishability into enhanced work extraction, cooling power, or charging speed. Developing an operational framework that treats indistinguishability as a controllable resource, equipped with measurable monotones directly connected to efficiency, output power, work, and cooling performance, could guide the design of next-generation quantum batteries, refrigerators, and hybrid thermal machines.

Ensembles of identical particles represent versatile systems for emerging quantum technologies, several of which have been reviewed in this *Colloquium*. It is worth mentioning that many-particle interference also enables quantum sampling problems believed to be intractable for classical computers (Aaronson and Arkhipov, 2011; Zhong *et al.*, 2020). For comprehensive discussions of quantum random sampling and its computational complexity, which relies on coherent interference of many indistinguishable photons, we refer the reader to Refs. (Brod *et al.*, 2019a; Hangleiter and Eisert, 2023). Recent fermionic analogs have shown that many-fermion interference, supplemented by suitable non-Gaussian resource states, can likewise provide quantum computational advantage (Hebenstreit *et al.*, 2019; Oszmaniec *et al.*, 2022). In both cases, partial distinguishability modifies the output statistics and degrades the expected advantage, motivating the advance of operational methods to quantify and certify genuine many-particle indistinguishability.

Developing a unified operational framework that quantitatively connects indistinguishability and exchange statistics to task-specific advantages remains an important goal. Progress in the preparation, control, and detection of identical particles will be essential for translating such a framework into scalable architectures, thereby broadening its applications across quantum computation, communication, and sensing. Extending these protocols from photonic systems to ultracold atoms, electrons, ions, phonons, and hybrid platforms will help identify universal advantages and enable architectures combining the complementary properties of different quantum carriers. Furthermore, controlled bosonic, fermionic, and anyonic statistics may also inspire new approaches to quantum simulation and to the design of quantum engines, refrigerators, and batteries. These developments could ultimately establish indistinguishability not merely as a fundamental consequence of particle identity, but as a controllable and exploitable resource for quantum information processing.

## Acknowledgments

R.L.F. acknowledges support by MUR (Ministero dell'Università e della Ricerca) through the following projects: PNRR Project ICON-Q – Partenariato Esteso NQSTI – PE00000023 – Spoke 2 – CUP: J13C22000680006, PNRR Project QUANTIP – Partenariato Esteso NQSTI – PE00000023 – Spoke 9 – CUP: E63C22002180006. J.S.X and C.F.L acknowledge the Innovation Program for Quantum Science and Technology (No. 2021ZD0301400 and No. 2021ZD0301200), the National Natural Science Foundation of China (No. 92365205 and No. W2411001), the USTC Major Frontier Research Program (No. LS2030000002) and the CAS Project for Young Scientists in Basic Research (YSBR-131). All the authors would like to thank Gabriele De Chiara, Mario Krenn, and Jyrki Piilo for fruitful feedback and insights.